\documentclass[journal,12pt,draftclsnofoot,onecolumn,english]{IEEEtran}

\usepackage[dvipsnames]{xcolor}
\usepackage{epsfig}
\usepackage{times}
\usepackage{float}
\usepackage{afterpage}
\usepackage{amsmath}
\usepackage{amstext,cite}
\usepackage{amssymb,bm}
\usepackage{latexsym}
\usepackage{color}
\usepackage{graphicx}
\usepackage{amsmath}
\usepackage{amsthm}
\usepackage{graphicx}
\usepackage[center]{caption}
\usepackage{pstricks}
\usepackage{caption}
\usepackage{subcaption} 
\usepackage{booktabs}
\usepackage{multicol}
\usepackage{lipsum}%
\usepackage[T1]{fontenc}
\usepackage{hyperref}
\usepackage{aecompl}
\usepackage{mathrsfs}
\usepackage{multirow}

\usepackage{tikz}
\usetikzlibrary{shapes.multipart}
\usetikzlibrary{fit}
\usetikzlibrary{shapes.geometric,positioning}
\usetikzlibrary{calc}
\usetikzlibrary{patterns.meta}
\usepackage{pgfplots}
\usetikzlibrary{plotmarks}
\usetikzlibrary{spy}

\usepackage{soul}
\usepackage{cancel}

\allowdisplaybreaks

\long\def\comment#1{}

\newcommand{\av}{{\mathbf a}}
\newcommand{\bv}{{\mathbf b}}

\newcommand{\dv}{{\mathbf d}}
\newcommand{\ev}{{\mathbf e}}

\newcommand{\gv}{{\mathbf g}}

\newcommand{\pv}{{\mathbf p}}
\newcommand{\qv}{{\mathbf q}}

\newcommand{\vv}{{\mathbf v}}
\newcommand{\xv}{{\mathbf x}}

\newcommand{\Am}{{\mathbf A}}
\newcommand{\Bm}{{\mathbf B}}

\newcommand{\Dm}{{\mathbf D}}
\newcommand{\Em}{{\mathbf E}}

\newcommand{\Gm}{{\mathbf G}}

\newcommand{\Id}{{\mathbf I}}

\newcommand{\Mm}{{\mathbf M}}

\newcommand{\Pm}{{\mathbf P}}
\newcommand{\Qm}{{\mathbf Q}}

\newcommand{\Ac}{{\mathcal A}}
\newcommand{\Bc}{{\mathcal B}}

\newcommand{\Gc}{{\mathcal G}}

\newcommand{\Ic}{{\mathcal I}}

\newcommand{\Lc}{{\mathcal L}}

\newcommand{\Pc}{{\mathcal P}}
\newcommand{\Qc}{{\mathcal Q}}

\newcommand{\Sc}{{\mathcal S}}
\newcommand{\Tc}{{\mathcal T}}

\newcommand{\Wc}{{\mathcal W}}

\newcommand{\qsf}{{\mathsf q}}
\newcommand{\rsf}{{\mathsf r}}

\newcommand{\Bsf}{{\mathsf B}}

\newcommand{\Lsf}{{\mathsf L}}
\newcommand{\Msf}{{\mathsf M}}
\newcommand{\Nsf}{{\mathsf N}}

\newcommand{\Rsf}{{\mathsf R}}

\newcommand{\Tbp}{{\mathrm{T}}}

\theoremstyle{definition}
\newtheorem*{rem*}{Remark}

\theoremstyle{plain}
\newtheorem{thm}{Theorem}[section]

\newtheorem{lem}{Lemma}

\newtheorem{rem}{Remark}

\providecommand{\definitionname}{Definition}

\newcommand{\Ksf}{{\mathsf K}}
\newcommand{\KsfActive}{{\magenta \mathsf A}}%

\title{On Coded Caching Systems with Offline Users, with and without Demand Privacy against Colluding Users}

\begin{document}

\author{
\IEEEauthorblockN{Yinbin Ma and Daniela Tuninetti \\}
\IEEEauthorblockA{University of Illinois Chicago, Chicago, IL 60607, USA \\ Email:\{yma52, danielat\}@uic.edu}
\thanks{The authors are with the Electrical and Computer Engineering Department of the University of Illinois Chicago, Chicago, IL 60607, USA (e-mail: yma52@uic.edu, danielat@uic.edu). 
Parts of this paper were presented in the 2022~\cite{ma2022coded} and in the 2023~\cite{ma2023demand} IEEE International Symposium on Information Theory (ISIT). 
This work was supported in part by NSF Awards 1910309 and 2312229.}
}
\maketitle

\IEEEpeerreviewmaketitle

\begin{abstract}
Coded caching is a technique that %
leverages locally cached contents at the end users to reduce the network's peak-time communication load.  
Coded caching has been shown to achieve significant performance gains compared to uncoded %
schemes and is thus considered a promising technique to boost performance in future networks by effectively trading off bandwidth for storage.
The original coded caching model introduced by Maddah-Ali and Niesen does not consider the case where some users involved in the placement phase, may be offline %
during the delivery phase. 
If so, the delivery may not start or it may be wasteful to perform the delivery with fictitious demands for the offline users. In addition, the active users may require their demand to be kept private.
This paper formally defines a coded caching system where some users are offline, and investigates the optimal performance with and without demand privacy against colluding users.
For this novel coded caching model with offline users, achievable and converse bounds are proposed. 
These bounds are shown to meet under certain conditions, and otherwise to be to within a constant multiplicative gap of one another.
In addition, the proposed achievable schemes have lower subpacketization and  lower load compared to baseline schemes (that trivially extend known schemes so as to accommodate for privacy) in some memory regimes. 
\end{abstract}

\begin{IEEEkeywords}
Coded caching with offline users;
Achievablity;
Demand privacy; Colluding users;
Optimality for small memory size;
Multiplicative constant gap.
\end{IEEEkeywords}

\section{Introduction}
\label{sec:intro}

Coded caching, first introduced by Maddah-Ali and Niesen in~\cite{maddah2014fundamental}, leverages locally cached contents at the users to reduce the communication load during peak-traffic times. A coded caching system has two phases. During the cache placement phase, the server populates the users' local caches, without knowing the users' future demands. During the delivery phase, the server broadcasts coded multicast messages to satisfy the users' demands. It is a promising technique as Maddah-Ali and Niesen show in~\cite{maddah2014fundamental} that a coded delivery reduce communication load significantly with a well-designed placement strategy compared to uncoded delivery.

The achievable scheme proposed in~\cite{maddah2014fundamental} (referred to as MAN in the following) has a combinatorial {\it uncoded cache placement} phase\footnote{Uncoded cache placement means that bits of the files are directly copied into the caches without any coding.} and a network coded delivery phase. In~\cite{yu2017exact}, an improved delivery was proposed (referred to as YMA in the following), which improves on the MAN delivery by removing those linearly dependent multicast messages that may occur when a file is requested by multiple users. The scheme with MAN placement with YMA delivery matches the converse bound derived in~\cite{wan2020index} under the constraint of uncoded placement~\cite{yu2017exact}; and it is otherwise order optimal to within a factor of two if no restrictions are imposed on the placement phase for the sake of deriving a converse bound~\cite{yu2018characterizing}. 
Wan {\em et al.} in~\cite{wan2020fundamental} extended the single-file retrieval model in~\cite{maddah2014fundamental} so as to include Scalar Linear Function Retrieval (SLFR), where users can demand a linear combination of the files. 
By generalizing the YMA scheme,~\cite{wan2020fundamental} proved that SLFR attains the same load as for single-file retrieval under uncoded placement. A general construction for SLFR with uncoded placement and linear coding for the delivery can be found in~\cite{ma2021general}.

For the the original coded caching model~\cite{maddah2014fundamental}, coded placement is known not only to strictly improve performance compared to uncoded placement, but to be exactly optimal in some regimes:
\cite{chen2016fundamental} shows how to achieve the cut-set bound in the small memory regime when there are more users than files; 
\cite{tian2018caching} proposes a scheme based on interference elimination for the case of more users than files;
\cite{gomez2018fundamental} shows an improved performance compared to~\cite{tian2018caching} in small memory regime when more users than files;
\cite{tian2018symmetry} derives the optimal performance for the case of two users (and any number of files), and a partial characterization for the case of two files (and any number of users);
\cite{ma2023coded} extends beyond~\cite{chen2016fundamental} the memory regime over which the optimal coded placement strategy is known, but it is restricted to the case when the number users and the number files are the same.

For converse bounds without any restrictions on the placement phase and that improve on the cut-set bound in~\cite{maddah2014fundamental}, in addition to those in~\cite{yu2018characterizing}, %
\cite{wang2018improved} proposes a converse bound where the worst-case and average-case loads are within a multiplicative gap of 2.315 and 2.507 to the YMA scheme respectively;
we note that~\cite{sengupta2017improved} derived a converse bound tighter than cut-set bound, and 
\cite{tian2018symmetry} proposed a computer-aided methodology to derive converse bounds based on Shannon-type inequalities (which is unfortunately limited to small instances of the problem, as the computational complexity of the resulting linear program scales doubly exponentially with the number of users).

A limitation of the above mentioned coded caching schemes, including the YMA scheme, is that users need to know all the demands in order to decode the transmitted multicast messages. This may infringe on user's privacy. Private Information Retrieval (PIR)~\cite{chor1995private} addresses the demand-privacy issue against multiple servers in a single-user system. Here we are interested in protecting {\it demand privacy} of each user against the remaining users. In~\cite{wan2020coded}, the demand privacy constraint was first introduced in the classical coded caching model.
Demand-private schemes include the {\it virtual-users} scheme in~\cite{engelmann2017content,kamath2019demand}, in which the server pretends to serve many more users than those actually present in the system\footnote{A construction of a demand-private scheme for a system with $K$ users and $N$ files is given in~\cite{kamath2020demand} based on a non-private scheme for a system with $KN$ users and $N$ files, assuming that users have local randomness available to them.}. 
In~\cite{wan2020coded}, %
the virtual-users scheme is shown to be order optimal to within a factor of 8;
in addition, a novel scheme that uses Maximum Distance Separable (MDS) codes for the placement phase is proved to be order optimal to within a factor of 2 for large memory regime.
In~\cite{kamath2020demand}, the exact demand-private memory-load tradeoff for 2~users and 2~files was found, where the virtual-users scheme with restricted demand type (see also Remark~\ref{rem:VUidea}) turns out to be optimal in the low memory regime, and the MDS-based scheme from~\cite{wan2020coded} in the large memory regime.
In~\cite{namboodiri2021optimal}, the virtual-users scheme with restricted demand type from~\cite{kamath2020demand} is generalized and shown to give an optimal tradeoff point in the very small memory regime, that is, matching the converse bound from~\cite{yan2021fundamental}.
Another demand-private scheme is the {\it privacy-keys} scheme in~\cite{yan2021fundamental}, where keys are pre-stored in the caches and used to `confuse' users so as to prevent them from gaining any knowledge about the demands of the remaining users; here placement is coded, with a linear code.
The virtual-users and privacy-keys schemes guarantee privacy even against colluding users, as defined in~\cite{yan2021fundamental} for the classical setting, and in~\cite{yapar2019optimality} for the Device-to-Device settings where the notion of colluding users was first considered.

Another practical concern in coded caching is that all users present during the placement phase must be active and synchronously send their demand to the server  before the delivery phase starts. %
In practical scenarios, some users may fail to send their demand to the server due to technical failures, or may have nothing to demand. The {\it asynchronous-demands} setting, already discussed in~\cite{maddah2014fundamental}, allows the server to start transmission as soon as the first demand arrives. Known schemes (as in~\cite{ghasemi2020asynchronous}, and references therein) however assume that all demands eventually arrive in finite time, otherwise the system fails to complete the delivery, i.e., the problem definition requires that all users present during the placement successfully decode their demanded file. We find the latter assumption too restrictive and thus propose a model where only the demands of the active users must be satisfied.
\cite{abolpour2023cache} further investigates the cache-aided multi-input single-output (MISO) model with dynamic user behavior, and maximize the degrees-of-freedom beyond the multiplexing gain by exploiting the global caching gain from coded caching. In their model, users can enter or leave the network at their will. During the placement phase each user is assigned a profile and all users with the same profile are populated the same cache content. In the delivery phase, the server transmits multicast and unicast messages together, depending on the presence of user profile. The drawbacks are 1) only uncoding placement is considered; 2) when the extreme case happens, such that all active users present with the same profile, only unicast messages are sent from server, and no performance gain from coded caching anymore.

{\bf The focus of this paper is on coded caching system with offline users where the demands of the active users need to be kept private against colluding users.} 
Performance with offline users was investigated in~\cite{liao2021fundamental}, where a scheme was proposed under the assumption that the number of active users is fixed and known; and in~\cite{yapar2019optimality} for the Device-to-Device model with random user activity, where an `outage event' (i.e., some active users may fail to decode) was allowed. 
In our work, we assume that the demands of the offline users never arrive (or arrive with infinite delay) and the demands of the remaining active users arrive synchronously; we do not allow for outages. We refer to this setting as {\it hotplug} coded caching\footnote{Hotplug is a computer system term that refers to a device that can be added or removed from the running system without having to restart the system.}.
We derive schemes for the hotplug setting with and without demand-privacy constraints for the active users. 
We shall see that a way to deal with offline users is to carefully design coded placement strategies based on MDS codes. We shall also show how to compose various linear codes so as to achieve both resiliency against offline users while guaranteeing demand privacy at the same time. {\bf In summary, to tackle various issues in coded caching system (such as offline users and privacy), we propose coded placement as a potential ``One Ring''\footnote{One Ring is among the most powerful artifacts in {\it The Lord of the Rings}.} solution.}

\subsection{Contributions}
In this paper, we show that coded placement not only benefits the load performance, but also preserves some advanced features including serving offline users and guaranteeing demand privacy.
We first formalize the hotplug coded caching problem, without or with demand privacy against colluding users. 
To tackle offline users, we propose three new schemes that allow the demands of the active users to be satisfied regardless of the set of offline users, where the number (but not the identity) of the offline users is assumed known at the time of placement, similarly to~\cite{liao2021fundamental}\footnote{The model is motivated by the law of large numbers for the case where the number of users, $K$, is large and each user is active with the same probability, $p$, independently at random; in such setting the number of active users does not fluctuate much compared to the average value $Kp$.
For the setting where the number of users is random, the solutions proposed in this paper can be used over multiple delivery rounds. 
}. Our schemes use coded cache placement.

For the case without privacy constraint, we show the following. 
\begin{enumerate}
\item
Our first new achievable scheme in Theorem~\ref{thm:HT1} exploits MDS codes in the placement phase and follows SLFR-fashion delivery phase, where coding is done within each file but not across files. 
We show that it achieves the optimal performance when the memory is small and the number of files is large. The matching converse is obtained from~\cite{yu2018characterizing}.
\item
Our second new achievable scheme in Theorem~\ref{thm:HT2} applies MDS-like coding to the coded placement of~\cite{chen2016fundamental}. This scheme achieves the optimal performance when the memory is small and there are less files than users. The matching converse is obtained from the cut-set bound in~\cite{maddah2014fundamental}.
\item
Our third new achievable scheme in Theorem~\ref{thm:HT3} also utilizes MDS codes in the placement phase. Unlike the first scheme, the third one needs large cache size to guarantee that non-trivial multicast messages exists. For each user in the set of users who benefit from a given multicast message, said user obtains additional `degrees of freedom' not present in its cache about the desired file, while the interference from non-demanded files can be removed thanks to the cached content.
The matching converse is obtained is the cut-set bound in~\cite{maddah2014fundamental}.
\end{enumerate}
For the memory regimes where we cannot prove exact optimality, we prove a constant gap result of two. 

To tackle the problem of demand privacy for the hotplug model, we adopt both the privacy-keys idea~\cite{yan2021fundamental} into our first and third schemes for the hotplug model without privacy (see Theorem~\ref{thm:HT1+PK} and Theorem~\ref{thm:HT3+PK}), as well as the virtual-users idea~\cite{kamath2020demand}. Regarding the virtual-users idea, our contribution is to extend the proof of~\cite{kamath2020demand} from the classical to the hotplug model; based on this, Theorem~\ref{thm:HT+VU} is an immediate consequence.
Our new schemes preserve demand privacy against colluding users and achieve better performance in the small memory regime and lower subpacketization compared to baseline schemes that trivially extended privacy-keys scheme or virtual-users scheme without offline users. 
As for the case without privacy, we can show exact optimality in the small and large memory regime, and a constant gap of less than six otherwise.

\subsection{Paper Outline}
The rest of the paper is organized as follows.
Section~\ref{sec: problem formation} formulates the hotplug problem with or without demand privacy against colluding users.
Section~\ref{sec:knownresults} summarizes related known results and extends them so as to provide baseline schemes for our considered settings.
Section~\ref{sec:mainwithoutprivacy} summarizes our main results for hotplug without privacy.
Section~\ref{sec:mainwithprivacy} summarizes our main results for hotplug with privacy.
Section~\ref{sec:examplehotplug} illustrates the key ideas for our novel schemes for hotplug without privacy by way of examples.
Section~\ref{sec:exampledemandprivacy} shows examples for the hotplug model with privacy.
Section~\ref{sec:NumExamples} provides some numerical examples. 
Section~\ref{sec:conclusion} concludes the paper.
Proof details can be found in Appendix.

\subsection{Notation Convention}
\label{sec: notation}
We adopt the following notation convention.
\begin{itemize}
    \item Calligraphic symbols denote sets, bold lowercase symbols %
    vectors, bold uppercase symbols matrices, and sans-serif symbols %
    system parameters.
    \item $\Mm[\Qc]$ denotes the submatrix of $\Mm$ obtained by selecting the rows indexed by $\Qc$.
    Similarly, $\dv[\Ic]$ is the subvector of $\dv$ obtained by selecting the elements indexed by $\Ic$.
    \item For an integer $b$, we let $[b] := \{1, \ldots, b\}$.
    \item For sets $\Sc$ and $\Qc$, we let $\Sc \setminus \Qc := \{k: k \in \Sc, k \notin \Qc\}$.
    \item For a vector $\dv$, $\mathsf{rank}(\dv)$ is the number of distinct elements in $\dv$. %
    \item For a matrix $\Dm$, $\mathsf{rank}(\Dm)$ is the rank of $\Dm$. %
    \item For a collection $\{Z_1, \ldots Z_n\}$ and a index set $\Sc \subseteq [n]$, we let $Z_\Sc := \{Z_i: i \in \Sc\}$.
    \item For a ground set $\Gc$ and an integer $t$, we let $\Omega_{\Gc}^{t} := \{ \Tc \subseteq \Gc : |\Tc| = t\}$. 
    \item For $\Tc \in \Omega_{\Gc}^t$, we let $\Tc_i$ be the $i$-th subset in $\Omega_{\Gc}^t$ in lexicographical order. For example, the sets in $\Omega_{\{1,2,3\}}^2$ are indexed as $\Tc_1 = \{1,2\}$, $\Tc_2 = \{1,3\}$, $\Tc_3 = \{2,3\}$.
    \item For integers $a$ and $b$, $\binom{a}{b}$ is the binomial coefficient, or zero if  $a \geq b \geq 0$ does not hold.
    \item The $i$-th standard basis vector is the vector $\ev_{i}$ that has only one non-zero component equal to 1 in position $i$.
  \end{itemize}

\section{Problem Formulation}
\label{sec: problem formation}

\subsection{Problem Formulation: Hotplug Model without Privacy}
\label{sec: problem formation for hotplug}

An $(\KsfActive,\Ksf,\Nsf)$ hotplug coded caching system consists of the following.
\begin{itemize}
    \item A central server stores $\Nsf$ files, denoted as $F_1, \cdots, F_\Nsf$.
    \item Each file has $\Bsf$ i.i.d. uniformly distributed symbols over a finite field $\mathbb{F}_\qsf$, where $\qsf$ is a prime-power number.
    \item The server communicates with $\Ksf$ users through an error-free shared link.
    \item Each user has a local cache that can contain up to $\Msf \Bsf$ symbols, where $\Msf \in [0, \Nsf]$.
    We refer to $\Msf$ as the {\it memory size}. The caches are denoted as $Z_1, \ldots, Z_\Ksf$.
    \item The server sends the signal $X$ to the users through the shared link, where $X$ has no more than $\Rsf \Bsf$ symbols, with $\Rsf\in [0, \Nsf]$.
    We refer to $\Rsf$ as the {\it load}.

    \item The system has a {\it placement phase} and a {\it delivery phase}.
    The placement phase occurs at a time when the server is still unaware of which users will be offline, and which files the active users will request.
    We assume that the server knows that exactly $\KsfActive$ users will be active, for some fixed $\KsfActive \in[\Ksf]$. 
    The delivery phase occurs after the active users have sent their demands to the server.

    Placement Phase:
The server populates the local caches as a function of the files, i.e.,
\begin{align}
    H(Z_k \mid F_{[\Nsf]}) = 0, \ \forall k \in [\Ksf].
    \label{eq:cacheplacemntNOprivacy}
\end{align}

    Delivery Phase:
User $k \in [\Ksf]$ demands the file indexed by $d_k \in [\Nsf]$.
Once the set of active users becomes known to the server, denoted by $\Ic \in \Omega_{[\Ksf]}^{\KsfActive}$, the server sends the message $X$ as a function of the files and of the demands of the active users, i.e., 
\begin{align}
    &H(X \mid \Ic, d_{\Ic}, F_{[\Nsf]}) = 0, \ \forall \Ic \in \Omega_{[\Ksf]}^{\KsfActive}.
    \label{eq:encodingNOprivacy}
\end{align}

    Decoding: Each active user must be able to decode its desired file with the help of its locally cached content and the delivery signal, i.e.,
\begin{align}
&H(F_{d_k} \mid d_k, Z_k, X) = 0, \ \forall k \in \Ic. %
    \label{eq:decodingNOprivacy}
\end{align}
 
    \item Performance:
For $\Msf \in [0, \Nsf]$, the minimum {\it worst-case load} (or simply load, for short in the following) is defined as
\begin{align}
&\Rsf^\star(\Msf) = \limsup_{\Bsf \rightarrow \infty} \ \min_{X,  Z_1, \ldots Z_\Ksf} \ \max_{\Ic,d_\Ic} \{\Rsf: 
\textrm{\small conditions in~\eqref{eq:cacheplacemntNOprivacy}, \eqref{eq:encodingNOprivacy} and~\eqref{eq:decodingNOprivacy} are satisfied}\}.
\label{eq:loadNOprivacy}
\end{align}

    \end{itemize}

We denote by $\Lsf(\Msf) \in \mathbb{Z}_{+}$ the {\it subpacketization level}, which is the smallest file-length $\Bsf$ needed to realize a given achievable scheme for memory size $\Msf$. The subpacketization level is a proxy for the implementation complexity of the proposed achievable scheme.

\subsection{Problem Formulation: Hotplug Model with Demand Privacy Against Colluding Users}
We next add the constraint of {\it demand privacy against colluding users} (or simply privacy, for short in the following) into the hotplug model introduced in Section~\ref{sec: problem formation for hotplug}. As in past work on coded caching with privacy~\cite{wan2020coded,yan2021fundamental} or security~\cite{yan2022robust}, we assume that each user~$j \in [\Ksf]$ has available some {\it local randomness} represented by the random variable (RV) $\tau_j$. We assume that $\tau_j$ is only known to the server and to user~$j \in [\Ksf]$, and is independent of other RVs. The local randomness RVs help to generate the cache contents and the transmit signal, and may be stored if needed for decoding.
Mathematically, in the placement phase the caches are filled as 
\begin{align}
    & H(Z_k \mid F_{[\Nsf]}, \tau_k) = 0, \ \forall k \in [\Ksf];
    \label{eq:cacheplacemntWITHprivacy}
\end{align}
and, in the delivery phase, for all $\Ic \in \Omega_{[\Ksf]}^{\KsfActive}$ and all $d_{\Ic} \in [\Nsf]^{\KsfActive}$, the transmit signal is
\begin{align}
    & H(X \mid \Ic, d_{\Ic}, F_{[\Nsf]}, \tau_{[\Ksf]} ) = 0, 
    \label{eq:encodingWITHprivacy}
\end{align}
the condition of correct decoding for the active users is %
\begin{align}
     & H(F_{d_k} \mid d_k, Z_k, X) = 0, \ \forall k \in \Ic, 
    \label{eq:decodingWITHprivacy}
\end{align}
and the privacy condition~\cite{yan2021fundamental} is
\begin{align}
   & I(\dv_{\Ic\setminus\Bc}; \, X, \dv_{\Bc}, Z_\Bc \mid {\Ic}, F_{[\Nsf]}) = 0, \ \forall \Bc \subseteq \Ic, \Bc \not=\emptyset,  
    \label{eq:privacyconstraint} 
\end{align}
where $\Bc$ represent the set of colluding active users
\footnote{The condition in~\eqref{eq:privacyconstraint} (i.e., with $\Bc \not=\emptyset$) implies $I(\dv_{\Ic}; \, X \mid {\Ic}, F_{[\Nsf]}) = 0$ (i.e., $\Bc=\emptyset$ can be included in~\eqref{eq:privacyconstraint}) by the argument in~\cite[Appendix A]{yan2021fundamental} with $[\Ksf]$ replaced by $\Ic$.}.
For $\Msf \in [0, \Nsf]$, the minimum load with privacy is 
\begin{align}
&\Rsf_{\text{\rm p}}^{\star}(\Msf) = \limsup_{\Bsf \rightarrow \infty} \ \min_{X,  Z_1, \ldots Z_\Ksf} \ \max_{\Ic,d_\Ic} \{\Rsf: 
\textrm{\small conditions in~\eqref{eq:cacheplacemntWITHprivacy}, \eqref{eq:encodingWITHprivacy}, \eqref{eq:decodingWITHprivacy} and~\eqref{eq:privacyconstraint} are satisfied}\},
\label{eq:loadWITHprivacy}
\end{align}
where the subscript `p' in~\eqref{eq:loadWITHprivacy} indicates that performance is under the privacy constraint in~\eqref{eq:privacyconstraint} as so as to distinguish it from~\eqref{eq:loadNOprivacy}. Clearly, $\Rsf^{\star} \leq \Rsf_{\text{\rm p}}^{\star}$\footnote{For the case $\Ksf = \KsfActive = \Nsf =2$ we have $\Rsf^{\star} < \Rsf_{\text{\rm p}}^{\star}$ for $\Msf \in (1/3,4/3)$~\cite{kamath2020demand,maddah2014fundamental}, thus imposing privacy strictly worsen performance in some memory regimes.}, and in both cases the load is a non-decreasing function of the number of active users $\KsfActive$.

\section{Known Schemes for Classical Model and Extensions to Baseline Schemes for Hotplug Model}
\label{sec:knownresults}
In this section we summarize relevant known results for the classical (i.e., not hotplug) coded caching model and provide straightforward extensions to the hotplug model, which will give us a set of baseline schemes to improve upon in subsequent sections.

In the following, the name of a baseline scheme is indicated as superscript among parentheses, and where the addition of the symbol `$+$' or `$\&$' indicate the scheme is an extension from the classical model to the hotplug model.

\subsection{The YMA Scheme for Classical Model without Privacy} 
\label{sec:YMA}
When $\KsfActive = \Ksf$, the hotplug model is equivalent to the classical coded caching model, for which a YMA-type scheme is optimal under the constraint of uncoded placement, both for single-file~\cite{yu2017exact} and with SLFR~\cite{wan2021optimal} demands. We describe next the YMA scheme with SLFR demands, as it will be needed to introduce the privacy-keys scheme in later sections.

\paragraph*{Placement Phase} 
Fix $t \in [0: \Ksf]$.
Partition each file into $\binom{\Ksf}{t}$ equal-size subfiles as
\begin{align}
    F_i = (F_{i,\Wc}  : \Wc \in \Omega_{[\Ksf]}^{t}),  \quad \forall i \in [\Nsf].
    \label{eq:MANsplit}
\end{align}
The subpacketization level is $\binom{\Ksf}{t}$.  
The cache content of user~$k$ is
\begin{align}
    Z_k = (F_{i,\Wc} : i \in [\Nsf], \Wc \in \Omega_{[\Ksf]}^{t}, k \in \Wc), \quad \forall k \in [\Ksf].
    \label{eq:MANacahe}
\end{align}
The needed memory size is
\begin{align}
	\Msf^\text{\rm(YMA)}_t := \Nsf \frac{\binom{\Ksf-1}{t-1}}{\binom{\Ksf}{t}} = \Nsf \frac{t}{\Ksf}.
	\label{eq:MANmemorysize}
\end{align}
The placement in~\eqref{eq:MANacahe} is referred to as {\it centralized placement} because it requires coordination among users during the placement phase~\cite{maddah2014fundamental}.

\paragraph*{SLFR Demands} 
We say that user~$k$ demands $\dv_k = [d_{k,1}, \ldots, d_{k,\Nsf}] \in \mathbb{F}_\qsf^{\Nsf}$ if it is interested in retrieving the linear function 
\begin{align}
	B_k := \sum_{n\in [\Nsf]} d_{k,n} F_n, \quad \forall k\in[\Ksf].
    \label{eq:SLFRdemand}
\end{align}
Given the file partitioning in~\eqref{eq:MANacahe}, we can partition~\eqref{eq:SLFRdemand} as $B_k = (B_{k,\Wc}  : \Wc \in \Omega_{[\Ksf]}^{t})$ where the {\it block} $B_{k,\Wc}$ is defined as
\begin{align}
	B_{k,\Wc} := \sum_{n\in [\Nsf]} d_{k,n} F_{n,\Wc}, \quad \forall k \in [\Ksf].
    \label{eq:SLFRblock}
\end{align}
The collection of all demand vectors forms the demand matrix $\Dm = [\dv_1; \ldots; \dv_\Ksf] \in \mathbb{F}_\qsf^{\Ksf \times \Nsf}$.
Note that for single-file retrieval, the demand vectors are restricted to be standard basis vectors, i.e., $\dv_k \in\{\ev_1, \ldots,\ev_\Nsf\}$ for all $k \in [\Ksf]$.

\paragraph*{MAN-type Multicast Messages} 
Given the demand matrix $\Dm$, the server constructs the multicast messages 
\begin{align}
    X_\Sc = \sum_{k \in \Sc} \alpha_{k, \Sc \setminus \{k\}} \, B_{k, \Sc \setminus \{k\}}, \ \forall \Sc \in \Omega_{[\Ksf]}^{t+1},
    \label{eq:MANmm}
\end{align}
where the {\it encoding coefficients} $\alpha_{k, \Sc \setminus \{k\}} \in \{+1,-1\}$  are selected as shown in~\cite{wan2021optimal} (for a more general construction, we refer the reader to~\cite{ma2021general}) and the blocks $B_{k, \Sc \setminus \{k\}}$ are defined in~\eqref{eq:SLFRblock}, for all $\Sc \in \Omega_{[\Ksf]}^{t+1}$ and $k \in [\Ksf]$.

\paragraph*{YMA-type Delivery Phase} 
Given the demand matrix $\Dm$, the server selects a {\it leader set} $\Lc \subseteq [\Ksf]$ such that $\mathsf{rank}( \Dm[\Lc] ) = \mathsf{rank}( \Dm )$, and sends
\begin{align}
    X = (\Lc, \, \Dm, \, X_\Sc: \Sc \in \Omega_{[\Ksf]}^{t+1}, \Sc \cap \Lc \neq \emptyset ).
    \label{eq:YMAmm}
\end{align}

\paragraph*{Decoding}
As shown in~\cite{wan2021optimal,ma2021general}, the delivery signal in~\eqref{eq:YMAmm} allows all users to successfully decode their requested SLFR. The key observation is that it is possible to choose the encoding coefficients in~\eqref{eq:MANmm} in such a way that the non-leader users, indexed by $[\Ksf] \setminus \Lc$, can locally reconstruct the not-sent multicast messages $X_\Ac,$ for all $\Ac \in \Omega_{[\Ksf] \setminus \Lc}^{t+1},$ from the delivery signal in~\eqref{eq:YMAmm}. 
Note that user $k \in \Sc$ can recover the missing block $B_{k, \Sc \setminus \{k\}}$ from $X_\Sc$ in~\eqref{eq:MANmm} by ``caching out'' $\sum_{u \in \Sc \setminus \{k\}} B_{u, \Sc \setminus \{u\}}$, which can be computed from $Z_k$ in~\eqref{eq:MANacahe}.

\paragraph*{Performance}
For the delivery signal in~\eqref{eq:YMAmm} with $|\Lc|$ leaders we have
\begin{align}
H(X) \leq |\Lc|\log_\qsf(\Ksf)+\Ksf\Nsf + \frac{ \binom{\Ksf}{t+1} - \binom{\Ksf-|\Lc|}{t}}{ \binom{\Ksf}{t} }\Bsf, %
\end{align}
thus, for very large $\Bsf$ and for worst case $|\Lc| = \mathsf{rank}(\Dm) = \min(\Ksf,\Nsf)$, the following is achievable.
\begin{thm}[YMA]
    \label{thm:YMA}
    For a $(\Ksf, \Nsf)$ classical coded caching system without privacy, 
    the lower convex envelope of the following points is achievable
\begin{align}
    ( \Msf^\text{\rm(YMA)}_t, \Rsf^\text{\rm(YMA)}_t )  = \biggl(
    \frac{t}{\Ksf}, %
    \frac{\binom{\Ksf}{t+1} - \binom{\Ksf - \min(\Ksf,\Nsf)}{t+1}}{\binom{\Ksf}{t}} \biggr), %
    \quad \forall t \in [0: \Ksf].
    \label{eq:performanceYMA}
\end{align}
The YMA scheme is optimal under the constraint of uncoded placement~\cite{yu2017exact}; otherwise, it is optimal to within a factor of 2~\cite{yu2018characterizing}.
\end{thm}

\subsection{Baseline Scheme~1 for Hotplug without Privacy: Extension of YMA to Hotplug}
\label{sec:YMA+hotplug}
A trivial extension of the centralized YMA scheme in~\eqref{eq:performanceYMA} to the hotplug model, i.e., $\KsfActive < \Ksf$, is to let the server assign as demand for the offline users the demand of the first leader user. Thus, the following is achievable for the hotplug model without privacy, by noting that in this case $|\Lc| \leq \min(\KsfActive,\Nsf)$, i.e., serving the ``fake'' demands of the offline users does not increase the size of the leader set for the active users.
\begin{thm}[Extension of YMA to Hotplug]
    \label{thm:extensionYMA}
    For a $(\KsfActive,\Ksf,\Nsf)$ hotplug system without privacy, 
    the lower convex envelope of the following points is achievable
    \begin{align}
        ( \Msf_t^\text{\rm(YMA)}, \Rsf^\text{\rm(YMA+)}_t ) = \biggl(
        \frac{t}{\Ksf}, %
        \frac{\binom{\Ksf}{t+1} - \binom{\Ksf - \min(\KsfActive,\Nsf)}{t+1}}{\binom{\Ksf}{t}} \biggr), 
        \quad \forall t \in [0: \Ksf].
    \label{eq:performanceYMA+}
    \end{align}
\end{thm}

\subsection{Baseline Scheme~2 for Hotplug without Privacy: Decentralized Placement}
\label{sec:DecentralizedPlacement+hotplug}
Decentralized placement refers to the case where users cache each symbol of the library in an i.i.d. fashion with probability $\mu:=\Msf/\Nsf \in[0,1]$~\cite{maddah2014decentralized,yu2017exact}. 
From~\cite[eq(20)]{yu2018characterizing}, for the classical model without privacy and leader set $\Lc$, the load with decentralized placement is 
$
    \frac{1-\mu}{\mu}\big( 1-(1-\mu)^{|\Lc|} \big). %
$
Similarly to Section~\ref{sec:YMA+hotplug}, a straghtforward extension of the decentralized scheme from the classical to the hotplug model is obtained by assigning ``fake'' demands to offline users in such a way as to leave the size of the leader set for the active users unchanged. 
\begin{thm}[Extension of decentralized scheme to hotplug]
    \label{thm:extensionYMAdecentralized}
    For a $(\KsfActive,\Ksf,\Nsf)$ hotplug system without privacy, 
    the following is achievable
    \begin{align}
        \Rsf^\text{\rm (decen+)}(\Msf) 
    = \frac{1-\mu}{\mu}\Big( 1-(1-\mu)^{\min(\KsfActive,\Nsf)} \Big),
    \quad \mu:=\Msf/\Nsf.      
    \label{eq:performanceDecentralized+} 
    \end{align}
\end{thm}
Notice: (i) $\Rsf^\text{\rm(YMA+)} \leq \Rsf^\text{\rm (decen+)}$~\cite{yu2018characterizing}; and (ii) $\Rsf^\text{\rm (decen+)}$ does not depend on $\Ksf$. %

\subsection{The PK Scheme for Classical Model with Privacy} %
\label{sec:YT}
A scheme that preserves privacy for the classic (i.e., not hotplug) coded caching model with single-file retrieval was introduced in~\cite[Theorem~1]{yan2021fundamental}.
The central idea was to use Privacy Keys (PK) to ``obfuscate'' the parts of a multicast message to the users who are not interested in it, while the intended user has a ``key'' to enable its extraction. The PK scheme works as follows.

\paragraph*{Placement Phase} 
Files are split is as in~\eqref{eq:MANsplit}.
In the following we shall use the notation $\Tbp(\av, \bv)$ to denote the ``bilinear product of the subfiles'' defined as
\begin{align}
    \Tbp(\av, \bv) := %
    \sum_{n \in [\Nsf]} \sum_{j\in\left[\binom{\Ksf}{t}\right]} a_n \, F_{n, \Wc_j} \, b_j \in \mathbb{F}_\qsf^{\Bsf/\binom{\Ksf}{t}}. 
    \label{eq:bilinearproduct}
\end{align}
for given vectors $\av \in \mathbb{F}_\qsf^{\Nsf}$ %
and $\bv \in \mathbb{F}_\qsf^{\binom{\Ksf}{t}}$. %
Note that we can express subfile $F_{i,\Wc_j}$ as $F_{i,\Wc_j} = \Tbp(\ev_{i}, \ev_{j}), i \in[\Nsf], j \in[\binom{\Ksf}{t}]$; with a slight abuse of notation, we shall sometimes write $F_{i,\Wc} = \Tbp(\ev_{i}, \ev_{\Wc})$, for $i \in [\Nsf], \ \Wc \in \Omega_{[\Ksf]}^t$.

For each user~$k, \ k\in [\Ksf]$, the server picks i.i.d uniformly at random the {\it key vector} $\pv_k$ from the set of all vectors whose entries sum to $\qsf -1$, i.e., $\pv_k \in \Pc_\Nsf$ where $\Pc_\Nsf$ is defined as
\begin{align}
    \Pc_\Nsf := \Big\{\pv=[p_1,\ldots,p_\Nsf] \in \mathbb{F}_\qsf^{\Nsf}: \sum_{i\in[\Nsf]} p_i = \qsf -1 \Big\}.
\label{eq:pkSet}
\end{align} 
The server populates the cache of user~user~$k$ as
\begin{subequations}
\begin{align}
    Z_k &= \left\{\Tbp(\ev_i, \ev_{\{k\}\cup\Qc}): i \in [\Nsf], \Qc \in \Omega_{[\Ksf] \setminus \{k\}}^{t-1}\right\} 
    \label{eq:place:man}
    \\ &\bigcup \ \left\{\Tbp(\pv_k, \ev_{\Wc}): \Wc \in \Omega_{[\Ksf]\setminus\{k\}}^{t}\right\}, \ \forall k \in [\Ksf],
    \label{eq:place:pk}
\end{align}
where~\eqref{eq:place:man} is the uncoded part of the cache as in~\eqref{eq:MANacahe}, while~\eqref{eq:place:pk} is the coded part corresponding to the privacy keys of the form $\Tbp(\pv_k, \ev_{\Wc})$ for all sets $\Wc$ that do not include $k$. 
\end{subequations}
Note that $\pv_k$ is the local randomness for user~$k, \ k\in [\Ksf]$, which is not stored directly in its cache.
The needed memory size is
\begin{align}
    \Msf^\text{\rm(PK)}_t := \Nsf \frac{\binom{\Ksf-1}{t-1}}{\binom{\Ksf}{t}} + \frac{\binom{\Ksf-1}{t}}{\binom{\Ksf}{t}} = 1 + \frac{t}{\Ksf}(\Nsf-1)\geq 1.
    \label{eq:YT:memory}
\end{align}
Note that %
$\Msf^\text{(YMA)}_t = \Nsf \frac{t}{\Ksf} \leq \Msf^\text{\rm(PK)}_t$ as users also store privacy keys.

\paragraph*{Delivery Phase}
User~$k$ has (single-file demand) {\it  demand vector} $\ev_{d_k}\in \mathbb{F}_\qsf^{\Nsf}$ (meaning that he is interested in retrieving the file with index $d_k\in [\Nsf]$). 
Associated to demand vector $\ev_{d_k}$ is the {\it query vector} 
\begin{align}
	\qv_k := \pv_k + \ev_{d_k} \in \mathbb{F}_\qsf^{\Nsf}, \ \forall k \in [\Ksf],
	\label{eq:QueryVect}
\end{align}
which can be thought of as a one-time pad of the demand vector $\ev_{d_k}$ by the key vector $\pv_k$.
The collection of all query vectors gives the query matrix $\Qm = [\qv_1; \ldots; \qv_\Ksf] \in \mathbb{F}_\qsf^{\Ksf \times \Nsf}$.

It is important to note that, 
by the definition in~\eqref{eq:pkSet}, the entries of the key vectors sum to $\qsf-1$; in addition,
the entries of the single-file demand vectors sum to $1$;
thus, the entries of the query vectors in~\eqref{eq:QueryVect} sum to zero in $\mathbb{F}_\qsf$, 
that is, by design query vectors are uniformly and independently distributed over an $\Nsf-1$ dimensional subspace in $\mathbb{F}_\qsf$.
This implies that $\mathsf{rank}(\Qm) \leq \min(\Ksf,\Nsf-1).$

Even though we consider single-file demands, the delivery is done as if we have SLFR demands according to the query vectors in~\eqref{eq:QueryVect}, that is, the server transmits 
\begin{align}
   X &= ( \Lc, \, \Qm, \,  X_\Sc: \Sc \in \Omega_{[\Ksf]}^{t+1}, \Sc \cap \Lc \neq \emptyset ), 
   \label{eq:YTmessages:X}
\end{align}
where $\Lc$ is the leader set for the query matrix $\Qm$, and the blocks in~\eqref{eq:SLFRblock} (in the multicast messages $X_\Sc$ in~\eqref{eq:MANmm}) are constructed according to the query matrix rather than the demand matrix.

\paragraph*{Decoding and Privacy}
The multicast message only useful for non-leader users (and not included in~\eqref{eq:YTmessages:X}) can be reconstructed locally as per the decoding procedure for the SLFR scheme.
\begin{subequations}
To understand how decoding proceeds once all multicast messages are locally available, we rewrite the  multicast message (which were designed according to the query matrix) useful for the users indexed by $\Sc$ as follows
\begin{align}
X_\Sc 
    &= \sum_{j \in \Sc} \alpha_{\Sc, j} \Tbp(\qv_j, \ev_{\Sc \setminus \{j\}})
    \\
    &= \alpha_{\Sc, v} \underbrace{ \Tbp(\ev_{d_v}, \ev_{\Sc \setminus \{v\}}) }_{\text{demanded by user~$v$}}
     + \alpha_{\Sc, v} \underbrace{ \Tbp(\pv_v, \ev_{\Sc \setminus \{v\}}) }_{\text{privacy key cached in $Z_v$}}
    \label{eq:YTmessages:S1}
    \\ 
    &+
    \sum_{j \in \Sc \setminus \{v\}}
    \alpha_{\Sc, j} \underbrace{\Tbp(\qv_j, \ev_{\Sc \setminus \{j\}})}_{\text{computable by user~$v$}}, 
    \quad \forall  v\in \Sc,
    \label{eq:YTmessages:S2}
\end{align}
where all terms in~\eqref{eq:YTmessages:S2} are computed locally from the cache content $Z_v$ in~\eqref{eq:place:man} and $\Qm$ in~\eqref{eq:YTmessages:X} as
\begin{align}
    \Tbp(\qv_j, \ev_{\Sc \setminus \{j\}}) = \sum_{n\in[\Nsf]} q_{j,n}\Tbp(\ev_n, \ev_{\Sc \setminus \{j\}}),
    \label{eq:YTmessages:S2evallocally}
\end{align}
\label{eq:YTmessages}
and $\Tbp(\ev_n, \ev_{\Sc \setminus \{j\}})$ is cached by the users indexed in $\Sc \setminus \{j\}$.
\end{subequations}
Therefore, user~$v\in\Sc$ can retrieve $\Tbp(\ev_{d_v}, \ev_{\Sc \setminus \{v\}})$ from $X_\Sc$ if $v\in\Sc$. This shows that all users can successfully decode their demanded file. 
The proof that the PK scheme is private can be found in~\cite{yan2021fundamental}, and hinges on the fact that the key vectors are independently and uniformly disturbed.

\paragraph*{Performance}
For very large $\Bsf$ and for the worst-case query matrix with $\mathsf{rank}(\Qm) = \min(\Ksf,\Nsf-1)$, the following is achievable.
\begin{thm}[PK]
    \label{thm:PK}
    For a $(\Ksf, \Nsf)$ classical coded caching system with privacy, 
the lower convex envelope of $(\Msf,\Rsf)=(0,\Nsf)$ with the following points is achievable 
\begin{align}
    (\Msf^\text{\rm(PK)}_t, \Rsf^\text{\rm(PK)}_t) = 
    \left( 1 + \frac{t}{\Ksf}(\Nsf-1),
    \frac{\binom{\Ksf}{t+1} - \binom{\Ksf - \min(\Ksf,\Nsf-1)}{t+1}}{\binom{\Ksf}{t}} \right),
    \quad t \in [0: \Ksf].
    \label{eq:YT:load}
\end{align}
The PK scheme is optimal to within a factor of 5.4606~\cite{yan2021fundamental}.
\end{thm}

\subsection{Baseline Scheme~1 for Hotplug with Privacy: Extension of PK to Hotplug}
\label{sec:PK+hotplug}
The same argument we used to the extend the YMA scheme from the classical to the hotplug model can be used to extend the PK scheme from the classical to the hotplug model.

\begin{thm}[Extension of PK to Hotplug] 
\label{thm:extensionPK}
    For a $(\KsfActive,\Ksf,\Nsf)$ hotplug model with privacy, 
    the lower convex envelope of $(\Msf,\Rsf)=(0,\Nsf)$ and the following points is achievable
    \begin{align}
        ( \Msf^\text{\rm(PK)}_t, \Rsf^\text{\rm(PK+)}_t) = \left( 1+\frac{t}{\Ksf}(\Nsf -1),%
         \frac{\binom{\Ksf}{t+1} - \binom{\Ksf - \min(\KsfActive,\Nsf-1)}{t+1}}{\binom{\Ksf}{t}}  \right),
    \quad \forall t \in [0: \Ksf].
    \label{eq:performancePK+}
    \end{align}
\end{thm}

Note that, in the low memory regime, the PK+ tradeoff is given by the segment connecting the trivial corner point
$(\Msf,\Rsf)=(0,\Nsf)$ to
$( \Msf^\text{\rm(PK)}_0, \Rsf^\text{\rm(PK+)}_0) = (1, \min(\KsfActive,\Nsf-1))$,
which may not be decreasing fast enough. Thus, we introduce next another baseline scheme.

\subsection{Baseline Schemes~2 and~3 for Hotplug with Privacy: Extension of VU to Hotplug}
\label{sec:VU+hotplug}
In~\cite{kamath2019demand}, privacy in the classical model was achieved by introducing the idea of virtual-users (VU). 
In the VU scheme, each physical user has associated to it a distinct set of $\Nsf$ {\it virtual-users}; 
each physical users picks independently and uniformly at random the cache content of one of its associated virtual users.
The trick in the delivery phase of the VU scheme is to assign demands to the $\Ksf\Nsf$ virtual users in such a way that each of the $\Nsf$ file is demanded exactly $\Ksf$ times\footnote{This special `restricted-type' demand vector does not in general coincide with the worst-case demand vector~\cite{tian2018symmetry}.}. It turns out that the performance of the VU scheme is the same as that of the YMA scheme with $\Ksf\Nsf$ users and $\Nsf$ files, i.e., in~\eqref{eq:performanceYMA} we replace $\Ksf$ with $\Ksf\Nsf$. 

Next, we extend the YMA and the decentralized schemes from the non-private classical model to the hotplug with privacy model by the the VU idea. By noting that $\min(\KsfActive\Nsf,\Nsf) = \Nsf = \min(\Ksf\Nsf,\Nsf)$, we conclude that the performance of the VU scheme for the classical and the hotplug models is the same. 
\begin{thm}[Extensions to hotplug with VU idea] 
\label{thm:extensionVU}
    For a $(\KsfActive,\Ksf,\Nsf)$ hotplug model with privacy, 
    the lower convex envelope of the following points is achievable 
    \begin{align}
        ( \Msf^\text{\rm(YMA\&VU)}_t, \Rsf^\text{\rm(YMA\&VU)}_t ) 
        =\left( \Nsf\frac{t}{\Ksf\Nsf}, \frac{\binom{\Ksf\Nsf}{t+1} - \binom{\Ksf\Nsf - \Nsf}{t+1}}{\binom{\Ksf\Nsf}{t}}  \right),
        \quad \forall t \in [0 : \Ksf \Nsf].  
    \label{eq:performanceYMAVU}
    \end{align}
    in addition, the following is achievable
    \begin{align}
        \Rsf^\text{\rm(decen\&VU)}(\Msf) 
    = \frac{1-\mu}{\mu}\Big( 1-(1-\mu)^{\Nsf} \Big).
    \quad \mu:=\Msf/\Nsf.      
    \label{eq:performanceDecentralizedVU} 
    \end{align}
\end{thm}
Notice: (i) the subpacketization level of the YMA\&VU scheme, given by $\binom{\Ksf\Nsf}{t}$ for $\Msf/\Nsf = t/(\Nsf \Ksf),$ may be huge and thus unaffordable in practical systems; and (ii) $\Rsf^\text{\rm(YMA\&VU)}$ and $\Rsf^\text{\rm(decen\&VU)}$ do not depend $\KsfActive$.

\begin{rem}\rm \label{rem:VUidea}
The VU idea is very powerful as one can readily derive a scheme for a $(\Ksf,\Nsf)$-classical-private model from a $(\Ksf\Nsf,\Nsf)$-classical-NONprivate model~\cite{kamath2020demand}.  The construction in~\cite[Theorem 4]{kamath2020demand} further leverages a critical observation: in the $(\Ksf\Nsf,\Nsf)$-classical-NONprivate model one considers all possible $[\Nsf]^{\Ksf\Nsf}$ demands, which are much more than the $[\Nsf]^{\Ksf}$ possible demands in the $(\Ksf,\Nsf)$-classical-private model. By studying a restricted $(\Ksf\Nsf,\Nsf)$-classical-NONprivate model where only a carefully selected subset of $[\Nsf]^{\Ksf}$ demands are allowed (see~\cite{kamath2020demand} for details), one can obtain a better load performance. The essence of in~\cite[Theorem 4]{kamath2020demand} is thus that: one can readily derive a scheme for a $(\Ksf,\Nsf)$-classical-private model from a $(\Ksf\Nsf,\Nsf)$-restricted-classical-NONprivate model. The challenge with this approach is that restricted-classical-NONprivate models have not been studied in the literature, except for~\cite{namboodiri2021optimal}.

The argument in~\cite[Theorem 4]{kamath2020demand} extends to the hotplug model: one can derive a scheme for a $(\KsfActive,\Ksf,\Nsf)$-hotplug-private model from a $(\KsfActive\Nsf,\Ksf\Nsf,\Nsf)$-restricted-hotplug-NONprivate model. The proof can be found in Appendix~\ref{app:VU for private hotplug from NONprivate hotplug}. 
The study of restricted-hotplug-NONprivate models is an interesting open problem. 
\hfill$\square$ \end{rem}

\section{Main Results for Hotplug without Privacy}
\label{sec:mainwithoutprivacy}
In this section we summarize our main results for the hotplug model without the constraint of demand privacy against colluding users.

\label{sec:main results}
\begin{thm}[New Scheme~1]%
    \label{thm:HT1}
    For a $(\KsfActive,\Ksf,\Nsf)$ hotplug system without privacy, 
    the lower convex envelope of the following point is achievable
    \begin{align}
        ( \Msf_t^\text{\rm HT1}, \Rsf^\text{\rm HT1}_t )  
        &=\biggl(
        \Nsf \frac{\binom{\Ksf-1}{t-1}}{\binom{\KsfActive}{t}}, 
        \frac{\binom{\KsfActive}{t+1} - \binom{\KsfActive - \min(\KsfActive,\Nsf)}{t+1}}{\binom{\KsfActive}{t}} \biggr),
    \quad \forall t \in [0: \KsfActive].
    \label{eq:performanceNEW1}  
    \end{align}
\end{thm}

\begin{rem}[Sketch of achievability for Theorem~\ref{thm:HT1}] \rm
    The proof that our first novel scheme attains the load in~\eqref{eq:performanceNEW1} can be found in Appendix~\ref{sec:NEW1achievablescheme}.  
    At a high level, we split each file into $\binom{\KsfActive}{t}$ equal-length subfiles and then code the subfiles with an MDS code of rate ${\binom{\KsfActive}{t}}/{\binom{\Ksf}{t}}$. %
    The placement of the MDS-coded symbols follows the MAN spirit and the delivery the YMA spirit.

By comparing $\Rsf^\text{\rm(YMA)}_t$ in~\eqref{eq:performanceYMA} for the classical coded caching system with $\KsfActive$ users (rather than $\Ksf$) %
with the load $\Rsf^\text{\rm HT1}_t$ in~\eqref{eq:performanceNEW1}, we notice they have the exact same expression; the difference is in the memory requirement, which is 
$\Msf/\Nsf %
= t/\KsfActive$ for the YMA scheme with $\KsfActive$ users and 
$\Msf/\Nsf %
= t/\Ksf \cdot {\binom{\Ksf}{t}}/{\binom{\KsfActive}{t}}$ for our first new scheme with $\KsfActive$ active users out of $\Ksf$ total users. In other words, we need more cache space (quantified by the inverse of the MDS code rate) in order to serve $\KsfActive$ online users and tolerate $\Ksf - \KsfActive$ offline users, compared to the classical YMA coded caching scheme for $\KsfActive$ users. Note that the two schemes have the same memory requirement for $t=1$.

For an example of Theorem~\ref{thm:HT1}, please see Sections~\ref{par:K'=N=2:M=1} and~\ref{par:K'=N=3:M=1}.
\hfill$\square$ \end{rem}

\begin{rem}[Optimality for Theorem~\ref{thm:HT1} in the small memory regime] \rm
Let $\rsf^\prime := \min(\KsfActive,\Nsf)$.
Consider the following corner points in Theorem~\ref{thm:HT1}
\begin{align}
( \Msf^\text{\rm HT1}_0, \Rsf^\text{\rm HT1}_0 ) &= (0, \rsf^\prime),
\label{eq:performanceNEWt=0}
    \\
( \Msf^\text{\rm HT1}_1, \Rsf^\text{\rm HT1}_1 ) &= 
\left(\frac{\Nsf}{\KsfActive},\rsf^\prime-\frac{\rsf^\prime(\rsf^\prime+1)}{2\KsfActive}\right).
\label{eq:performanceNEW1:t=1}
\end{align}
The segment connecting the points~\eqref{eq:performanceNEWt=0} and~\eqref{eq:performanceNEW1:t=1} (achievable by memory sharing) outperforms the baseline scheme YMA+ in~\eqref{eq:performanceYMA+}. Furthermore, it is optimal when the number of files is sufficiently large enough, as proved in Theorem~\ref{thm:optimality}.
\hfill$\square$ \end{rem}

\begin{thm}[New Scheme~2]%
    \label{thm:HT2}
    For a $(\KsfActive,\Ksf,\Nsf)$ hotplug system without privacy, 
    if $\Ksf \geq \KsfActive \geq \Nsf$, the following point is achievable 
    \begin{align}
    ( \Msf^\text{\rm HT2}, \Rsf^\text{\rm HT2} ) = \biggl( \frac{1}{\KsfActive}, \Nsf\left(1-\frac{1}{\KsfActive}\right) \biggr). 
    \label{eq:performanceNEW2}
    \end{align}
\end{thm}

\begin{rem}[Sketch of achievability for Theorem~\ref{thm:HT2}] \rm
The proof of the achievabilty of~\eqref{eq:performanceNEW2} can be found in Appendix~\ref{sec:proofofHT2}. In this scheme, we first code the files together, and then we apply another level of MDS coding before the placement. The general delivery has two steps: first we `unlock' the cache contents of the active users similarly to in~\cite{chen2016fundamental}, and then we perform a sequence of YMA-type deliveries to subsets of active users requesting the same file.

For an example of Theorem~\ref{thm:HT2}, please see Sections~\ref{par:K'=N=2:M=1/2} and~\ref{par:K'=N=3:M=1/3}.
\hfill$\square$ \end{rem}

\begin{rem}[Optimality for Theorem~\ref{thm:HT2} in the small memory regime] \rm    
    As shown in Theorem~\ref{thm:optimality}, our second new scheme is optimal for $\Ksf \geq \KsfActive \geq \Nsf$ in the memory regime $\Msf \leq \frac{1}{\KsfActive}$ as it meets the cut-set bound $\Rsf^\star(\Msf) \geq \Nsf(1-\Msf)$.
\hfill$\square$ \end{rem}

\begin{thm}[New Scheme~3]%
\label{thm:HT3} 
    For a $(\KsfActive,\Ksf,\Nsf)$ hotplug system without privacy, the following point is achievable for $\KsfActive \geq 3$ 
\begin{align}
    (\Msf^\text{\rm HT3}, \Rsf^\text{\rm HT3}) = \left(\Nsf \frac{\KsfActive-1}{\KsfActive}, \frac{1}{\KsfActive} \right).
    \label{eq:performanceNEW3}
\end{align}
\end{thm}

\begin{rem}[Sketch of achievability for Theorem~\ref{thm:HT3}] \rm
The proof of the achievability of~\eqref{eq:performanceNEW3} can be found in Appendix~\ref{sec:proofofFLEX} as the special case of  the following scheme for $\Lsf_t = \KsfActive$ and $t=\KsfActive-1$.
\begin{lem} \label{lem:FLEX}
    For a $(\KsfActive,\Ksf,\Nsf)$ hotplug system without privacy, 
    the lower convex envelope of the following points is achievable:
    let the subpacketization $\Lsf_t$ be the largest positive integer which satisfies $\Lsf_t < \binom{\KsfActive-1}{t-1} \frac{t}{t-1}$,
    then \begin{align}
        (\Msf_t^\text{\rm FLEX}, \Rsf_t^\text{\rm FLEX}) = \left(\Nsf\frac{\binom{\KsfActive-1}{t-1}}{\Lsf_t}, \frac{\binom{\KsfActive}{t+1} - \binom{\KsfActive-\min(\KsfActive,\Nsf)}{t+1}}{\Lsf_t}\right), \quad \forall t \in [2: \KsfActive-1].
        \label{eq:performanceFLEX}
    \end{align}
\end{lem}
For the scheme in Lemma~\ref{lem:FLEX}, we construct an MDS generator matrix of dimension $\Lsf_t \times \Ksf\binom{\KsfActive-1}{t-1}$, then partition it into $\Ksf$ disjoint parts (one per user) where each part is used to populate the cache of a user. Then, we construct the multicast messages within the linear span of the caches. Users can eliminate the `interference' on non-demanded files and thus recover their desired information. 
Note that the constraint on the subpacketization level implies $\Msf_t^\text{\rm FLEX} > \Nsf (t-1)/t$. 

For an example of Theorem~\ref{thm:HT3}, please see Section~\ref{par:K'=N=3:M=2}.
\hfill$\square$ \end{rem}

\begin{rem}[Optimality for Theorem~\ref{thm:HT3} in the large memory regime] \rm    
As shown in Theorem~\ref{thm:optimality}, our third new scheme is optimal in the memory regime $\Msf \geq \Nsf \frac{\KsfActive-1}{\KsfActive}$ as it meets the cut-set bound $\Rsf^\star(\Msf) \geq 1-\Msf/\Nsf$.
\hfill$\square$ \end{rem}

\bigskip
Next we provide performance guarantees for the three new scheme we have introduced.
As a converse bound, we can use {\it any} converse result for the classical coded caching system without privacy with $\KsfActive$ users and $\Nsf$ files; this is so because the performance of our hotplug system cannot be better than that of a system in which the server knows a priori which set of $\KsfActive$ users will be active, and does the optimal placement and delivery for those users. With this type of converse bounds, we can show the following optimality result. The detailed proof can be found in Appendix~\ref{sec:converseproof}. 

\begin{thm}[Optimality Guarantees] 
\label{thm:optimality} 
For a $(\KsfActive,\Ksf,\Nsf)$ hotplug system without privacy we have the following optimality guarantees.
\begin{enumerate}

\item\label{item:opt:r=1}
When $\min(\KsfActive,\Nsf)=1$, $\Rsf^\text{\rm (YMA+)}$ is optimal. 

\item\label{item:opt:gap is constant}
$\Rsf^\text{\rm (YMA+)}$ is at most a factor 2 from optimal.

\item\label{item:opt:K'=2N=2}
When $\Ksf \geq \KsfActive = 2$ and $\Nsf = 2$, the optimal scheme has two non-trivial corner points: $(\Msf_1^\text{\rm HT1},\Rsf_1^\text{\rm HT1})=(1,1/2)$ and $(\Msf^\text{\rm HT2},\Rsf^\text{\rm HT2})=(1/2,1)$.

\item\label{item:opt:K'=2N>=3}
When $\Ksf \geq \KsfActive = 2$ and $\Nsf \geq 3$, the only non-trivial optimal corner point is $(\Msf_1^\text{\rm HT1},\Rsf_1^\text{\rm HT1})=(\Nsf/2,1/2)$. 

\item\label{item:opt:NlargeMsmall-firstsegment}
When $\Nsf \geq \KsfActive (\KsfActive + 1)/2$, the corner point $(\Msf_1^\text{\rm HT1},\Rsf_1^\text{\rm HT1})=(\Nsf/\KsfActive,(\KsfActive-1)/2)$ is optimal.
Memory sharing between this point and the trivial corner point $(0,\KsfActive)$ gives optimality for $\Msf \leq \Nsf/\KsfActive$ when $\Nsf \geq \KsfActive (\KsfActive + 1)/2$.

\item\label{item:optNsmallMsmall-firstsegment}
When $\Nsf \leq \KsfActive$, the corner point $(\Msf_1^\text{\rm HT2},\Rsf_1^\text{\rm HT2})=(1/\KsfActive, \Nsf(1-1/\KsfActive))$ is optimal.
Memory sharing between this point and the trivial corner point $(0,\Nsf)$ gives optimality for $\Msf \leq 1/\KsfActive$ when $\Nsf \leq \KsfActive$.

\item\label{item:opt:lastsegment}
When $\KsfActive \geq 3$, the pont $(\Msf^\text{\rm HT3}, \Rsf^\text{\rm HT3}) = \left(\Nsf \frac{\KsfActive-1}{\KsfActive}, \frac{1}{\KsfActive} \right)$ is optimal. 
Memory sharing between this point and the trivial corner point $(\Nsf,0)$ gives optimality for $\Msf \geq \Nsf(1-1/\KsfActive)$ when $\KsfActive \geq 3$.

\end{enumerate}
\end{thm}
Notice: Theorem~\ref{thm:optimality} provides a tight characterization for $\KsfActive=2$ (see Item~\ref{item:opt:K'=2N=2} and Item~\ref{item:opt:K'=2N>=3}) but not for $\min(\KsfActive,\Nsf)=2$,  as even the classical coded caching model with $\Nsf=2$ files is only partially solved~\cite{tian2018symmetry}. %

\begin{rem}[Does the optimal hotplug performance only depend on $\KsfActive$?] \rm
\label{rem:should opt depend on Kprime?}
It is interesting to note that the exact optimality results in Theorem~\ref{thm:optimality} do not depend on $\Ksf$ (the total number of users) but only on $\KsfActive$ (the total number of {\it active} users). It is not obvious that this should be the case in general. It is an intriguing open question to prove, or disprove, that the optimal hotplug performance depends on $\KsfActive$ but not on $\Ksf$.
\hfill$\square$ \end{rem}

\section{Main Results for Hotplug with Privacy}
\label{sec:mainwithprivacy}
In this section we summarize our main results for the hotplug model with the constraint of demand privacy against colluding users.

\begin{thm}[Privacy-Keys idea with Theorem~\ref{thm:HT1}] \label{thm:HT1+PK}
    For a $(\KsfActive,\Ksf,\Nsf)$ hotplug model with privacy,
    the lower convex envelope of $(\Msf,\Rsf)=(\Nsf,0)$ and the following points is achievable 
    \begin{align}
        ( \Msf^\text{\rm HT1+PK}_t, \Rsf^\text{\rm HT1+PK}_t ) = \left( 
        \frac{\Nsf \binom{\Ksf-1}{t-1}+ \left[\binom{\KsfActive}{t}- \binom{\Ksf-1}{t-1}\right]^+}{\binom{\KsfActive}{t}}, %
        \frac{\binom{\KsfActive}{t+1} - \binom{\KsfActive - \min(\KsfActive,\Nsf-1)}{t+1}}{\binom{\KsfActive}{t}}  \right),
        \quad \forall t \in [0 : \KsfActive].
    \label{eq:HT1+PK}
    \end{align}
\end{thm}

\begin{rem}[Sketch of achievability for Theorem~\ref{thm:HT1+PK}] \rm
Theorem~\ref{thm:HT1+PK} combines ideas from the hotplug achievable scheme without privacy in Theorem~\ref{thm:HT1} with the PK scheme in Theorem~\ref{thm:PK}. The proof is provided in Appendix~\ref{sec:proofHT1-PK}. The main idea is to apply an MDS code of rate ${\binom{\KsfActive}{t}}/{\binom{\Ksf}{t}}$ to the subfiles, after file partitioning with subpacketization level $\binom{\KsfActive}{t}$, to get a load that is the same as if the system only had $\KsfActive$ overall users (i.e., load as in~\eqref{eq:YT:load} with $\Ksf$ replaced by $\KsfActive$) which comes at a  price of storing MDS-coded privacy keys, i.e., 
\begin{align}
\Msf^\text{\rm HT1+PK}_t
\geq 1+\frac{t}{\KsfActive} \frac{\binom{\Ksf-1}{t-1}}{\binom{\KsfActive-1}{t-1}} ( \Nsf - 1)
\geq 1+\frac{t}{\KsfActive} ( \Nsf - 1),
\quad \forall t \in [0 : \KsfActive].
\end{align}
Moreover, %
as we shall see in the numerical examples, Theorem~\ref{thm:HT1+PK} has a lower subpacketization level than Theorem~\ref{thm:extensionPK} while maintaining a competitive performance.

For an example of Theorem~\ref{thm:HT1+PK}, please see Sections~\ref{par:K'=N=3:M=5/3 HT1+PK t=1} and~\ref{par:K'=N=3:M=15/3 HT1+PK t=2}.
$\hfill\square$ \end{rem}

\begin{thm}[Privacy-Keys ideas with Theorem~\ref{thm:HT3}] 
\label{thm:HT3+PK} 
For a $(\KsfActive,\Ksf,\Nsf)$ hotplug model with privacy, the following point is achievable
\begin{align}
    (\Msf^\text{\rm HT3+PK}, \Rsf^\text{\rm HT3+PK}) = 
    \left(\Nsf - \frac{\Nsf-1}{\KsfActive}, \frac{1}{\KsfActive}\right).
    \label{eq:HT3+PK}
\end{align}
\end{thm}

\begin{rem}[Sketch of achievability for Theorem~\ref{thm:HT3+PK}] \rm
The proof of Theorem~\ref{thm:HT3+PK} is provided in Appendix~\ref{sec:proofHT3-PK}. 
Similarly to Theorem~\ref{thm:HT3}, we first construct an MDS generator matrix of dimension $\Ksf(\KsfActive-1) \times \KsfActive$, partition it into $\Ksf$ disjoint parts (one for each user), and use it populate the caches. Each user needs to cache one addition privacy key, with memory cost $1/\KsfActive$. Then the server sends one multicast message such that each active user can eliminate the interference from files demanded by other users, and extract its desired information. We have the same load performance as $\Rsf^\text{\rm HT3}=1/\KsfActive$, however the memory size is larger by an additive term $1/\KsfActive$.

For an example of Theorem~\ref{thm:HT3+PK}, please see Section~\ref{par:K'=N=3:M=7/3 HT3+PK}.
$\hfill\square$ \end{rem}

Theorem~\ref{thm:HT1+PK} and Theorem~\ref{thm:HT3+PK} adopt the privacy-keys idea into our newly proposed achievable schemes for hotplug without privacy. Next we present achievable tradeoff points that leverage the virtual-users ideas.
Two tradeoff points attain optimal performance in the small and large, respectively, memory regime.

\begin{thm}[Virutal-Users idea to achieve privacy in hotplug and optimality results] 
\label{thm:HT+VU}
    For a $(\KsfActive, \Ksf, \Nsf)$ hotplug system with privacy,
    the following points are achievable
    \begin{align}
        (\Msf^\text{\rm HT1\&VU}, \Rsf^\text{\rm HT1\&VU})  &=
        \left(\frac{1}{\KsfActive},\Nsf-\frac{\Nsf+1}{2\KsfActive}\right), 
        \label{eq:performanceht1+vu}
        \\
        (\Msf^\text{\rm HT\&VU}, \Rsf^\text{\rm HT+VU}) &= 
        \left(\frac{1}{\KsfActive(\Nsf-1)+1}, \Nsf\left(1-\frac{1}{\KsfActive(\Nsf-1)+1}\right)\right),
        \label{eq:performancehtNONE+vu}
        \\
        (\Msf^\text{\rm HT3\&VU}, \Rsf^\text{\rm HT3\&VU}) &= 
        \left(\Nsf \left(1 - \frac{1}{\KsfActive\Nsf}\right) , \ \frac{1}{\KsfActive\Nsf}\right).
        \label{eq:performanceht3+vu}
    \end{align}
    Memory sharing between the point in~\eqref{eq:performancehtNONE+vu} and the trivial point $(0,\Nsf)$ provides the optimal tradeoff for $\Msf \leq \frac{1}{\KsfActive(\Nsf-1)+1}$.
    Memory sharing between the point in~\eqref{eq:performanceht3+vu} and the trivial point $(\Nsf,0)$ provides the optimal tradeoff for $\Msf \geq \Nsf \left(1 - \frac{1}{\KsfActive\Nsf}\right)$. 
\end{thm}

\begin{rem}[Proof Sketch for Theorem~\ref{thm:HT+VU}] \rm
The achievablity results are an immediate application of the general result proved in Appendix~\ref{app:VU for private hotplug from NONprivate hotplug} that a scheme for a $(\KsfActive, \Ksf, \Nsf)$ hotplug system with privacy can be obtained from scheme for a $(\Nsf\KsfActive, \Nsf\Ksf, \Nsf)$ hotplug system without privacy. 
Note that all tradeoffs points in Theorem~\ref{thm:HT+VU} only depend on $\KsfActive$ and $\Nsf$, and not on $\Ksf$ (see also Remark~\ref{rem:should opt depend on Kprime?}).

The schemes HT1\&VU in~\eqref{eq:performanceht1+vu} and HT3\&VU in~\eqref{eq:performanceht3+vu} are Theorem~\ref{thm:HT1} (for $t=1$) and Theorem~\ref{thm:HT3}, respectively, used with parameters $(\KsfActive\Nsf, \Ksf\Nsf, \Nsf)$. The load in~\eqref{eq:performanceht3+vu} is optimal as it meets the non-private cut-set bound $\Rsf^\star(\Msf) \geq 1-\Msf/\Nsf$.

The scheme HT\&VU in~\eqref{eq:performancehtNONE+vu} extends the optimal classical virtual-users scheme in~\cite{namboodiri2021optimal} to the hotplug model. %
Note that HT\&VU is better than HT2\&VU (i.e., HT2\&VU is the scheme obtained from Theorem~\ref{thm:HT2} with parameters $(\KsfActive\Nsf, \Ksf\Nsf, \Nsf)$ which gives an achievable scheme with privacy by the virtual-users trick) as they both meet the converse bound $\Rsf_{\text{\rm p}}^\star(\Msf) \geq \Nsf(1-\Msf)$, i.e., see~\eqref{eq:yt-converse} for $\ell=\Nsf$, but the cache size in~\eqref{eq:performancehtNONE+vu} is larger than $\frac{1}{\Nsf\KsfActive}$ (from Theorem~\ref{thm:HT2} with parameters $(\KsfActive\Nsf, \Ksf\Nsf, \Nsf)$).

For an example of Theorem~\ref{thm:HT+VU}, please see Sections~\ref{par:K'=N=3:M=1/7 HT+VU} and~\ref{par:HT+VU again}.
\hfill$\square$ \end{rem}

\begin{rem}[Converse bound for hotplug with privacy] \rm
\label{rem:converse}
A trivial converse bound for the hotplug system with privacy can be obtained by considering the converse bound in~\cite[Theorem~3]{yan2021fundamental} (for demand demand privacy with colluding users in the classical setting) by replacing $\Ksf$ with $\KsfActive$. The reasoning is that we cannot do better than knowing at the time of placement which $\KsfActive$ users will be active, and derive the optimal performance of these active users.  The lower bound from~\cite[Theorem~3]{yan2021fundamental} is as follows
    \begin{align}
    \Rsf_{\text{\rm p}}^\star(\Msf) \geq \max_{\ell\in[\Nsf]}\Big\{ \ell+\frac{(\Nsf-\ell) \min\{\ell+1,\KsfActive\} }{(\Nsf-\ell)+\min\{\ell+1,\KsfActive\}}-\ell \ \Msf\Big\},
    \label{eq:yt-converse}
    \end{align}
which is known to be good in the low memory regime~\cite{yan2021fundamental}, i.e., note $\Rsf_{\text{\rm p}}^\star(0) \geq \Nsf.$ In Section~\ref{sec:exampledemandprivacy} and Section~\ref{sec:NumExamples}, we combine the bound in~\eqref{eq:yt-converse} (with privacy) together with~\cite[Theorem 2 and Theorem 4]{yu2018characterizing} (without privacy) as a converse bound for our plots, and for the gap result next.
$\hfill\square$ \end{rem}

\begin{thm}[Gap result for hotplug with prvacy]\label{thm:gap hotplug+privacy}  
For a $(\KsfActive, \Ksf, \Nsf)$ hotplug system with privacy,
(a) when $\KsfActive \geq \Nsf$, we have $\Rsf^\text{\rm (YMA\&VU)}  \leq 2.00884 \ \Rsf_{\text{\rm p}}^\star$; and
(b) when $\KsfActive < \Nsf$, we have  $\Rsf^\text{\rm(PK+)} \geq  5.4606 \ \Rsf^\star_\text{\rm p}$.
\end{thm}
The proof of Theorem~\ref{thm:gap hotplug+privacy} is provided in Appendix~\ref{app:constantgaponhotplugprivacy}.

\section{Examples for Hotplug without Privacy}
\label{sec:examplehotplug}

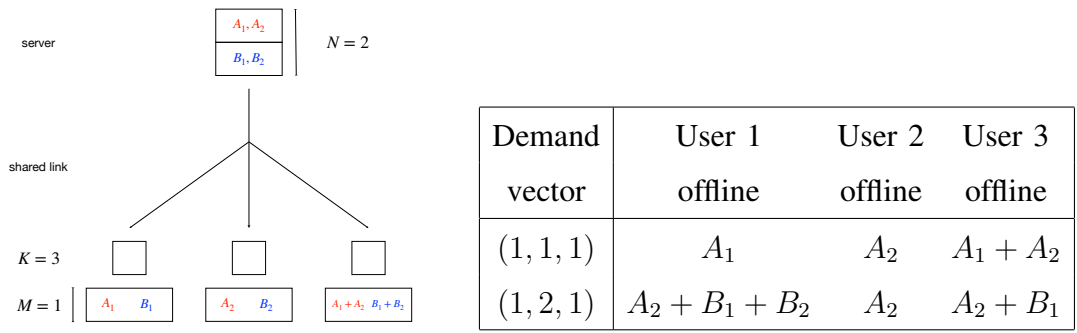
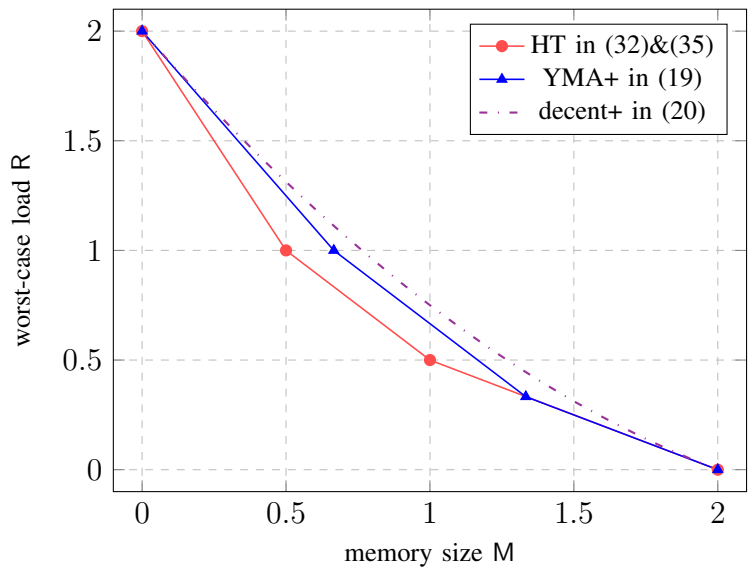
\begin{figure}
   \centering 

    \begin{subfigure}{0.4\textwidth}
    \centering 
   \includegraphics[width=\textwidth]{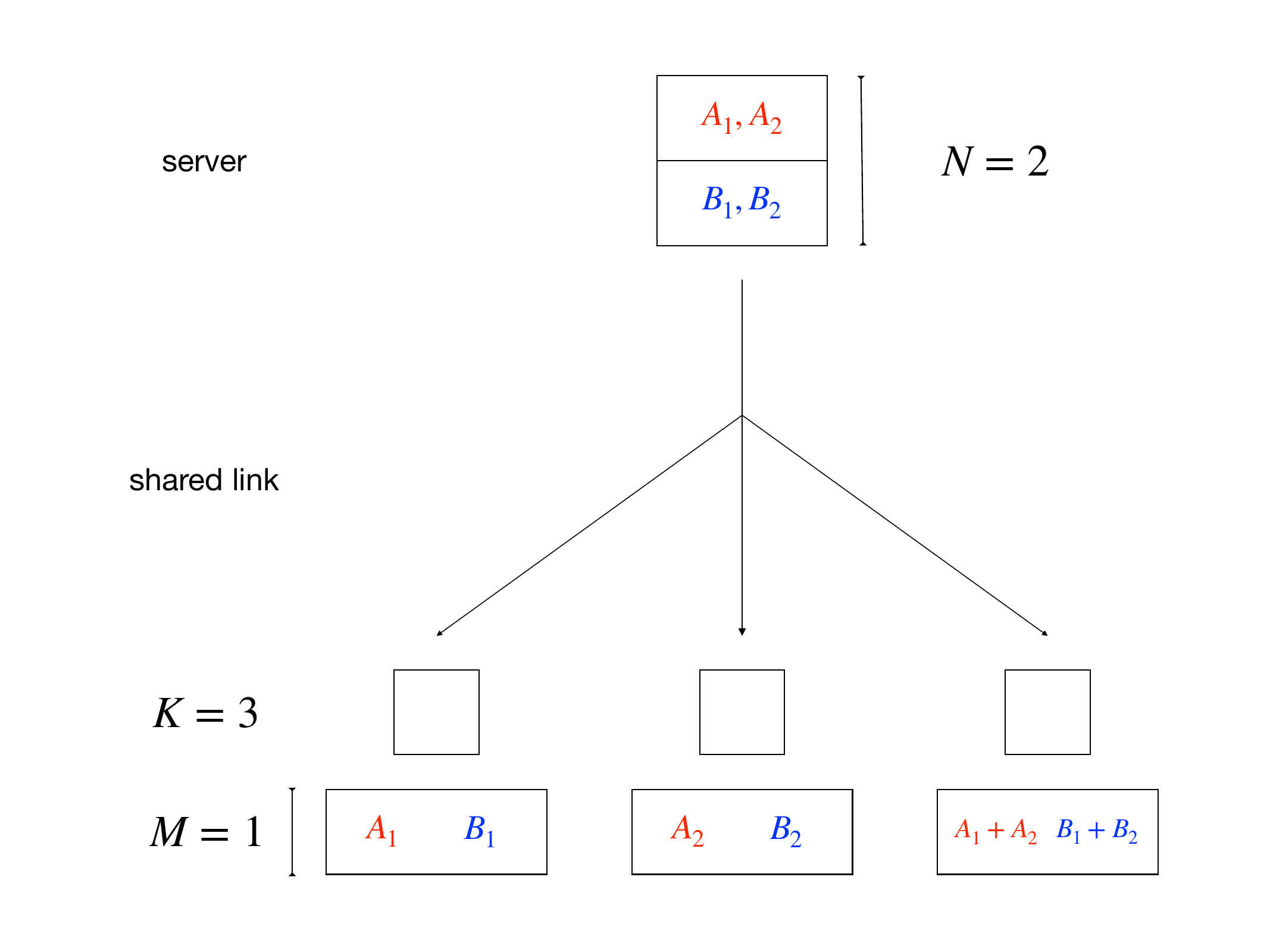}
    \caption{\small Cache placement for memory size $\Msf=1$. The third user caches the two parity bits.} 
    \label{fig: caches (3,2,2)}
    \end{subfigure}
\begin{subfigure}{0.5\textwidth}
\centering 
\begin{align*}
\begin{array}{|c|c c c|}
\hline
\text{Demand}      & \text{User~1} & \text{User~2} & \text{User~3} \\
\text{vector}      & \text{offline}& \text{offline}& \text{offline}\\
\hline
(1,1,1) & A_1 & A_2 &  A_1 + A_2 \\
(1,2,1) & A_2 + B_1 + B_2 & A_2 &  A_2 + B_1 \\
\hline
\end{array}
\end{align*}
\caption{\small Delivery for memory size $\Msf=1$ for two different demand vectors as a function of which user is offline.
}
\label{fig: signals (3,2,2)}
\end{subfigure}
\medskip
\begin{subfigure}[t]{0.8\textwidth}
    \centering 
    \begin{tikzpicture}
    \begin{axis}[
        xmin=-0.1, xmax=2.1,
        ymin=-0.1, ymax=2.1,
        xtick distance=0.5, 
        legend entries={ 
            \small HT in \eqref{eq:performanceNEW1}\&\eqref{eq:performanceNEW2},
            \small YMA+ in~\eqref{eq:performanceYMA+},
            \small decent+ in~\eqref{eq:performanceDecentralized+}
        },
        height=8cm,
        width=10cm,
        grid=major,
        grid style=dashed,
        legend pos=north east,
        xlabel={\small memory size $\Msf$},
        ylabel={\small worst-case load $\Rsf$}
    ]
    \addplot[color=red!70, mark=*, semithick] coordinates {
        (0,2) (0.5, 1) (1, 0.5) (2,0)
    };
    \addplot[color=blue, mark=triangle*, semithick] coordinates {
        (0,2) (0.666, 1) (1.333, 0.333) (2,0)
    };
    \addplot[color=violet!80, loosely dashdotted, thick] table [x=x, y=y] {data/decentralized2.data};
    \end{axis}
    \end{tikzpicture}
    \caption{\small Memory-load tradeoffs.}
    \label{fig: tradeoffs (3,2,2)}
    \end{subfigure}

    \caption{\small Memory-load tradeoffs for the hotplug system with $\Nsf=2$ files, $\KsfActive=2$ active users, and $\Ksf=3$ total users. 
    The converse bound is achievable by HT schemes in~\eqref{eq:performanceNEW1} and~\eqref{eq:performanceNEW2} for any $\Ksf \geq \KsfActive=2$ with $\Nsf=2$, and the optimal performance does not depend on $\Ksf$.}
    \label{fig: hotplug (3,2,2)}
\vspace*{-5mm}
\end{figure}

\subsection{Optimal Schemes for $\Ksf \geq \KsfActive = \Nsf =2$}
\label{sec:K'=N=2 NO privacy}
We aim to show the achievability of the two non-trivial corner points of the optimal memory-load tradefoff for the classical coded caching setting with two users and two files~\cite{maddah2014fundamental}, which is a converse bound for the hotplug system with $\Ksf \geq \KsfActive = \Nsf = 2$.
To prove the achievability of the non-trivial corner points $(1/2,1)$ and $(1,1/2)$ (in addition to the trivial points $(2,0)$ and $(0,2)$) we proceed as follows.
We first combine the coded placement idea of~\cite{maddah2014fundamental} with our MDS coded placement and attain the performance of HT2 scheme in Theorem~\ref{thm:HT2}, which achieves the point $(1/2,1)$ by using MDS coded placement (where coding is only within each file).
We then derive the performance of Theorem~\ref{thm:HT1} with our MDS coded placement to show the achievability of the point $(1,1/2)$.

The case $\Ksf = 2$ (classical model without privacy) was solved in~\cite{maddah2014fundamental}, so it will not be discussed here.
Next, we go into the proof details for the case of $\Ksf=3$ users only, which is the simplest case that highlights the novelty of our new schemes.
The general case $\Ksf\geq 3$ follows from the proofs in Appendix~\ref{sec:NEW1achievablescheme} and Appendix~\ref{sec:proofofHT2}.

\subsubsection{Case $\Ksf=3$ and $\Msf=1/2$: HT2 Scheme in Theorem~\ref{thm:HT2}}
\label{par:K'=N=2:M=1/2}
Our first new scheme with MDS-coded placement attains only one corner point on the converse bound from~\cite{maddah2014fundamental} (given by the optimal performance of the classical setting without privacy with two users and two files). 
In~\cite{maddah2014fundamental} it was shown that the point $(\Msf,\Rsf)=(1/2,1)$ can be achieved by coded placement in the classical setting without privacy with two users and two files. We next combine the idea of~\cite{maddah2014fundamental} with our MDS coded placement idea to show that $(\Msf,\Rsf)=(1/2,1)$ is achievable for $\Ksf=3$. 

Placement Phase:
The $\Nsf = 2$ files are partitioned into two equal-size subfiles;
after partitioning, we treat the files as column vectors of length 2 over a suitably large finite field as follows
\begin{align*}
    &F_1 =[A_1; A_2] =: \Am, \\ 
    &F_2 =[B_1; B_2] =: \Bm,  
\end{align*}
The cache contents are obtained by coding the files as follows
\begin{align*}
    &Z_1 = A_1+B_1 = \gv_1 (\Am+\Bm), \ \text{where} \ \gv_1:=[1,0], \\
    &Z_2 = A_2+B_2 = \gv_2 (\Am+\Bm), \ \text{where} \ \gv_2:=[0,1], \\
    &Z_3 = A_1+A_2 + B_1+B_2 = \gv_3 (\Am+\Bm), \ \text{where} \ \gv_3:=[1,1].
\end{align*}
\label{eq:codedMANplacemnt}
Note that here we have two levels of coding. First we combine all the files together as $\Am+\Bm$, and we split this sum into two symbols $[A_1+B_1; A_2+B_2]$.  Then, we code the two symbols with a rate $2/3$ MDS code with generator matrix $[\gv_1;\gv_2;\gv_3]$ so as to obtain three coded symbols.
Finally, we place one coded symbol in each cache.

Delivery Phase:
Out of the $\Ksf=3$ users, $\KsfActive = 2$ users will be active during the delivery.
When the pair of active users requests the same file, the server transmits the requested file, and the load is $1$ file.
Otherwise, for the pair of active users $(i,j)$ with $d_i=1,\ d_j=2, \ i\not=j,$ the signal sent is the server is
\begin{align*}
X = ( \gv_j \Am, \ \gv_i \Bm ).
\end{align*}

Decoding:
User $i$, who requested file $\Am$,  does
\begin{align*}
\begin{bmatrix} Z_i-\gv_i \Bm \\ \gv_j \Am \\ \end{bmatrix} 
= \begin{bmatrix} \gv_i   \\ \gv_j \\ \end{bmatrix}  \Am.
\end{align*}
Since the matrix $[\gv_i; \gv_j ]$ is $2 \times 2$ and full rank matrix, user $i$ can invert it and recover file $\Am=[A_1; A_2]$.
Similarly for user $j$, who requested file $\Bm$. 

Performance:
We can serve any pair of users, regardless of the demands, with load $\Rsf^\text{\rm HT2}=1$.

\subsubsection{Case $\Ksf=3$ and $\Msf=1$: HT1 Scheme in Theorem~\ref{thm:HT1}}
\label{par:K'=N=2:M=1}
In Fig.~\ref{fig: hotplug (3,2,2)} we show the cache contents and the delivery signals for memory size $\Msf=1$ and $\Ksf=3$ users. 

Placement Phase:
The files are partitioned into two equal-size subfiles as $F_1=[A_1, A_2]$ and $F_2=[B_1, B_2]$. 
The subfiles of each file are coded with an MDS code of rate $2/3$. The cache contents are 
\begin{align*}
    Z_1 &= [A_1, B_1], \\ 
    Z_2 &= [A_2, B_2], \\ 
    Z_3 &= [A_1+A_2, B_1+B_2].
\end{align*}
Note that the third user caches the parity symbols. 
Fig.~\ref{fig: caches (3,2,2)} shows the cache contents. 

Delivery Phase:
Regardless of which user is active and what the other two users demand, each active user must receive the missing half of the demanded file. Fig.~\ref{fig: signals (3,2,2)} gives the signals sent by the server according to Theorem~\ref{thm:HT1}, for two different demand vectors as a function of which user is offline; all the other demand vectors can be dealt similarly. 

Decoding:
Consider, for example, the case where user~$1$ is offline, user~$2$ wants $F_2=[B_1, B_2]$, and user~$3$ wants $F_1=[A_1, A_2]$.
The server sends $X=A_2+B_1+B_2$. 
User~$2$ has $[A_2, B_2]$ in its cache and misses $B_1$; by $X-A_2-B_2$, it recovers $B_1$.
User~$3$ has $A_1+A_2$ in its cache and misses another independent linear combination of $[A_1, A_2]$; by $X-(B_1+B_2)$ he recovers $A_2$; then, with $A_2$ and the cached content he recovers $A_1$.
All other cases are dealt similarly.

Performance:
The load is $\Rsf_1^\text{\rm HT1}=1/2$. 

\subsubsection{Overall Performance}
In Fig.~\ref{fig: tradeoffs (3,2,2)} the red solid line shows the memory-load tradeoff attained by HT1 in~\eqref{eq:performanceNEW1} and HT2 in~\eqref{eq:performanceNEW2}, which is the lower convex envelope of the corner points $(1/2,1)$ and $(1,1/2)$, achieved by our novel schemes, with the trivial corner points $(0, 2)$ and $(2, 0)$; this optimal tradeoff is actually achievable for any $\Ksf\geq 3$ and only depends on $\KsfActive = \Nsf = 2$.
The blue line represents the memory-load tradeoff of YMA+ scheme in Theorem~\ref{thm:extensionYMA} with $(\KsfActive, \Ksf, \Nsf) = (2, 3, 2)$, where the demand of offline user is chosen the same as the first active user.
For comparison, we also added the performance of the decentralized coded caching scheme in violet dashed line, given by~\eqref{eq:performanceDecentralized+}; the decentralized performance does not depend on $\Ksf$ and is an upper bound for the centralized performance for any $\Ksf$. 
This show that load savings are possible when the system is aware that only $\KsfActive = 2$ users out of $\Ksf\geq 3$ can be active.

\subsection{Schemes for $\Ksf \geq \KsfActive = \Nsf = 3$}
\label{sec:K'=N=3}
We consider now the hotplug system with $\KsfActive = 3$ active users and $\Nsf = 3$ files. 
We shall show the achievability of our HT's schemes in Theorem~\ref{thm:HT1}, in Theorem~\ref{thm:HT2}, and in Theorem~\ref{thm:HT3}.

The classical model with three users, i.e., $\Ksf =3$, and three files was addressed in~\cite{tian2018symmetry}, so we do not repeat it here. 
The achievable scheme in~\cite{tian2018symmetry} for three users and three files gives us the achievability for our hotplug with $\Ksf=\KsfActive =3$ users, while the converse in~\cite{tian2018symmetry} for three users and three files gives us a converse bound for any $\Ksf \geq \KsfActive =3$.
We consider here in detail only the case of $\Ksf = 4$ users. 
The general case $\Ksf \geq 4$ follows the proofs in Appendix~\ref{sec:NEW1achievablescheme}, Appendix~\ref{sec:proofofHT2} and Appendix~\ref{sec:proofofFLEX}.

\subsubsection{Case $\Ksf=4$ and $\Msf=1/3$: HT2 Scheme in Theorem~\ref{thm:HT2}}
\label{par:K'=N=3:M=1/3}
We can achieve $\Rsf^\text{\rm HT2}=2$ for memory $\Msf=1/3$, similarly to Section~\ref{par:K'=N=2:M=1/2}.

\subsubsection{Case $\Ksf=4$ and $\Msf = 1$: HT1 Scheme in Theorem~\ref{thm:HT1}}
\label{par:K'=N=3:M=1}
We can achieve $\Rsf^\text{\rm HT1}=1$ for memory $\Msf=1$, similarly to Section~\ref{par:K'=N=2:M=1}.

\subsubsection{Case $\Ksf=4$ and $\Msf = 2$: HT3 Scheme in Theorem~\ref{thm:HT3}}
\label{par:K'=N=3:M=2}
We now explain how scheme HT3 works. Scheme HT3 was not needed in the previous example.
 
Placement Phase:
We partition each files into three equal length sub-files and treat the files as column vectors of length 3 over a suitably large finite field as follows
\begin{align*}
    &F_1 = [F_{1,1}, F_{1,2}, F_{1,3}], \\
    &F_2 = [F_{2,1}, F_{2,2}, F_{2,3}], \\ 
    &F_3 = [F_{3,1}, F_{3,2}, F_{3,3}].
\end{align*}
User~$k$ is assigned a $2 \times 3$ generator matrix 
\begin{align*}
\Gm_k = \begin{bmatrix} \gv_{k,1} \\ \gv_{k,2} \end{bmatrix} : 
    \gv_{k,i} = [1, \ 2k+i-2, \ (2k+i-2)^2], \quad \forall i\in[2], \forall k \in [4].
\end{align*}
Equivalently, if we concatenate all $\{\gv_{k,i}: i \in [2], k \in [4]\}$ together, we obtain a Vandermonde matrix of dimension $8 \times 3$.
The cache contents are as follows 
\begin{align*}
Z_k = \big[\Gm_{k} F_n:  n \in [\Nsf]\big], \quad \forall k \in [\Ksf].
\end{align*}

Delivery Phase:
WLOG, assume the active user triple is $\Ic = \{1,2,3\}$ and their demands are $\dv[\Ic]=[d_1, d_2, d_3]$.
For this set of active users, the generator matrixes are
\begin{align*}
    \Gm_1 = \begin{bmatrix}
        1 & 1 & 1 \\
        1 & 2 & 4
    \end{bmatrix}, \ \Gm_2 = \begin{bmatrix}
        1 & 3 & 9 \\
        1 & 4 & 16
    \end{bmatrix}, \ \Gm_3 = \begin{bmatrix}
        1 & 5 & 25 \\
        1 & 6 & 36
    \end{bmatrix}.
\end{align*}
Define the vectors
\begin{align*}
\vv_{1} = [2, 9, 39] = [-1, 3] \Gm_2 = [3, -1] \Gm_3, \\
\vv_{2} = [2, 7, 17] = [-3, 5] \Gm_1 = [5, -3] \Gm_3, \\
\vv_{3} = [2, 5, 11] = [-1, 3] \Gm_1 = [3, -1] \Gm_2, 
\end{align*}
which have the following properties
\begin{itemize}
    \item $[\Gm_i; \vv_{i}]$ is a $3 \times 3$ full rank matrix, for all $i \in [3]$; and
    \item $\vv_{i}$ lies in the linear span of $\Gm_j$ for every $j \in [3] \setminus \{i\}$. 
\end{itemize}
The server sends the message $X$ as follows,
\begin{align*}
    X &= \sum_{j\in[3]} \vv_{j} F_{d_j}. 
\end{align*}
From the point of view of active user~$1$ we can write
\begin{align*}
    X &= \vv_{1} F_{d_1} + \vv_{2} F_{d_2} + \vv_{3} F_{d_3}
    \\&= \underbrace{\vv_{1} F_{d_1}}_{\text{demanded by user }1} 
    + \ [-3, 5] \underbrace{\Gm_1 F_{d_2}}_{\text{cached by user }1} 
    + \ [-1, 3] \underbrace{\Gm_1 F_{d_3}}_{\text{cached by user }1},
\end{align*}
so user~$1$ can recover $\vv_{1} F_{d_1}$, and further restore $F_{d_1}$ as $[\Gm_1;\vv_1]$ is full rank.
Similarly for the other two active users. 

Performance:
We can serve any triplet of users, regardless of their demands, with load $\Rsf^\text{\rm HT3}=1/3$.

\subsubsection{Overall Performance}
Fig.~\ref{fig: tradeoffs (4,3,3)} shows the memory-load tradeoffs for the hotplug system with $\KsfActive=\Nsf=3$.
In Fig.~\ref{fig: tradeoffs (4,3,3)}, the lower convex envelop of our three optimal corner points is shown by the red sold line, which do not depend on $\Ksf$. 
The converse in Fig.~\ref{fig: tradeoffs (4,3,3)}, in dash-dotted grey line, is derived from~\cite[Fig.~5]{tian2018symmetry}, which matches the lower convex envelop of our  corner points except for $\Msf\in(1/3,1)$ which is open even in the classical setting~\cite{tian2018symmetry}.
The solid blue line is the YMA+ scheme in~\eqref{eq:performanceYMA+} for $\Ksf=4$ users, while the violet dotted line is decen+ scheme in~\eqref{eq:performanceDecentralized+} also for $\Ksf=4$ users.

\begin{figure}
    \centering 
    \begin{tikzpicture}
        \begin{axis}[
            xmin=-0.0, xmax=3.0,
            ymin=-0.0, ymax=3.0,
            legend entries={ 
                \small HT in~\eqref{eq:performanceNEW1} \eqref{eq:performanceNEW2}~\eqref{eq:performanceNEW3},
                \small YMA+ in~\eqref{eq:performanceYMA+},
                \small decen+ in~\eqref{eq:performanceDecentralized+},
                \small Converse from~\cite{tian2018symmetry}
            },
            height=8cm,
            width=10cm,
            grid=major,
            grid style=dashed,
            legend pos=north east,
            xlabel={\small memory size $\Msf$},
            ylabel={\small worst-case load $\Rsf$}
        ]
        \addplot[color=red!70, mark=*, semithick] coordinates {
            (0,3) (0.333, 2) (1, 1) (2, 0.333) (3,0)
        };
        \addplot[color=blue, mark=triangle*, semithick] coordinates {
            (0,3) (0.75, 1.5) (1.5, 0.666) (2.25, 0.25) (3,0)
        };
        \addplot[color=violet!80, loosely dashdotted, thick] table [x=x, y=y] {data/decentralized3.data};
        \addplot[color=gray, densely dashdotted, thick] coordinates {
            (0,3) (0.333, 2) (0.666, 1.333) (1, 1) (2, 0.333) (3,0)
        };
        \end{axis}
    \end{tikzpicture}
    \caption{\small Memory-load tradeoffs for the hotplug system with $\Nsf=3$ files, $\KsfActive=3$ active users, and $\Ksf=4$ total users. 
    The converse is achievable for any $\Ksf \geq \KsfActive=3$ with $\Nsf=3$, except for $\Msf\in(1/3,1)$ which is open even in the classical setting~\cite{tian2018symmetry}. The performance of our scheme does not depend on $\Ksf$.}
    \label{fig: tradeoffs (4,3,3)}
\end{figure}
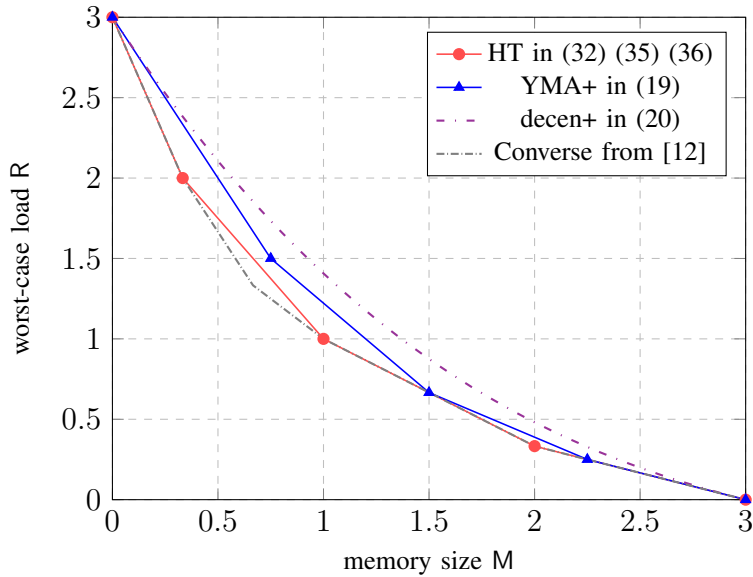

\section{Example for Hotplug Model with Demand Privacy}
\label{sec:exampledemandprivacy}

\subsection{Uncoded Scheme} 
Each user caches the first $\Bsf\Msf/\Nsf$ symbols of each file, and the server sends the remaining $\Bsf(1-\Msf/\Nsf)$ symbols of each file regardless of the demands; the attained tradeoff is $\Rsf^\text{\rm(uncoded)}(\Msf)=\Nsf-\Msf$; this scheme is clearly private and the users can compute locally any functions, not just downloading single files. Thus, we attain the trivial corner points $(\Msf,\Rsf)=(0,\Nsf)=(0, 3)$ and $(\Msf,\Rsf)=(\Nsf,0)=(3, 0)$.
This holds for any set of parameters and does not depend on  $\Ksf$ or $\KsfActive$.

\subsection{Schemes for $(\KsfActive,\Ksf,\Nsf) = (3,6,3)$ with privacy}
We provide an example to illustrate our new proposed schemes for the $(\KsfActive,\Ksf,\Nsf) = (3,6,3)$ hotplug model with demand privacy against colluding users. %

\subsubsection{HT1+PK Scheme in Theorem~\ref{thm:HT1+PK}: Case $t=0$}
\label{par:K'=N=3:M=1 HT1+PK t=0}
Placement Phase:
For $t=0$ there is no file split and no MDS coding. Each user~$j\in[6]$ only caches a privacy key
\begin{align*}
Z_j = T(\pv_j,1)= \sum_{n\in[3]} p_{j,n} F_n, \quad \forall  j\in[6],
\end{align*} 
where $\pv_j =[p_{j,1},p_{j,2},p_{j,3}]$  is sampled i.i.d. uniformly at random from the set of vectors such that $\sum_{n\in[3]} p_{j,n} = \qsf-1$. 
The memory size is $\Msf=1$.

Delivery Phase:
For active user set $\Ic=\{a,b,c\}$ and corresponding demands $\dv[\Ic]=(d_a, d_b, d_c)$, we let $\qv_j = \pv_j + \ev_{d_j}$ for every $j \in \Ic$; note that the entries of each $\qv_j$ sum to zero in $\mathbb{F}_\qsf$.
Then the server sends 
\begin{align*}
    X_{\{a\}} &= \Tbp(\qv_a,1),
    \\
    X_{\{b\}} &= \Tbp(\qv_b,1),
\end{align*}
as well as the query matrix $\Qm[\Ic]$ and $\Ic$.
We note that there exist two coefficients, say $\alpha_a$ and $\alpha_b$, from $\mathbb{F}_\qsf$ such that $\qv_c = \alpha_a \qv_a + \alpha_2 \qv_b$ (because the matrix $[\qv_a; \qv_b; \qv_c]$ has rank 2 on $\mathbb{F}_\qsf$), so that
\begin{align}
    X_{\{c\}} &= \Tbp(\qv_c, 1) = \alpha_a \Tbp(\qv_a, 1) + \alpha_b \Tbp(\qv_b, 1),
\end{align}
i.e., $X_{\{c\}}$ is linearly dependent to $X_{\{a\}}$ and $X_{\{b\}}$, and thus need not be sent.

Decoding:
From the delivery and the cache content, each user $j\in\Ic$ has its privacy key $\Tbp(\pv_j, 1)$, thus it can extract the desired file $\Tbp(\ev_{d_j}, 1) = F_{d_j}$.
Thus, having decoded the whole library, the delivery is obviously private. 

Performance:
The cache size is $\Msf=1$ and the load is $\Rsf_{\text{\rm p}}=2$. 
Note that in this case $\Rsf^\text{\rm(uncoded)}(1)=2$ as well, so we do not need to use our scheme; we have introduce it here to show a case where some multicast messages need not be sent.

\subsubsection{HT1+PK Scheme in Theorem~\ref{thm:HT1+PK}: Case $t=1$}
\label{par:K'=N=3:M=5/3 HT1+PK t=1}
Placement Phase:
Partition each file as
\begin{align*}
    F_n &= [F_{n,1}, F_{n,2}, F_{n,3}]
      \\&= [ \Tbp(\ev_n,\ev_{\{1\}}), \Tbp(\ev_n,\ev_{\{2\}}), \Tbp(\ev_n,\ev_{\{3\}}) ], \quad \forall n \in [3].
\end{align*}
Let $\Gm = [\gv_{\ell}: \ell \in [6]],$ be the generator matrix of an MDS code of rate $3/6$ , i.e., any $3$ vectors $\gv$'s are linear independent. 
Sample  $\pv_j \in \mathbb{F}_\qsf^3$ i.i.d. uniformly at random from the set of vectors such that $\sum_{n\in[3]} p_{j,n} = \qsf-1$, for each user~$j \in [6]$.
The caches are populated as
\begin{align*}
    Z_j &= \{\pv_j\} \cup \{\Tbp(\ev_n, \gv_{j}): n \in [3]\} 
    \\&\bigcup \{\Tbp(\pv_j, \gv_{\xi_{j,\ell}}): \ell \in [2]\}, \quad \forall j \in [6],
\end{align*}
where $\{\xi_{j,1}, \xi_{j,2}\}$ are two indices different from $j$. %
The terms of the form $\Tbp(\pv_j, \gv_{\xi_{j,\ell}})$ are privacy keys that helps in the decoding process.
The memory size is $\Msf=5/3$.

Note that, by definition of $\Gm$, its $3 \times 3$ sub-matrix $[\gv_{j}; \gv_{\xi_{j,1}}; \gv_{\xi_{j,2}}]$ is full rank for all $j \in [6]$.
This fact is important as it allows users to locally compute certain privacy keys as follows.
Consider user~$1$. It has cached $\pv_1$ as well as $\Tbp(\ev_n,\gv_1)$ for $n\in[3]$. With these, he can compute 
\begin{align*}
    \Tbp(\pv_1,\gv_1)  = \sum_{n\in[3]} p_{1,n} \Tbp(\ev_n,\gv_1).
\end{align*} 
With $\Tbp(\pv_1,\gv_1)$ and the cached privacy keys $\Tbp(\pv_1,\gv_{\xi_{1,1}})$ and $\Tbp(\pv_1,\gv_{\xi_{1,2}})$, it can compute 
\begin{align*}
\Tbp(\pv_1,\gv_u)=  a_{u,1} \Tbp(\pv_1,\gv_1) + a_{u,2} \Tbp(\pv_1,\gv_{\xi_{1,1}}) + a_{u,3} \Tbp(\pv_1,\gv_{\xi_{1,2}}), \quad \forall  u\in[6]\setminus\{1,\xi_{1,1}, \xi_{1,2}\},
\end{align*}
since $[\gv_{1}; \gv_{\xi_{1,1}}; \gv_{\xi_{1,2}}]$ if full rank by the MDS property and thus we can express $\gv_u = a_{u,1} \gv_1 + a_{u,2} \gv_{\xi_{1,1}} + a_{u,3} \gv_{\xi_{1,2}} \in \mathbb{F}_\qsf^3$ for some choice of $[a_{u,1},a_{u,2},a_{u,3}]\in \mathbb{F}_\qsf^3$.
In other words, user $u$ has all bilinear combinations of the form $\Tbp(\pv_u, \cdot)$ and $\Tbp(\cdot,\gv_u), \ u\in[6]$.

Delivery Phase:
As an example, consider active user set $\Ic = \{4,5,6\}$ and demands $\dv[\Ic]=[1,2,3]$.
Let $\qv_j = \ev_{d_j} + \pv_j, j\in\Ic$. 
The server sends 
\begin{align*}
  & X_{\{4,5\}} = \Tbp(\qv_4, \gv_5) + \Tbp(\pv_5, \gv_4) = \Tbp(\ev_{1} + \pv_4, \gv_5) + \Tbp(\ev_{2} + \pv_5, \gv_4), 
\\& X_{\{4,6\}} = \Tbp(\qv_4, \gv_6) + \Tbp(\pv_6, \gv_4) = \Tbp(\ev_{1} + \pv_4, \gv_6) + \Tbp(\ev_{3} + \pv_6, \gv_4), 
\\& X_{\{5,6\}} = \Tbp(\pv_5, \gv_6) + \Tbp(\qv_6, \gv_5) = \Tbp(\ev_{2} + \pv_5, \gv_6) + \Tbp(\ev_{3} + \pv_6, \gv_5), 
\end{align*}
as well as the query matrix $\Qm[\Ic]$ and the active user set $\Ic$.

Decoding:
We now show each how each active user can correctly decode its desired subfiles. User~$4$ misses $\Tbp(\ev_1, \gv_5)$ (in $X_{\{4,5\}}$) and $\Tbp(\ev_1, \gv_6)$ (in $X_{\{4,6\}}$). Since it has all bilinear combinations of the form $\Tbp(\pv_4, \cdot)$ and $\Tbp(\cdot,\gv_4)$, it can recover $\Tbp(\ev_1, \gv_5)$ (from $X_{\{4,5\}}$) and $\Tbp(\ev_1, \gv_6)$ (from $X_{\{4,6\}}$). Finally, with $\Tbp(\ev_1, \gv_j) : j\in\Ic,$ it can compute $F_1$ since $\{\gv_j : j\in\Ic\}$ are linearly independent by the MDS property.
Similarly for the other users.

Performance:
We showed that $(\Msf,\Rsf) = (5/3, 3/3)$ is achievable.

\subsubsection{HT1+PK Scheme in Theorem~\ref{thm:HT1+PK}: Case $t=2$}
\label{par:K'=N=3:M=15/3 HT1+PK t=2}
Placement Phase:
Partition the files into 3 parts as
\begin{align*}
    F_n &= [F_{n,\{1,2\}}, F_{n,\{1,3\}}, F_{n,\{2,3\}}]
      \\&=[ \Tbp(\ev_n,\ev_{\{1,2\}}), \Tbp(\ev_n, \ev_{\{1,3\}}), \Tbp(\ev_n, \ev_{\{2,3\}})], \quad \forall n \in [3].
\end{align*}
Consider the MDS generator matrix  $\Gm = [\gv_{\Tc}: \Tc \in \Omega_{[6]}^2]$ of rate $3/15$, i.e., any $3$ vectors $\gv$'s are linear independent.
Sample $\pv_j \in \mathbb{F}_\qsf^3$ i.i.d. uniformly at random from the set of vectors such that $\sum_{i=1}^3 p_{j,i} = \qsf-1$, for each user~$j, j \in [6]$.
The placement is
\begin{align*}
    Z_j = \{ \pv_j \} \cup \{\Tbp(\ev_n, \gv_{\{j,u\}}): n \in [3],  u\in[6]\setminus\{j\} \}, \quad j \in [6].
\end{align*}
As for the case $t=1$, each user~$j, j \in [6]$, compute the all privacy keys of the form $\Tbp(\pv_j,\gv_{\Tc}) : \Tc \in \Omega_{[6]}^2$, and $\Tbp(\cdot,\gv_{\{j\}\cup\Wc}): \Wc \in \Omega_{[6]\setminus\{j\}}^1$.

Delivery Phase:
As an example, consider active user set $\Ic = \{4,5,6\}$ and demands $\dv[\Ic]=[1,2,3]$.
Here $t+1=|\Ic|=\KsfActive$.
The server sends the active set $\Ic$, the query vector $\qv_j = \ev_{d_j} + \pv_j, j\in\Ic$, and 
\begin{align*}
X_{\Ic} 
  &= \sum_{j\in\Ic} \Tbp(\qv_j,\gv_{\Ic\setminus\{j\}})
\\&= \Tbp(\ev_1+\pv_4,\gv_{\{5,6\}}) + \Tbp(\ev_2+\pv_5,\gv_{\{4,6\}})  + \Tbp(\ev_3+\pv_6,\gv_{\{4,5\}}).
\end{align*}

Decoding:
Each active user decodes its desired file similarly to the case $t=1$, namely 
user~4 recovers $\Tbp(\ev_1,\gv_{\{5,6\}})$, and
user~5 recovers $\Tbp(\ev_2,\gv_{\{4,6\}})$, and
user~6 recovers $\Tbp(\ev_3,\gv_{\{4,5\}})$.
Then, with the cached content and leveraging the MDS property, each active user recovers its desired file.

Performance:
This shows that $(\Msf,\Rsf) = (15/3,1/3)$ is achievable. Here $5 = \Msf > \Nsf = 3$, i.e., each user caches more than the whole library.
The point to include this case here was to show a case where users need not store extra privacy keys.

\subsubsection{HT3+PK Scheme in Theorem~\ref{thm:HT3+PK}}
\label{par:K'=N=3:M=7/3 HT3+PK}
Placement Phase:
Partition each file into three equal size parts as
\begin{align*}
    F_n &= [F_{n,\{1,2\}}, F_{n,\{1,3\}}, F_{n,\{2,3\}}]
      \\&= [ \Tbp(\ev_n,\ev_{\{1,2\}}), \Tbp(\ev_n, \ev_{\{1,3\}}), \Tbp(\ev_n, \ev_{\{2,3\}}) ], \quad \forall n \in [3].
\end{align*}

Consider the MDS generator matrix  $\Gm = [\gv_{k, i}: k \in [6], i \in [2]]$ of rate $3/12$, i.e., any $3$ vectors $\gv$'s are linear independent.
Sample  $\pv_j \in \mathbb{F}_\qsf^3$ i.i.d. uniformly at random from the set of vectors such that $\sum_{i=1}^3 p_{j,i} = \qsf-1$, for each user~$j \in [6]$.
 
We select a vector $\xi_j$ that it is linear independent of $\{\gv_{j,1}, \gv_{j,2}\}$. 
Then we populate the cache of user $j$ as follows
\begin{subequations}
\begin{align}
    Z_j &= \{\Tbp(\ev_n, \gv_{j,i}): n \in [3], i \in [2]\} \label{eq:newpkscheme2examplecache1} \\
    & \ \cup \{\Tbp(\pv_j, \xi_{j})\} \cup \{\pv_j\}, \quad j\in[6]. \label{eq:newpkscheme2examplecache2}
\end{align}
\end{subequations}
For example, we can choose $\Gm$ be a Vandermonde matrix and $\xi$ be a vector which independent of $\Gm$ and set
\begin{align*}
    \Gm = \begin{bmatrix}
        1 & 1 & 1 & \ldots & 1 \\
        1 & 2 & 3 & \ldots & 12 \\
        1 & 4 & 9 & \ldots & 12^2
    \end{bmatrix}^T, \
    \xi_1 = \xi_2 = \ldots \xi_6 = [1, 13, 13^2].
\end{align*}
Note that each user $j, j \in [6]$, despite only caching one privacy key $\Tbp(\pv_j, \xi_j)$, can compute locally with the help of its cached content in~\eqref{eq:newpkscheme2examplecache1} all bilinear combinations of the form $\Tbp(\pv_j, \cdot)$; it can also compute all bilinear combinations of the form $\Tbp(\cdot, \gv_{j,1})$ and $\Tbp(\cdot, \gv_{j,2})$.

Delivery Phase: 
As an example, consider active user set $\Ic$ and demands $\dv[\Ic]$. %
Let $\{\phi^{(\Ic)}_{j}: j \in [\Ic]\}$ be a set of vectors of length $3$ such that:
\begin{itemize}
    \item for $k \in \Ic \setminus \{j\}$, $\phi^{(\Ic)}_{j}$ lies in the linear span of $\{\gv_{k,1}, \gv_{k,2}\}$; and
    \item the matrix $[ \phi^{(\Ic)}_{j}; \gv_{j,1}; \gv_{j,2}]$ is full rank for all $j \in [\Ic]$.
\end{itemize}
For example, for $\Ic= \{1,2,3\}$ we can choose
\begin{align*}
    &\phi^{(\{1,2,3\})}_{1} = [2, 9, 39] = - \gv_{2,1} + 3 \gv_{2,2} = 3 \gv_{3,1} -  \gv_{3,2}, \\
    &\phi^{(\{1,2,3\})}_{2} = [2, 7, 17] = -3 \gv_{1,1} + 5 \gv_{1,2} = 5 \gv_{3,1} -3 \gv_{3,2}, \\ 
    &\phi^{(\{1,2,3\})}_{3} = [2, 5, 11] = - \gv_{1,1} + 3 \gv_{1,2} = 3 \gv_{2,1} -1\gv_{2,2}.
\end{align*}
Let $\qv_j = \ev_{d_j} + \pv_j, j\in\Ic$.
The server sends $\Qm[\Ic]$ and $\Ic$ and
\begin{align*}
    X_\Ic
      &= \sum_{j \in \Ic} \Tbp(\qv_j, \phi^{(\Ic)}_{j})
    \\&= \Tbp(\ev_{d_j}, \phi^{(\Ic)}_{j}) + \text{``terms known at user~$j$''}, \quad \forall j\in\Ic.
\end{align*}

Decoding:
The sent multicast message allows each active user to recover $\Tbp(\ev_{d_j}, \phi^{(\Ic)}_{j}), j \in\Ic.$
Indeed, for $\Ic= \{1,2,3\}$ and user~$1$, we can write
\begin{align*}
    X_{\{1,2,3\}}
      &= \Tbp(\qv_1, \phi^{(\{1,2,3\})}_{1}) + \Tbp(\qv_2, \phi^{(\{1,2,3\})}_{2}) + \Tbp(\qv_1, \phi^{(\{1,2,3\})}_{3}) 
    \\&= \Tbp(\ev_{d_1} + \pv_1, \phi^{(\{1,2,3\})}_{1}) + \Tbp(\qv_2, -3 \gv_{1,1} + 5 \gv_{1,2}) + \Tbp(\qv_3, - \gv_{1,1} + 3 \gv_{1,2})
    \\&= \Tbp(\ev_{d_1}, \phi^{(\{1,2,3\})}_{1})  + \text{``terms known at user~1''},
\end{align*}
so it recovers $\Tbp(\ev_{d_1}, \phi^{(\{1,2,3\})}_{1})$.
User~$1$ also has 
$\Tbp(\ev_{d_1}, \gv_{1,1})$ and
$\Tbp(\ev_{d_1}, \gv_{1,2})$, and thus by the MDS property, it can recover file $F_{d_1}$.

Performance:
The point $(\Msf, \Rsf)=(7/3, 1/3)$ is achievable.

\subsubsection{HT\&VU Scheme in Theorem~\ref{thm:HT+VU}}
\label{par:K'=N=3:M=1/7 HT+VU}
Placement Phase: 
Partition each file into $7$ parts as,
\begin{align*}
    F_n &= (F_{n,\ell}: \ell \in [7]) = (\Tbp(\ev_n, \ev_\ell): \ell \in [7]), \ \forall n \in [3].
\end{align*}
Consider the MDS generator matrix $\Gm = [\gv_{j}: j \in [18]]$ of rate $7/18$, i.e., any $7$ vectors $\gv$'s are linear independent.
Sample the scalar $p_j \in [3]$ i.i.d. uniformly at random for each user~$j \in [6]$.

The placement is,
\begin{align*}
    Z_j = \{p_j\} \cup \{\sum_{n=1}^{3} \Tbp(\ev_n, \gv_{3(j-1)+p_j})\}, \ \forall j \in [6].
\end{align*}
 
Delivery Phase: 
Let $\Psi: [3]^3 \rightarrow [3]^3$ denote the {\it cyclic shift operator}, such that $\Psi(1,2,3)=(3,1,2)$.
Let us denote a vector $\mathbb{I}_3:=(1,2,3)$, and define the subtraction under modular $3$ as $\ominus$, for example $1 \ominus 1 = 0, 1 \ominus 2 = 2$.
Consider a active user set $\Ic = \{j_1, j_2, j_3\} \in \Omega_{6}^{3}$ and their demands $d_{\Ic} \in \{d_{j_1}, d_{j_2}, d_{j_3}\}$, the server creates an vector of length $3$ for each user $j \in \Ic$ as
\begin{align}
\widetilde{\dv}_j
:= \Psi^{p_{j}\ominus d_{j}}(\mathbb{I}_3), \label{eq:expendeddemands}
\end{align}
where the operator $\Psi^i$ denotes the $i$-times cyclic shift operator, which is the operator $\Psi$ applied $i$ times.
We define the expended active user set $\widetilde{\Ic}$ of size $9$ as,
\begin{align}
    \widetilde{\Ic} := \left((j-1)\Nsf + \ell: j \in \Ic, \ell \in [3]\right), \label{eq:expendedactiveuser}
\end{align}
and 
\begin{align}
p_\Ic \ominus d_\Ic := \left( p_{j_1} \ominus d_{j_1}, p_{j_2} \ominus d_{j_2}, p_{j_3}\ominus d_{j_3} \right), \label{eq:onetimepaddemands}
\end{align}
Therefore, the server sends $(\widetilde{\Ic}, p_\Ic \ominus d_\Ic)$, and
\begin{align*}
    X = \left\{\Tbp(\ev_n, \gv_{j}): j \in \widetilde{\Ic}, \ell \in [3], n \in [3] \setminus \{\tilde{d}_{j,\ell}\} \right\},
\end{align*}
where $\tilde{d}_{j,\ell}$ is the $\ell$-th element of $\widetilde{\dv}_j$.

Decoding: User $j \in [\Ic]$, who knows $\{\Tbp(\ev_n, \gv_{3(j-1)+p_j}): n \in [3] \setminus \{d_j\}\}$ from $X$, can ``unlocks'' $\Tbp(\ev_{d_j}, \gv_{3(j-1)+p_j})$ from its cache.
Next, for every $i \in \Ic$ and $\ell \in [3]$ such that $\tilde{d}_{i,\ell} \neq d_j$, the bilinear combination $\Tbp(\ev_{d_j}, \gv_{3(i-1)+\ell})$ is contained in $X$, in total $6$.
Since user~$j$ knows $7$ independent coded subfiles and the MDS rate is $7/18$,
it can restore $F_{d_j}$.

Performance: The point $(\Msf, \Rsf) = (1/7, 18/7)$ is achievable.

\subsubsection{HT1\&VU and HT3\&VU Schemes in Theorem~\ref{thm:HT+VU}}
\label{par:HT+VU again}
These tradeoffs is attained from non-private HT1 and HT3 Schemes respectively in Theorem~\ref{thm:HT3} with parameters $(\widetilde{\KsfActive}, \widetilde{\Ksf}, \Nsf) = (9,18,3)$ and a restricted demands. Each real user $j \in [6]$, together with 2 extra virtual users forms a group. The server populates the cache content for all users including virtual ones, then the user uniformly chooses one content from each group at random. In the delivery phase, once the real users sent their demands, the server transmits an expended demands and expended active user set, as shows in~\eqref{eq:expendeddemands} and~\eqref{eq:expendedactiveuser} respectively. Then it transmits the multicast messages generated by expended active set and demands given by the non-private hotplug scheme. The examples for HT1 and HT3 schemes are referred to Section~\ref{sec:examplehotplug}.

Performance: The point $(\Msf, \Rsf) = (1/3, 14/6)$ and $(\Msf, \Rsf) = (24/9, 1/9)$ are achievable respectively.

\subsubsection{Overall Performance}
Fig.~\ref{fig: memory-tradeoff for (6,3,3)} compares the performance of  
the PK+ scheme in Theorem~\ref{thm:extensionPK}, 
the YMA\&VU scheme in Theorem~\ref{thm:extensionVU}, and 
the new schemes, i.e., the line labeled `HT+PK' is the convex hull of the corner points from Theorems~\ref{thm:HT1+PK} and~\ref{thm:HT3+PK}, and the three tradeoffs labeled `HT\&VU' from Theorem~\ref{thm:HT+VU}. 
The converse bound is derived as detailed in Remark~\ref{rem:converse}.

This is a case where our HT+PK schemes outperforms the PK+ scheme but overall YMA\&VU is better.
Two tradeoff derived from HT\&VU and HT3\&VU schemes respectively, are optimal, and the tradeoff derived from HT1\&VU has a similar performance with YMA\&VU.
In Section~\ref{sec:NumExamples} we show that this is not always the case. 
We choose this case as an example as it is the smallest example where we can clearly and simply show all the new aspects of the proposed schemes.

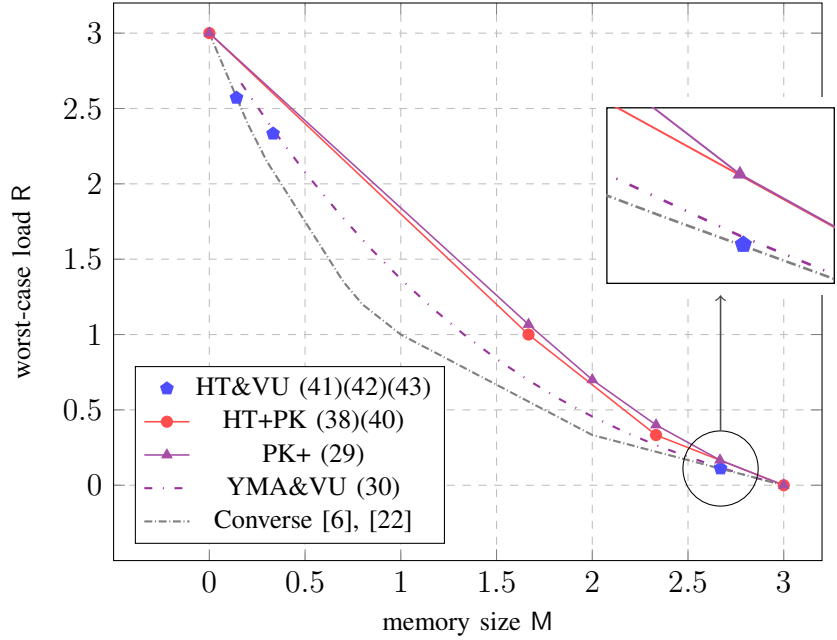
\begin{figure}
    \centering
    \begin{tikzpicture}[every pin/.style={fill=white}]
        \begin{axis}[
            xmin=-0.5, xmax=3.2,
            ymin=-0.5, ymax=3.2,
            xtick={0,0.5,...,3},
            ytick={0,0.5,...,3},
            legend entries={ 
                \small HT\&VU~\eqref{eq:performanceht1+vu}\eqref{eq:performancehtNONE+vu}\eqref{eq:performanceht3+vu},
                \small HT+PK~\eqref{eq:HT1+PK}\eqref{eq:HT3+PK},
                \small PK+~\eqref{eq:performancePK+},
                \small YMA\&VU~\eqref{eq:performanceYMAVU},
                \small Converse~\cite{yu2018characterizing,yan2021fundamental}
            },
            xtick distance=0.5, ytick distance=0.5,
            height=9cm,
            width=11cm,
            grid=major,
            grid style=dashed,
            legend pos=south west,
            xlabel={\small memory size $\Msf$},
            ylabel={\small worst-case load $\Rsf$},
        ]
        \addplot[only marks, color=blue!70, %
        mark size=2.5pt, mark=pentagon*] coordinates {
            (0.142, 2.571) (0.333, 2.333) (2.67, 0.111) 
        };
        \addplot[color=red!70, mark=*, semithick] coordinates {
            (0.0, 3.0) (1.667, 1.0) (2.333, 0.333) (3.0, 0.0)
        };
        \addplot[color=violet!70, mark=triangle*, semithick] coordinates {
            (0.0, 3.0) (1.667, 1.067) (2.0, 0.7) (2.333, 0.4) (2.667, 0.167) (3.0, 0.0)
        };
        \addplot[color=violet!80, loosely dashdotted, thick] table [x=x, y=y] {data/virtusers3.data};
        \addplot[color=gray, densely dashdotted, thick] table [x=x, y=y] {data/converseprivacy3.data};
        \coordinate (spypoint) at (axis cs:2.67, 0.111);
        \end{axis}
        \node[pin={[pin edge={->,black!70,semithick},pin distance=1.8cm]90:{%
        \begin{tikzpicture}[baseline,trim axis left,trim axis right,scale=1.25]
            \begin{axis}[
                    tiny,ticks=none,
                    xmin=2.55,xmax=2.75,
                    ymin=0.08,ymax=0.22,
                    yticklabels={,,},
                    xticklabels={,,}
                ]
                \addplot[only marks, color=blue!70, %
                mark size=2.5pt, mark=pentagon*] coordinates {
                    (0.142, 2.571) (0.333, 2.333) (2.67, 0.111) 
                };
                \addplot[color=red!70, mark=*, semithick] coordinates {
                    (0.0, 3.0) (1.667, 1.0) (2.333, 0.333) (3.0, 0.0)
                };
                \addplot[color=violet!70, mark=triangle*, semithick] coordinates {
                    (0.0, 3.0) (1.667, 1.067) (2.0, 0.7) (2.333, 0.4) (2.667, 0.167) (3.0, 0.0)
                };
                \addplot[color=violet!80, loosely dashdotted, thick] table [x=x, y=y] {data/virtusers3.data};
                \addplot[color=gray, densely dashdotted, thick] table [x=x, y=y] {data/converseprivacy3.data};
            \end{axis}
        \end{tikzpicture}%
        }},draw,circle,minimum size=1cm] at (spypoint) {};
    \end{tikzpicture}
    \caption{\small Memory-load tradeoffs for the hotplug model with privacy when $(\KsfActive,\Ksf,\Nsf) = (3, 6, 3)$.}
    \label{fig: memory-tradeoff for (6,3,3)}
\end{figure}

\section{Numerical Evaluations}
\label{sec:NumExamples}
In this Section we provide some numerical examples, to illustrate the performance of our new schemes. 
For the hotplug model without privacy, the schemes we mention in this section include Theorem~\ref{thm:HT1} and Theorem~\ref{thm:HT3}; 
the converse is derived from~\cite[Theorems~2 and~4]{yu2018characterizing}. Theorem~\ref{thm:HT2} is used in examples with $\KsfActive \geq \Nsf$.
For the hotplug model with  privacy, the schemes in this section include Theorem~\ref{thm:HT1+PK} and Theorem~\ref{thm:HT3+PK}; 
the converse is derived as detailed in Remark~\ref{rem:converse}.

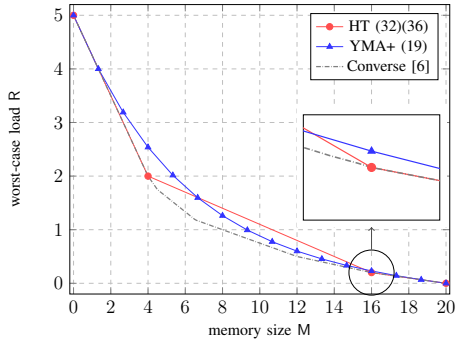
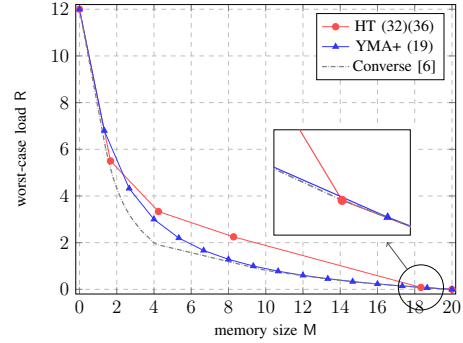
\begin{figure}
    \centering
    \begin{subfigure}[b]{0.45\textwidth}
        \centering 
        \scalebox{0.6}{
        \begin{tikzpicture}[every pin/.style={fill=white}]
        \begin{axis}[
            xmin=-0.2, xmax=20.2,
            ymin=-0.2, ymax=5.2,
            legend entries={
                \small HT~\eqref{eq:performanceNEW1}\eqref{eq:performanceNEW3},
                \small YMA+~\eqref{eq:performanceYMA+},
                \small Converse~\cite{yu2018characterizing}
            },
            height=8cm,
            width=10cm,
            grid=major,
            grid style=dashed,
            legend pos=north east,
            xlabel={\small memory size $\Msf$},
            ylabel={\small worst-case load $\Rsf$}
        ]
        \addplot[color=red!70, mark=*, semithick] coordinates {
            (0.0, 5.0) (4.0, 2.0) (16.0, 0.2) (20.0, 0.0)
        };
        \addplot[color=blue!80, mark=triangle*, semithick] coordinates {
            (0.0, 5.0) (1.333, 4.0) (2.667, 3.19) (4.0, 2.538) (5.333, 2.015) (6.667, 1.597) (8.0, 1.262) (9.333, 0.993) (10.667, 0.776) (12.0, 0.6) (13.333, 0.455) (14.667, 0.333) (16.0, 0.231) (17.333, 0.143) (18.667, 0.067) (20.0, 0.0)
        };
        \addplot[color=gray, densely dashdotted, thick] table [x=x, y=y] {data/converseht5.data};
        \coordinate (spypoint) at (axis cs:16.0, 0.2);
        \end{axis}
        \node[pin={[pin edge={->,black!70,semithick},pin distance=0.5cm]90:{%
        \begin{tikzpicture}[baseline,trim axis left,trim axis right,scale=1.25]
            \begin{axis}[
                    tiny,ticks=none,
                    xmin=15.5,xmax=16.5,
                    ymin=0.1,ymax=0.3,
                    yticklabels={,,},
                    xticklabels={,,}
                ]
                \addplot[color=red!70, mark=*, semithick] coordinates {
                    (0.0, 5.0) (4.0, 2.0) (16.0, 0.2) (20.0, 0.0)
                };
                \addplot[color=blue!80, mark=triangle*, semithick] coordinates {
                    (0.0, 5.0) (1.333, 4.0) (2.667, 3.19) (4.0, 2.538) (5.333, 2.015) (6.667, 1.597) (8.0, 1.262) (9.333, 0.993) (10.667, 0.776) (12.0, 0.6) (13.333, 0.455) (14.667, 0.333) (16.0, 0.231) (17.333, 0.143) (18.667, 0.067) (20.0, 0.0)
                };
                \addplot[color=gray, densely dashdotted, thick] table [x=x, y=y] {data/converseht5.data};
            \end{axis}
        \end{tikzpicture}%
        }},draw,circle,minimum size=1cm] at (spypoint) {};
        \end{tikzpicture}
        }
        \caption{\small Case $(\KsfActive,\Ksf,\Nsf) = (5,15,20)$.} 
        \label{fig: memory-tradeoff for (15,5,20)}
    \end{subfigure}
    \begin{subfigure}[b]{0.45\textwidth}
        \centering 
        \scalebox{0.6}{
        \begin{tikzpicture}[every pin/.style={fill=white}]
        \begin{axis}[
            xmin=-0.2, xmax=20.2,
            ymin=-0.2, ymax=12.2,
            legend entries={ 
                \small HT~\eqref{eq:performanceNEW1}\eqref{eq:performanceNEW3},
                \small YMA+~\eqref{eq:performanceYMA+},
                \small Converse~\cite{yu2018characterizing}
            },
            height=8cm,
            width=10cm,
            grid=major,
            grid style=dashed,
            legend pos=north east,
            xlabel={\small memory size $\Msf$},
            ylabel={\small worst-case load $\Rsf$}
        ]
        \addplot[color=red!70, mark=*, semithick] coordinates {
            (0.0, 12.0)  (1.667, 5.5)  (4.242, 3.333)  (8.273, 2.25)  (18.333, 0.083)  (20.0, 0.0)
        };
        \addplot[color=blue!80, mark=triangle*, semithick] coordinates {
            (0.0, 12.0)  (1.333, 6.8)  (2.667, 4.324)  (4.0, 3.0)  (5.333, 2.2)  (6.667, 1.667)  (8.0, 1.286)  (9.333, 1.0)  (10.667, 0.778)  (12.0, 0.6)  (13.333, 0.455)  (14.667, 0.333)  (16.0, 0.231)  (17.333, 0.143)  (18.667, 0.067)  (20.0, 0.0)
        };
        \addplot[color=gray, densely dashdotted, thick] table [x=x, y=y] {data/converseht12.data};
        \coordinate (spypoint) at (axis cs:18.333, 0.083);
        \end{axis}
        \node[pin={[pin edge={->,black!70,semithick},pin distance=0.5cm]95:{%
        \begin{tikzpicture}[baseline,trim axis left,trim axis right,scale=1.25]
            \begin{axis}[
                    tiny,ticks=none,
                    xmin=17.833,xmax=18.833,
                    ymin=0.05,ymax=0.15,
                    yticklabels={,,},
                    xticklabels={,,}
                ]
                \addplot[color=red!70, mark=*, semithick] coordinates {
            (0.0, 12.0)  (1.667, 5.5)  (4.242, 3.333)  (8.273, 2.25)  (18.333, 0.083)  (20.0, 0.0)
                };
                \addplot[color=blue!80, mark=triangle*, semithick] coordinates {
                    (0.0, 12.0)  (1.333, 6.8)  (2.667, 4.324)  (4.0, 3.0)  (5.333, 2.2)  (6.667, 1.667)  (8.0, 1.286)  (9.333, 1.0)  (10.667, 0.778)  (12.0, 0.6)  (13.333, 0.455)  (14.667, 0.333)  (16.0, 0.231)  (17.333, 0.143)  (18.667, 0.067)  (20.0, 0.0)
                };
                \addplot[color=gray, densely dashdotted, thick] table [x=x, y=y] {data/converseht12.data};
            \end{axis}
        \end{tikzpicture}%
        }},draw,circle,minimum size=1cm] at (spypoint) {};
        \end{tikzpicture}
        }
        \caption{\small Case $(\KsfActive,\Ksf,\Nsf) = (12,15,20)$.}
        \label{fig: memory-tradeoff for (12,15,20)}
    \end{subfigure}
    \caption{\small Memory-load tradeoffs for the hotplug system with $(\Ksf, \Nsf) = (15, 20)$ and different values of $\KsfActive$.}
    \label{fig: memory-tradeoff for hotplug and baseline}
\end{figure}

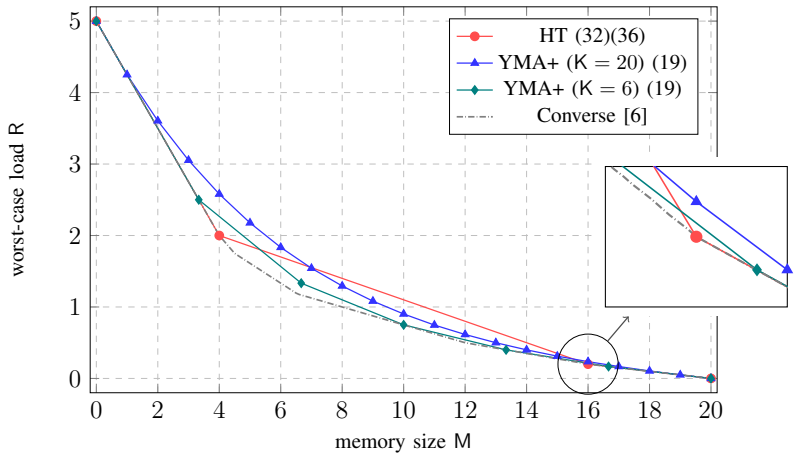
\begin{figure}
    \centering 
    \scalebox{0.8}{
        \begin{tikzpicture}[every pin/.style={fill=white}]
        \begin{axis}[
            xmin=-0.2, xmax=20.2,
            ymin=-0.2, ymax=5.2,
            legend entries={ \small HT~\eqref{eq:performanceNEW1}\eqref{eq:performanceNEW3},
                \small YMA+ ($\Ksf=20$)~\eqref{eq:performanceYMA+},
                \small YMA+ ($\Ksf=6$)~\eqref{eq:performanceYMA+},
                \small Converse~\cite{yu2018characterizing}
            },
            height=8cm,
            width=12cm,
            grid=major,
            grid style=dashed,
            legend pos=north east,
            xlabel={\small memory size $\Msf$},
            ylabel={\small worst-case load $\Rsf$}
        ]
        \addplot[color=red!70, mark=*, semithick] coordinates {
            (0.0, 5.0) (4.0, 2.0) (16.0, 0.2) (20.0, 0.0)
        };
        \addplot[color=blue!80, mark=triangle*, semithick] coordinates {
            (0.0, 5.0) (1.0, 4.25) (2.0, 3.605) (3.0, 3.053) (4.0, 2.58) (5.0, 2.177) (6.0, 1.834) (7.0, 1.542) (8.0, 1.294) (9.0, 1.082) (10.0, 0.902) (11.0, 0.747) (12.0, 0.615) (13.0, 0.5) (14.0, 0.4) (15.0, 0.312) (16.0, 0.235) (17.0, 0.167) (18.0, 0.105) (19.0, 0.05) (20.0, 0.0)
        };
        \addplot[color=teal, mark=diamond*, semithick] coordinates {
            (0.0, 5.0) (3.333, 2.5) (6.667, 1.333) (10.0, 0.75) (13.333, 0.4) (16.667, 0.167) (20.0, 0.0)
        };
        \addplot[color=gray, densely dashdotted, thick] table [x=x, y=y] {data/converseht5.data};
        \coordinate (spypoint) at (axis cs:16,0.2);
        \end{axis}
        \node[pin={[pin edge={->,black!70,semithick},pin distance=0.3cm]80:{%
        \begin{tikzpicture}[baseline,trim axis left,trim axis right,scale=1.25]
            \begin{axis}[
                    tiny,ticks=none,
                    xmin=15,xmax=17,
                    ymin=0.13,ymax=0.27,
                    yticklabels={,,},
                    xticklabels={,,}
                ]
                \addplot[color=red!70, mark=*, semithick] coordinates {
            (0.0, 5.0) (4.0, 2.0) (16.0, 0.2) (20.0, 0.0)
        };
                \addplot[color=blue!80, mark=triangle*, semithick] coordinates {
                    (0.0, 5.0) (1.0, 4.25) (2.0, 3.605) (3.0, 3.053) (4.0, 2.58) (5.0, 2.177) (6.0, 1.834) (7.0, 1.542) (8.0, 1.294) (9.0, 1.082) (10.0, 0.902) (11.0, 0.747) (12.0, 0.615) (13.0, 0.5) (14.0, 0.4) (15.0, 0.312) (16.0, 0.235) (17.0, 0.167) (18.0, 0.105) (19.0, 0.05) (20.0, 0.0)
                };
                \addplot[color=teal, mark=diamond*, semithick] coordinates {
                    (0.0, 5.0) (3.333, 2.5) (6.667, 1.333) (10.0, 0.75) (13.333, 0.4) (16.667, 0.167) (20.0, 0.0)
                };
                \addplot[color=gray, densely dashdotted, thick] table [x=x, y=y] {data/converseht5.data};
            \end{axis}
        \end{tikzpicture}%
    }},draw,circle,minimum size=1cm] at (spypoint) {};
        \end{tikzpicture}
    }
    \caption{\small Memory-load tradeoffs for the hotplug system for with $(\KsfActive, \Nsf) = (5, 20)$ and various values of $\Ksf$.}
    \label{fig: memory-tradeoff among various k}
\end{figure}

\subsection{Hotplug without Privacy}
\indent $\bullet$ 
{Case $(\KsfActive, \Nsf) = (5, 20)$:}
Fig.~\ref{fig: memory-tradeoff among various k} shows the memory-load tradeoff for the case $(\KsfActive, \Nsf) = (5, 20)$ and various $\Ksf$.
The performance of HT schemes and of the converse bound does not depend on the value of $\Ksf$, while that of the YMA+ scheme worsen as $\Ksf$ increases. The last segment is optimal as HT3 scheme matches the converse bound.

\indent $\bullet$ 
{Case $(\Ksf, \Nsf) = (15, 20)$:}
Fig.~\ref{fig: memory-tradeoff for hotplug and baseline} shows the memory-load tradeoff for two different values of $\KsfActive$ for fixed $(\Ksf, \Nsf) = (15, 20)$. For $\Msf \in [0, \Nsf / \KsfActive]$ in Fig.~\ref{fig: memory-tradeoff for (15,5,20)}, the HT1 scheme with MDS coded placement in Theorem~\ref{thm:HT1} outperforms the YMA+ scheme and is exactly optimal in the small memory regime. The last segment is optimal as HT3 scheme matches the converse bound.

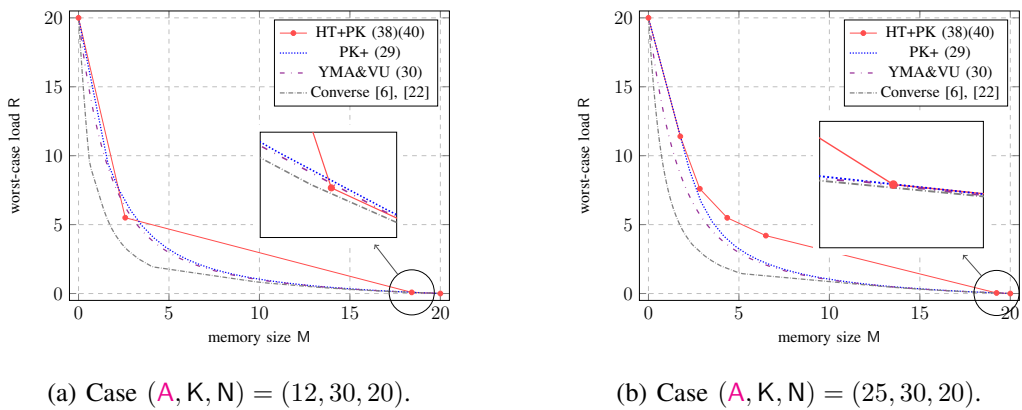
\begin{figure}
    \centering
    \begin{subfigure}[t]{0.45\textwidth}
        \centering 
        \scalebox{0.6}{
        \begin{tikzpicture}[every pin/.style={fill=white}]
        \begin{axis}[
            xmin=-0.5, xmax=20.5,
            ymin=-0.5, ymax=20.5,
            legend entries={ \small HT+PK~\eqref{eq:HT1+PK}\eqref{eq:HT3+PK},
                \small PK+~\eqref{eq:performancePK+},
                \small YMA\&VU~\eqref{eq:performanceYMAVU},
                \small Converse~\cite{yu2018characterizing,yan2021fundamental}
            },
            height=8cm,
            width=10cm,
            grid=major,
            grid style=dashed,
            legend pos=north east,
            xlabel={\small memory size $\Msf$},
            ylabel={\small worst-case load $\Rsf$}
        ]
        \addplot[color=red!70, mark=*, mark size=1.5pt, semithick] coordinates {
            (0.0, 20.0) (2.583, 5.5) (18.417, 0.083) (20.0, 0.0)
        };
        \addplot[color=blue, densely dotted, thick] coordinates {
            (0.0, 20.0) (1.633, 9.4) (2.267, 7.457) (2.9, 5.996) (3.533, 4.887) (4.167, 4.036) (4.8, 3.375) (5.433, 2.854) (6.067, 2.436) (6.7, 2.097) (7.333, 1.817) (7.967, 1.583) (8.6, 1.385) (9.233, 1.214) (9.867, 1.067) (10.5, 0.937) (11.133, 0.824) (11.767, 0.722) (12.4, 0.632) (13.033, 0.55) (13.667, 0.476) (14.3, 0.409) (14.933, 0.348) (15.567, 0.292) (16.2, 0.24) (16.833, 0.192) (17.467, 0.148) (18.1, 0.107) (18.733, 0.069) (19.367, 0.033) (20.0, 0.0)
        };
        \addplot[color=violet!80, loosely dashdotted, thick] table [x=x, y=y] {data/virtusers12.data};
        \addplot[color=gray, densely dashdotted, thick] table [x=x, y=y] {data/converseprivacy12.data};
        \coordinate (spypoint) at (axis cs:18.417, 0.083);
        \end{axis}
        \node[pin={[pin edge={->,black!70,semithick}, distance=0.4cm]100:{%
        \begin{tikzpicture}[baseline,trim axis left,trim axis right,scale=1.25]
            \begin{axis}[
                    tiny,ticks=none,
                    xmin=18,xmax=18.8,
                    ymin=0.05,ymax=0.12,
                    yticklabels={,,},
                    xticklabels={,,}
                ]
                \addplot[color=red!70, mark=*, mark size=1.5pt, semithick] coordinates {
            (0.0, 20.0) (2.583, 5.5) (18.417, 0.083) (20.0, 0.0)
                };
                \addplot[color=blue, densely dotted, thick] coordinates {
                    (0.0, 20.0) (1.633, 9.4) (2.267, 7.457) (2.9, 5.996) (3.533, 4.887) (4.167, 4.036) (4.8, 3.375) (5.433, 2.854) (6.067, 2.436) (6.7, 2.097) (7.333, 1.817) (7.967, 1.583) (8.6, 1.385) (9.233, 1.214) (9.867, 1.067) (10.5, 0.937) (11.133, 0.824) (11.767, 0.722) (12.4, 0.632) (13.033, 0.55) (13.667, 0.476) (14.3, 0.409) (14.933, 0.348) (15.567, 0.292) (16.2, 0.24) (16.833, 0.192) (17.467, 0.148) (18.1, 0.107) (18.733, 0.069) (19.367, 0.033) (20.0, 0.0)
                };
                \addplot[color=violet!80, loosely dashdotted, thick] table [x=x, y=y] {data/virtusers12.data};
                \addplot[color=gray, densely dashdotted, thick] table [x=x, y=y] {data/converseprivacy12.data};
            \end{axis}
        \end{tikzpicture}%
    }},draw,circle,minimum size=1cm] at (spypoint) {};
        \end{tikzpicture}
    }
    \caption{Case $(\KsfActive,\Ksf,\Nsf) = (12, 30, 20)$.} 
    \label{fig: memory-tradeoff for (30,12,20)}
    \end{subfigure}
    \begin{subfigure}[t]{0.45\textwidth}
        \centering 
        \scalebox{0.6}{
        \begin{tikzpicture}[every pin/.style={fill=white}]
        \begin{axis}[
            xmin=-0.5, xmax=20.5,
            ymin=-0.5, ymax=20.5,
            legend entries={ 
                \small HT+PK~\eqref{eq:HT1+PK}\eqref{eq:HT3+PK},
                \small PK+~\eqref{eq:performancePK+},
                \small YMA\&VU~\eqref{eq:performanceYMAVU},
                \small Converse~\cite{yu2018characterizing,yan2021fundamental}
            },
            height=8cm,
            width=10cm,
            grid=major,
            grid style=dashed,
            legend pos=north east,
            xlabel={\small memory size $\Msf$},
            ylabel={\small worst-case load $\Rsf$}
        ]
        \addplot[color=red!70, mark=*, mark size=1.5pt, semithick] coordinates {
            (0.0, 20.0) (1.76, 11.4) (2.837, 7.6) (4.354, 5.493) (6.488, 4.2) (19.24, 0.04) (20.0, 0.0)
        };
        \addplot[color=blue, densely dotted, thick] coordinates {
            (0.0, 20.0) (2.267, 8.954) (2.9, 6.669) (3.533, 5.183) (4.167, 4.163) (4.8, 3.428) (5.433, 2.875) (6.067, 2.444) (6.7, 2.1) (7.333, 1.818) (7.967, 1.583) (8.6, 1.385) (9.233, 1.214) (9.867, 1.067) (10.5, 0.938) (11.133, 0.824) (11.767, 0.722) (12.4, 0.632) (13.033, 0.55) (13.667, 0.476) (14.3, 0.409) (14.933, 0.348) (15.567, 0.292) (16.2, 0.24) (16.833, 0.192) (17.467, 0.148) (18.1, 0.107) (18.733, 0.069) (19.367, 0.033) (20.0, 0.0)
        };
        \addplot[color=violet!80, loosely dashdotted, thick] table [x=x, y=y] {data/virtusers25.data};
        \addplot[color=gray, densely dashdotted, thick] table [x=x, y=y] {data/converseprivacy25.data};
        \coordinate (spypoint) at (axis cs:19.24, 0.04);
        \end{axis}
        \node[pin={[pin edge={->,black!70,semithick},pin distance=0.35cm]100:{%
        \begin{tikzpicture}[baseline,trim axis left,trim axis right,scale=1.5]
            \begin{axis}[
                    tiny, ticks=none,
                    xmin=19.15,xmax=19.35,
                    ymin=0,ymax=0.08,
                ]
                \addplot[color=red!70, mark=*, mark size=1.5pt, semithick] coordinates {
                    (0.0, 20.0) (1.76, 11.4) (2.837, 7.6) (4.354, 5.493) (6.488, 4.2) (19.24, 0.04) (20.0, 0.0)
                };
                \addplot[color=blue, densely dotted, thick] coordinates {
                    (0.0, 20.0) (2.267, 8.954) (2.9, 6.669) (3.533, 5.183) (4.167, 4.163) (4.8, 3.428) (5.433, 2.875) (6.067, 2.444) (6.7, 2.1) (7.333, 1.818) (7.967, 1.583) (8.6, 1.385) (9.233, 1.214) (9.867, 1.067) (10.5, 0.938) (11.133, 0.824) (11.767, 0.722) (12.4, 0.632) (13.033, 0.55) (13.667, 0.476) (14.3, 0.409) (14.933, 0.348) (15.567, 0.292) (16.2, 0.24) (16.833, 0.192) (17.467, 0.148) (18.1, 0.107) (18.733, 0.069) (19.367, 0.033) (20.0, 0.0)
                };
                \addplot[color=violet!80, loosely dashdotted, thick] table [x=x, y=y] {data/virtusers25.data};
                \addplot[color=gray, densely dashdotted, thick] table [x=x, y=y] {data/converseprivacy25.data};
            \end{axis}
        \end{tikzpicture}%
    }},draw,circle,minimum size=1cm] at (spypoint) {};
        \end{tikzpicture}
        }
        \caption{Case $(\KsfActive,\Ksf,\Nsf) = (25, 30, 20)$.}
        \label{fig: memory-tradeoff for (30,25,20)}
    \end{subfigure}
    \caption{Memory-load tradeoffs for the hotplug model with private demands when $(\Ksf, \Nsf) = (30, 10)$ and various $\KsfActive$.}
    \label{fig: memory-tradeoff for hotplugprivacy and baseline}
\end{figure}
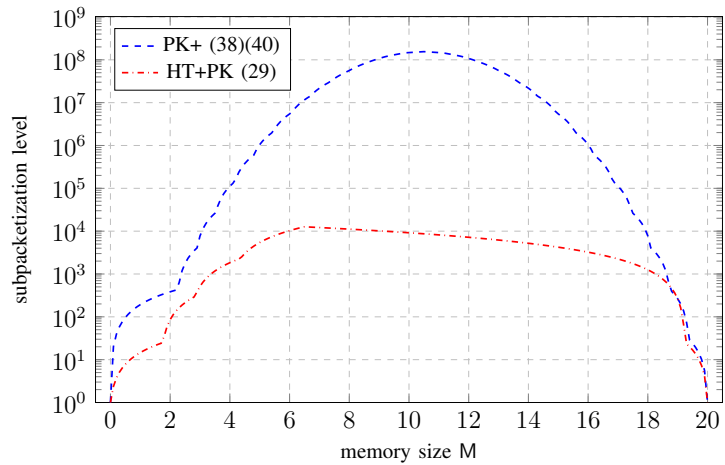
\begin{figure}
    \centering 
    \scalebox{0.8}{
        \begin{tikzpicture}
        \begin{semilogyaxis}[
            xmin=-0.5, xmax=20.5,
            ymin=1,
            scaled y ticks=base 10:10,
            legend entries={ \small PK+~\eqref{eq:HT1+PK}\eqref{eq:HT3+PK},
                \small HT+PK~\eqref{eq:performancePK+}
            },
            height=8cm,
            width=12cm,
            grid=major,
            grid style=dashed,
            legend pos=north west,
            xlabel={\small memory size $\Msf$},
            ylabel={\small subpacketization level}
        ]
        \addplot[color=blue, dashed, thick] table [x=x, y=y] {data/subpacketization.data};
        \addplot[color=red, dashdotted, thick] table [x=x, y=z] {data/subpacketization.data};
        \end{semilogyaxis}
        \end{tikzpicture}
        }
    \caption{Subpacketization level for the hotplug system with private demands when $(\KsfActive,\Ksf,\Nsf) = (25, 30, 20)$.}
    \label{fig: sublevel for (30,25,20)}
\end{figure}

\subsection{Hotplug with Privacy}
\indent $\bullet$ %
Fig~\ref{fig: memory-tradeoff for hotplugprivacy and baseline} shows the memory-load tradeoffs for two values of $\KsfActive$ for fixed $(\Ksf, \Nsf) = (30, 20)$.
In Fig~\ref{fig: memory-tradeoff for (30,12,20)}, with $\KsfActive = 12$, HT+PK schemes have good load performance in both small and large memory regimes.
In Fig~\ref{fig: memory-tradeoff for (30,25,20)}, with $\KsfActive = 20$, HT+PK schemes have good load performance only in small memory regime.
We do not provide the HT\&VU tradeoffs from Theorem~\ref{thm:HT+VU} as they are very closed to two trivial corner points, i.e. $(0, \Nsf)$ and $(\Nsf, 0)$.

\indent $\bullet$ Fig.~\ref{fig: sublevel for (30,25,20)} shows the subpacketization level for the case $(\KsfActive,\Ksf,\Nsf) = (30, 25, 10)$. %
The subpacketization level of HT+PK scheme is better than the PK+ scheme; %
we did not plot the subpacketization level of the YMA\&VU scheme since it is doubly exponential in the number of files, and thus way off the shown scale.

\section{Conclusion}
\label{sec:conclusion}
In this paper, we introduced the novel hotplug coded caching model to address a practical limitation of the original coded caching system, namely, to allow the server to start the delivery phase for a subset of active users, while the remaining users are offline. We proposed new achievable schemes exploit MDS codes for the placement phase, and are exactly optimal in some regimes, and optimal to within a factor of 2 otherwise.
This work shows that load savings are possible when the system is aware that only a subset of users will be active. 
Interestingly, when optimality can be be shown, the optimal performance only depends on the number active users and not on the total number of users; we are tempted to conjecture that this is true in general.

We also considered the hotplug model with demand privacy against colluding users. New achievable schemes that combine ideas from privacy-keys, virtual-users, and MDS coding are proposed. The performance of the new schemes is better in general than baseline schemes, and attain lower subpacketization levels. 
In the low and large memory regime we can prove exact optimality, and a constant gap otherwise.

Current work includes further extending exact optimality results, including to scalar linear file retrieval, and to different statistical models for the users' activity.

\appendices

\section{Proof of Theorem~\ref{thm:HT1}}
\label{sec:NEW1achievablescheme}

\paragraph*{Placement Phase} 
Fix $t \in [0: \KsfActive]$ and partition each file into $\binom{\KsfActive}{t}$ equal-size subfiles as 
\begin{align}
    F_i =  ( F_{i,\Wc} : \Wc \in \Omega_{[\KsfActive]}^{t} ),  \quad \forall i \in [\Nsf].
\end{align}
Then, for every $i \in [\Nsf]$, we treat the subfiles of each file as the information symbols of an MDS code with generator matrix $\Gm$ of dimension $\binom{\Ksf}{t} \times \binom{\KsfActive}{t}$, i.e., any $\binom{\KsfActive}{t}$ rows are linearly independent over $\mathbb{F}_{\qsf}$, where ${\qsf}$ is a sufficiently large prime number. 
The MDS-coded symbols are %
    \begin{align}
    \begin{bmatrix}
            C_{i,\Tc_1} \\ C_{i, \Tc_2} \\ \vdots \\ C_{i, \Tc_{\binom{\Ksf}{t}}}
        \end{bmatrix}
        &=\Gm \
        \begin{bmatrix}
            F_{i, \Wc_1} \\ F_{i, \Wc_2} \\ \vdots \\ F_{i, \Wc_{\binom{\KsfActive}{t}}}
        \end{bmatrix}, \quad \forall i \in [\Nsf].
    \end{align}

The cache contents are
\begin{align}
    Z_k = ( C_{i,\Tc}: i \in [\Nsf], \Tc \in \Omega_{[\Ksf]}^{t}, k \in \Tc ), \quad \forall k \in [\Ksf].
    \label{eq:NEWcache}
\end{align}
Thus the memory size is $\Msf^\text{\rm HT1}_t$ as in Theorem~\ref{thm:HT1}, which is the cache required by the MAN scheme for $\Ksf$ users divided by the rate of the MDS code used to `pre-code' each file.

\paragraph*{Delivery Phase} 
For any set of active users indexed by $\Ic \in \Omega_{[\Ksf]}^{\KsfActive}$ with demands $d_\Ic = [d_{i_1}, d_{i_2}, \ldots, d_{i_{\KsfActive}}]$, the server forms the  the following multicast signals %
\begin{align}
    X_\Sc &= \sum_{k \in \Sc} C_{d_k, \Sc \setminus \{k\}}, \quad \forall \Sc\in \Omega_{\Ic}^{t+1}.
    \label{eq:NEWmm}
\end{align}
If the server were to broadcast all the multicast signals in~\eqref{eq:NEWmm}, %
the load would be $\binom{\KsfActive}{t+1}/\binom{\KsfActive}{t}$. 
Let $r = \mathsf{rank}\big(d_{\Ic}\big)$; there are $\binom{\KsfActive - r}{t+1}$ out of $\binom{\KsfActive}{t+1}$ redundant multicast signals in~\eqref{eq:NEWmm}, which need not be sent (akin to the YMA delivery). For the largest possible $r = \rsf^\prime = \min(\KsfActive,\Nsf)$, the load is $\Rsf^\text{\rm HT1}_t$ as in Theorem~\ref{thm:HT1}.

\paragraph*{Correctness} 
By leveraging the received multicast signal and the local cache content, that is, user $k\in\Ic$ knows $\big\{C_{d_k, \Qc}: \Qc \in \Omega_{\Ic}^{t} \big\}$, 
user $k\in\Ic$ has the following system of equations %
    \begin{align}
    \begin{bmatrix}
            C_{d_k,\Qc_1} \\ C_{d_k, \Qc_2} \\ \vdots \\ C_{d_k, \Qc_{\binom{\KsfActive}{t}}}
        \end{bmatrix}
        &= \Gm[\Omega_{\Ic}^{\KsfActive}] \
        \begin{bmatrix}
            F_{d_k, \Wc_1} \\ F_{d_k, \Wc_2} \\ \vdots \\ F_{d_k, \Wc_{\binom{\KsfActive}{t}}}
        \end{bmatrix},
        \label{eq: mds recover}
    \end{align}
where $\Gm[\Omega_{\Ic}^{\KsfActive}]$ is square and invertible by the properties the MDS code; thus, each active user obtains its desired file.

\section{Proof of Theorem~\ref{thm:HT2}}
\label{sec:proofofHT2}

We start with the details of the case of two files to easy the reader into the notation.
We aim to show the achievability of the corner point
\begin{align}
(\Msf,\Rsf) = (1/\KsfActive, \ 2(1-1/\KsfActive)),
\end{align}
which satisfies with equality the cut-set bound in Lemma~\ref{lem: converse lemma MAN} with $s=2$, namely  $\Rsf \geq 2(1-\Msf)$, thereby showing that the segment connecting the trivial corner point $(\Msf,\Rsf) = (0,2)$ with this corner point is optimal.

\paragraph*{Placement Phase}The caches are populated as
\begin{align}
Z_k = \mathbf{G}_{k} (F_1+F_2), \ \forall k\in[\Ksf],
\end{align}
where $\mathbf{G}_{k}$ is the `cache-coding' matrix of user $k$ which is of dimension $\Bsf/\KsfActive \times \Bsf$, thus $\Msf = 1/\KsfActive$.

\paragraph*{Delivery Phase}
Consider a demand vector with $n_1$ active users demanding file $F_1$ and $n_2$ active users demanding file $F_2$, with $n_1+n_2=\KsfActive \geq 2$.
When $n_1=0$ (or $n_2=0$), the server sends $F_1$ (or $F_2$) which has load $\Rsf=1 \leq 2(1-1/\Ksf)$.

Next we consider the case where both $n_1$ and $n_2$ are strictly positive.
Let $\Ic_1 \subseteq \Omega_{[\Ksf]}^{n_1}$ be the set of users demanding file $F_1$
and $\Ic_2 \subseteq \Omega_{[\Ksf]\setminus \Ic_1}^{n_2}$ be the set of users demanding file $F_2$.
The delivery has two steps.
\begin{enumerate}

\item
In the first step, the server's transmissions aim to `decode' the caches of the active users 
\begin{align}
X_\text{step1} = 
[&\mathbf{G}_{j} F_1 : \forall j\in \Ic_2, \\
 &\mathbf{G}_{i} F_2 : \forall i\in \Ic_1
]. 
\end{align}
There are $n_1+n_2=\KsfActive$ sub-messages in $X_\text{step1}$, each of size  $\Bsf/\KsfActive$.
The net result of this first step is that the active users have now a `decoded' cache containing
\begin{align}
Z_i^\prime &= (\mathbf{G}_{u} F_1, \ \forall u\in \Ic_2 \cup \{i\}), \ \forall i\in \Ic_1,
\\
Z_j^\prime &= (\mathbf{G}_{u} F_2, \ \forall u\in \Ic_1 \cup \{j\}), \ \forall j\in \Ic_2.
\end{align}

\item
In the second step, the server creates MAN-type multicast messages to serve pairs of active users requesting the same file. For any two users in $\Ic_1$ (or in $\Ic_2$), we face a classical MAN problem with $(t,r)=(1,1)$ where each subfile is cached exclusively by one user and all the users request the same file. Thus, with $u^\star_n = \min\{u : u \in \Ic_n\}$ being the `leader' user for file $n\in[2]$, we have
\begin{align}
X_\text{step2} = 
[&\mathbf{G}_{u^\star_1} F_1 + \mathbf{G}_{i} F_1 : \forall i\in \Ic_1\setminus\{u^\star_1\}, \\
 &\mathbf{G}_{u^\star_2} F_2 + \mathbf{G}_{j} F_2 : \forall j\in \Ic_2\setminus\{u^\star_2\}
]. 
\end{align}
There are $n_1-1+n_2-1=\KsfActive-2$ sub-messages in $X_\text{step2}$, each of size  $\Bsf/\KsfActive$.
\end{enumerate}

In total the server has sent $\KsfActive-2+\KsfActive=2(\KsfActive-1)$ sub-messages, each of size $\Bsf/\KsfActive$. 
The load is thus $\Rsf=2(1-1/\KsfActive)$, as claimed.

\paragraph*{Correctness}
We still need to show that each active user can decode its demanded file.
At the end of the delivery phase, each active user $k \in \Ic_1 \cup \Ic_2$ (recall $ |\Ic_1 \cup \Ic_2| = \KsfActive$) has the following set of equations
\begin{align}
\underbrace{[
 \mathbf{G}_{u} : \forall u \in \Ic_1 \cup \Ic_2
]}_\text{$\KsfActive \frac{\Bsf}{\KsfActive} \times \Bsf$ matrix}
\underbrace{F_{d_k}}_\text{$\Bsf \times 1$ vector}
\end{align}
which can be inverted if the collection of cache-encoding matrices $\{ \mathbf{G}_{1}, \mathbf{G}_{2}, \ldots \mathbf{G}_{\Ksf}\}$ has the following MDS-like property: every $[\mathbf{G}_{\Sc} : \forall \Sc \in \Omega_{[\Ksf]}^{\KsfActive}]$ is full rank. Such matrices exists.

\bigskip
We aim to show the achievability of the corner point
\begin{align}
(\Msf,\Rsf) = (1/\KsfActive, \ \Nsf(1-1/\KsfActive)).
\end{align}

\paragraph*{Placement Phase}
The caches are populated as
\begin{align}
Z_k = \mathbf{G}_{k} (F_1+F_2+\ldots+F_{\Nsf}), \ \forall k\in[\Ksf],
\end{align}
where $\mathbf{G}_{k}$ is the `cache-coding' matrix of user $k$ which is of dimension $\Bsf/\KsfActive \times \Bsf$, thus $\Msf = 1/\KsfActive$.

\paragraph*{Delivery Phase}
Consider a demand vector with $n_j$ active users demanding file $F_j$ for $j\in[\Nsf]$, with $n_1+n_2+\ldots+n_{\Nsf}=\KsfActive \geq \Nsf$.
When at least one of the $n_j$ is zero, the server sends all the demanded files, which has load $\Rsf \leq \Nsf-1 \leq \Nsf(1-1/\Ksf)$.

Next we consider the case where all $n_j$'s are strictly positive.
Let $\Ic_j \subseteq \Omega_{[\Ksf]}^{n_j}$ be the set of active users demanding file $F_j$, where the $\Ic_j$'s are disjoint and $|\Ic_j|=n_j$ for $j\in[\Nsf]$.
The delivery has two steps.
\begin{enumerate}

\item
In the first step, the server's transmissions aim to `decode' the caches of the active users 
\begin{align}
X_\text{step1} = 
[ 
&\mathbf{G}_{u} F_n : \forall u\in \Ic_1, \ n\in[\Nsf]\setminus\{1\}, \\
&\mathbf{G}_{u} F_n : \forall u\in \Ic_2, \ n\in[\Nsf]\setminus\{2\}, \\
&\vdots \\
&\mathbf{G}_{u} F_n : \forall u\in \Ic_{\Nsf}, \ n\in[\Nsf]\setminus\{\Nsf\}
]. 
\end{align}
There are $\KsfActive(\Nsf-1)$ sub-messages in $X_\text{step1}$, each of size  $\Bsf/\KsfActive$.

Let $\Ic = \cup_{j\in[\Nsf]} \Ic_j$, with $|\Ic|=n_1+n_2+\ldots+n_{\Nsf}=\KsfActive$.
The net result of this first step is that the active users have now an `unlocked' cache containing
\begin{align}
Z_i^\prime &= (\mathbf{G}_{u} F_\ell, \ \forall u\in (\Ic \setminus \Ic_\ell) \cup\{i\}),  \forall i\in \Ic_\ell, \ \ell\in[\Nsf],
\end{align}

\item
In the second step, the server creates MAN-type multicast messages to serve pairs of active users requesting the same file. 
For any two users in $\Ic_n$, we face a classical MAN problem where each subfile is cached exclusively by one user in $\Ic_n$ and all the users in $\Ic_n$ request the same file $F_n$, for $n\in[\Nsf]$.
Thus, with $u^\star_n = \min\{u : u \in \Ic_n\}$ being the `leader' user for file $n\in[\Nsf]$, we have
\begin{align}
X_\text{step2} = 
[&\mathbf{G}_{u^\star_1} F_1 + \mathbf{G}_{j} F_1 : \forall j\in \Ic_1\setminus\{u^\star_1\}, \\
 &\mathbf{G}_{u^\star_2} F_2 + \mathbf{G}_{j} F_2 : \forall j\in \Ic_2\setminus\{u^\star_2\}, \\
 &\vdots  \\
 &\mathbf{G}_{u^\star_{\Nsf}} F_{\Nsf} + \mathbf{G}_{j} F_{\Nsf} : \forall j\in \Ic_{\Nsf}\setminus\{u^\star_{\Nsf}\}
]. 
\end{align}
There are $\KsfActive-\Nsf$ sub-messages in $X_\text{step2}$, each of size  $\Bsf/\KsfActive$.
\end{enumerate}

In total the server has sent $\KsfActive(\Nsf-1) + \KsfActive-\Nsf =\Nsf(\KsfActive-1)$ sub-messages, each of size $\Bsf/\KsfActive$. 
The load is thus $\Nsf(1-1/\KsfActive)$, as claimed.

\paragraph*{Correctness}
At the end of the delivery phase, each active user $k \in \Ic$ has the following set of equations
\begin{align}
\underbrace{[
 \mathbf{G}_{u} : \forall u \in \Ic
]}_\text{$\KsfActive \frac{\Bsf}{\KsfActive} \times \Bsf$ matrix}
\underbrace{F_{d_k}}_\text{$\Bsf \times 1$ vector}
\end{align}
which can be inverted if the collection of cache-encoding matrices $\{ \mathbf{G}_{1}, \mathbf{G}_{2}, \ldots \mathbf{G}_{\Ksf}\}$ has the following MDS-like property: every $[\mathbf{G}_{\Sc} : \forall \Sc \in \Omega_{[\Ksf]}^{\KsfActive}]$ is full rank. Such matrices exists on a large enough finite field.

\section{Proof of Lemma~\ref{lem:FLEX}}
\label{sec:proofofFLEX}
Given a $(\KsfActive,\Ksf,\Nsf)$ hotplug system, let $t \in [2: \KsfActive-1]$, and fix $\Lsf_t$ such that 
\begin{align}
    \Lsf_t < \binom{\KsfActive-1}{t-1} \frac{t}{t-1}.
    \label{eq:subpacketizationconstraint}
\end{align}

\paragraph*{Cache Placement}
For each user $k \in [\Ksf]$, we define $\Em_k \in \mathbb{F}_\qsf^{\binom{\KsfActive-1}{t-1} \times \Lsf_t}$ as the encoding matrix for user $k$. The cache $Z_k$ is populated from $\Em_k$ as
\begin{align}
    Z_k = ( (\Em_k \otimes \Id_{\Bsf / \Lsf_t}) F_i: i \in [\Nsf] ), \quad \forall k \in[\Ksf].
\end{align}
Thus, the memory size $\Msf = \Nsf \binom{\KsfActive-1}{t-1}/\Lsf_t > (t-1)/t$. We require the concatenation of all encoding matrices, $[\Em_1; \ldots; \Em_{\Ksf}]$ to be a MDS matrix of size $\Ksf \binom{\KsfActive-1}{t-1}  \times \Lsf_t$.

\paragraph*{Delivery Phase}
The server knows the active user set $\Ic$ and the demands $d_{\Ic}$; 
it transmits the multicast messages
\begin{align}
    X_\Sc = \sum_{j \in \Sc} \alpha_{\Sc, j} \left(\pv_{j, \Sc \setminus \{j\}} \otimes \Id_{\Bsf / \Lsf_t}\right) F_{d_j}, \forall \Sc \in \Omega_{\Ic}^{t+1}, 
    \label{eq:multicastmessage}
\end{align} 
where $\alpha_{\Sc, j} \in \mathbb{F}_\qsf$ is a coefficient chosen as in~\cite{ma2021general,wan2021optimal}, $\pv_{j, \Sc \setminus \{j\}} \in \mathbb{F}_\qsf^{\Lsf_t}$ is a `decoding vector.' We denote the decoding matrix for user $j \in [\Ic]$ as $\Pm_j$, which is a concatenation of $\{\pv_{j, \Tc}: \Tc \in \Omega_{[\Ic] \setminus \{j\}}^{t} \}$. Then $\Pm_j$ is of dimension $\binom{\KsfActive-1}{t} \times \Lsf_t$.

We require that every decoding vector $\pv_{j, \Tc}$ is in the span of $\Em_{k}$ when $k \in \Tc$, i.e.
\begin{align}
    \pv_{j, \Tc} \in \{\Em_k^T \lambda: \lambda \in \mathbb{F}_\qsf^{\binom{\KsfActive-1}{t-1}}\}, \forall k \in \Tc.
\end{align}
We will prove that such decoding vectors exist later.

\paragraph*{Correctness}
We can rewrite $X_\Sc$ in~\eqref{eq:multicastmessage} as follows: for $\ell \in \Sc$
\begin{align}
    X_\Sc &= \alpha_{\Sc, \ell} \left(\pv_{\ell, \Sc \setminus \{\ell\}} \otimes \Id_{\Bsf / \Lsf_t}\right) F_{d_\ell} + \underbrace{\sum_{j \in \Sc\setminus\{\ell\}} \alpha_{\Sc, j} \left(\pv_{j, \Sc \setminus \{j\}} \otimes \Id_{\Bsf / \Lsf_t}\right) F_{d_j}}_{\text{can be computed by user $\ell$ as every }\pv_{j,\Sc\setminus\{j\}} \in \mathsf{Span}(\Em_{\ell})}.
\end{align}
Thus, for $j \in \Ic$, $\{\Em_j; \Pm_j\} \in \mathbb{F}_\qsf^{\binom{\KsfActive}{t} \times \Lsf_t}$ is known by user $j$. Since $\{\Em_1, \ldots, \Em_\Ksf\}$ is a MDS matrix, if $\{\Em_j; \Pm_j\}$ s full rank $F_{d_j}$ can be recovered by user $j$.

\paragraph*{Existence of Decoding Vectors} WLOG, let $\Tc = [t]$, then we can find $\{\lambda_j \in \mathbb{F}_\qsf^{\binom{\KsfActive-1}{t-1}}: j \in \Tc\}$, such that
\begin{align}
    \Em_1^T \lambda_1 = \Em_2^T \lambda_2 = \Em_3^T \lambda_3 \ldots = \Em_{t}^T \lambda_{t},
\end{align}
i.e. we have a linear system equation as follows,
\begin{align}
    \underbrace{
    \begin{bmatrix}
        (\Em_{1})^T & -(\Em_{2})^T  &  &  & \\
        \vdots & & \ddots & \\
        (\Em_{1})^T & & & -(\Em_{t})^T 
    \end{bmatrix}
    }_{:= \Gm, \ \mathsf{dim}(\Gm) = \Lsf_t (t-1) \times \binom{\KsfActive-1}{t-1} t}
    \begin{bmatrix}
        \lambda_{1} \\ \vdots \\ \lambda_{t}
    \end{bmatrix}
    = 0.
\end{align}
Each row of $\Gm$ is linearly independent as the concatenation of $\{\Em_j: j \in [t]\}$ is a MDS code. 
To ensure $\{\lambda_1;\ldots;\lambda_t\} \neq 0$, $\Gm$ must be a fat matrix, i.e. 
\begin{align}
\Lsf_t (t-1) < \binom{\KsfActive-1}{t-1} t \, \Longleftrightarrow \, \Lsf_t < \binom{\KsfActive-1}{t-1} \frac{t}{t-1},
\end{align}
which already holds in~\eqref{eq:subpacketizationconstraint}. Thus, such decoding vectors do exist.

\section{Proof of Theorem~\ref{thm:HT1+PK}}\label{sec:proofHT1-PK}
We continue to use the notation of bilinear product in~\eqref{eq:bilinearproduct}. 

\paragraph*{File Partition} 
Fix $t \in [\KsfActive]$ and partition the files into $\Lsf_t=\binom{\KsfActive}{t}$ equal-length subfiles. 
Then, code the subfiles of each file with the MDS generator matrix $\Gm$, where any $\binom{\KsfActive}{t}$ columns of $\Gm$ are linearly independent: 
for all $n\in[\Nsf]$
\begin{align}
    &C_{n,\Tc_\ell} 
    := \sum_{j\in\left[\binom{\KsfActive}{t}\right]}  F_{n, \Wc_j} \ g_{\ell,j}
    = \Tbp(\ev_n, \gv_{\ell}) \in \mathbb{F}_\qsf^{\Bsf/\binom{\KsfActive}{t}},
    \label{eq: mds subfiles}
    \\
    &\gv_{\ell} 
    := (g_{\ell,1}, \ldots, g_{\ell,\binom{\KsfActive}{t}})^T \in \mathbb{F}_\qsf^{\binom{\KsfActive}{t}}, 
    \ \forall \ell\in\left[\binom{\Ksf}{t}\right], 
    \\
    &\Gm
    :=[\gv_{1}; \cdots; \gv_{\binom{\Ksf}{t}}]\in \mathbb{F}_\qsf^{\binom{\KsfActive}{t} \times \binom{\Ksf}{t}}.
\end{align}
Note the slight abuse of notation in~\eqref{eq: mds subfiles}: the definition of $T(\cdot, \cdot)$ in~\eqref{eq:bilinearproduct} assumes subpacketization level $\binom{\Ksf}{t}$, while here the subpacketization level is $\binom{\KsfActive}{t}$; we decided not to introduce a new notation in~\eqref{eq: mds subfiles} as we essentially use $T(\cdot, \cdot)$ to denote a linear combination of subfiles whose subpacketization level is clear from the context.

\paragraph*{Cache Placement}
For each user~$j \in [\Ksf]$, the server generates an i.i.d uniformly random vector $\pv_j \in \mathbb{F}_\qsf^\Nsf$ such that $\sum_{i=1}^\Nsf p_{j,i} = \qsf-1$, and places it into $Z_j$, together with the MAN-like content 
\begin{align}
\{\Tbp(\ev_n, \gv_\Tc): n \in [\Nsf], \Tc \in \Omega_{[\Ksf]}^{t}, j \in \Tc\}. 
\label{eq:LMDC:MAN-like}
\end{align}
To guarantee privacy, we also need to store (inspired by the PK scheme in~\eqref{eq:place:pk}) the privacy keys $\{\Tbp(\pv_j, \gv_\Wc) : \Wc \in \Omega_{[\Ksf]\setminus\{j\}}^{t} \}$ in the cache of user~$j \in [\Ksf]$. It would therefore seem that we need to store $\Nsf\binom{\Ksf-1}{t-1} + \binom{\Ksf-1}{t}$ subfiles as in the PK scheme. The privacy keys are however MDS coded, so we only need to store $\binom{\KsfActive}{t}$ of them to recover all possible $\binom{\Ksf}{t}$. Now, from the MAN-like content in~\eqref{eq:LMDC:MAN-like} and the local randomness $\pv_j$, we can locally compute (similarly to~\eqref{eq:YTmessages:S2evallocally}) %
\begin{align}
\Tbp(\pv_j, \gv_\Tc) = \sum_{n\in[\Nsf]} p_{j, n}\Tbp(\ev_n, \gv_\Tc) : \Tc \in \Omega_{[\Ksf]}^{t}, j \in \Tc,
\label{eq:LMDC:from MAN-like}
\end{align}
this implies that if $\binom{\Ksf-1}{t-1} \geq \binom{\KsfActive}{t}$ we are good; otherwise we need to store additional $\binom{\KsfActive}{t} - \binom{\Ksf-1}{t-1}$ privacy keys that are linearly independent of those in~\eqref{eq:LMDC:from MAN-like}.
\begin{subequations}
The memory size is thus
\begin{align}
    \Msf^\text{\rm(new)}_t 
    &=\begin{cases} 
    \Nsf \frac{\binom{\Ksf-1}{t-1}}{\binom{\KsfActive}{t}} &  \binom{\Ksf-1}{t-1} \geq \binom{\KsfActive}{t} \\
    1+\frac{t}{\KsfActive}\frac{\binom{\Ksf-1}{t-1}}{\binom{\KsfActive-1}{t-1}}(\Nsf-1) &  \binom{\Ksf-1}{t-1} < \binom{\KsfActive}{t} 
    \end{cases}.
\end{align}
\label{eq:LMDC:M}
\end{subequations}
Let $\eta_t := \left[\binom{\KsfActive}{t}- \binom{\Ksf-1}{t-1}\right]^+$.
The cache  for user~$j \in [\Ksf]$ is 
\begin{subequations}
\begin{align}
Z_j &=  \left\{\Tbp(\ev_i, \gv_{\{j\}\cup\Qc}): i \in [\Nsf], \Qc \in \Omega_{[\Ksf] \setminus \{j\}}^{t-1}\right\}  
\label{eq:LMDC:ZjMAN}
\\ & \cup \ \left\{\Tbp(\pv_j, \xi_{j,\ell}): \ell \in \left[\eta_t\right]\right\} \cup \{\pv_j\},
\label{eq:LMDC:ZjPK}
\end{align}
\label{eq:LMDC:Zjall}
\end{subequations}
where $\{\xi_{j,\ell}: \ell \in [\eta_t]\}$ is a subset of $\{\gv_{\Tc}: \Tc \in \Omega_{[\Ksf]\setminus \{j\}}^{t}\}$ that is linearly independent w.r.t. $\{\gv_{\{j\}\cup\Qc}: \Qc \in \Omega_{[\Ksf] \setminus \{j\}}^{t-1}\}$.

\paragraph*{Delivery Phase}
Given the active users indexed by $\Ic \in \Omega_{[\Ksf]}^{\KsfActive}$ %
where user~$j \in \Ic$ demands $F_{d_j}$, %
we let $\qv_j = \pv_j + \ev_{d_j}$ as in the PK scheme. 
As the sum entries of $\ev_{d_j}$ is always 1, the sum entries of $\qv_j$ thus is 0.
Let $\Qm = [\qv_i: i \in \Ic] \in \mathbb{F}_\qsf^{\KsfActive}$, the maximal rank of $\Qm$ is $\min(\KsfActive, \Nsf-1)$.
The server sends 
\begin{align}
  X &=  \{X_\Sc: \Sc \in \Omega_{\Ic}^{t+1} \} \cup \{ \Qm, \Ic \}.
\label{eq:LayerMDSmessages:X}
\end{align} 
  
\paragraph*{Proof of Correctnes}
\begin{subequations}
We write
\begin{align}
  X_\Sc &= \sum_{j \in \Sc} \alpha_{j, \Sc \setminus \{j\}} \Tbp(\qv_j, \gv_{\Sc \setminus \{j\}}) \\
&= \alpha_{v,\Sc \setminus \{v\}} \left(\underbrace{ \Tbp(\ev_{d_v}, \gv_{\Sc \setminus \{v\}}) }_{\text{demanded by user~$v$}}
     + \underbrace{ \Tbp(\pv_v, \gv_{\Sc \setminus \{v\}}) }_{\text{from cache $Z_v$ in~\eqref{eq:LMDC:ZjPK}}} \right)
\\
    &+ \sum_{j \in \Sc \setminus \{v\}} \alpha_{j, \Sc \setminus \{j\}} \underbrace{\Tbp(\qv_j, \gv_{\Sc \setminus \{j\}})}_{\text{from cache $Z_v$ in~\eqref{eq:LMDC:ZjMAN} and $\Qm$ in~\eqref{eq:LayerMDSmessages:X}}}, \ \forall  v\in \Sc.
\end{align}
\label{eq:LayerMDSmessages}
\end{subequations}
Note that $\alpha_{j, \Sc \setminus \{j\}} \in \{+1, -1\}$ are encoding coefficients as shown in~\cite{wan2021optimal}.
As in the PK scheme, some multicast messages in~\eqref{eq:LayerMDSmessages:X} need not be sent (similarly to the PK scheme), %
resulting in the load 
\begin{align}
    \Rsf^\text{\rm HT1+PK}_t = \bigl[\binom{\KsfActive}{t+1} - \binom{\KsfActive - \min(\KsfActive, \Nsf-1)}{t+1}\bigr] /\binom{\KsfActive}{t}.
\end{align}

\paragraph*{Proof of Privacy}
We prove that our scheme satisfies the privacy constraint in~\eqref{eq:privacyconstraint}. Fix $\Bc \subseteq \Ic$, then
\begin{align*}
    & I(d_{\Ic \setminus \Bc};X, d_{\Bc}, Z_{\Bc} \mid {\Ic}, F_{[\Nsf]}) 
    \\&\stackrel{\rm{(a)}}{=} I(d_{\Ic \setminus \Bc}; \qv_{\Ic}, d_\Bc, Z_\Bc \mid {\Ic}, F_{[\Nsf]}) %
    \\&\stackrel{\rm{(b)}}{=} I(d_{\Ic \setminus \Bc}; \qv_{\Ic}, d_\Bc, \pv_\Bc \mid {\Ic}, F_{[\Nsf]}) %
    \\&\stackrel{\rm{(c)}}{=} I(d_{\Ic \setminus \Bc}; \qv_{\Ic \setminus \Bc}, d_\Bc, \pv_\Bc \mid {\Ic}, F_{[\Nsf]}) %
    \\&= I(d_{\Ic \setminus \Bc}; d_\Bc, \pv_\Bc \mid {\Ic}, F_{[\Nsf]}) 
       + I(d_{\Ic \setminus \Bc}; \qv_{\Ic \setminus \Bc} \mid d_\Bc, \pv_\Bc, {\Ic}, F_{[\Nsf]}) %
    \\&\stackrel{\rm{(d)}}{=} 0
       + I(d_{\Ic \setminus \Bc}; \qv_{\Ic \setminus \Bc} \mid d_\Bc, \pv_\Bc, {\Ic}, F_{[\Nsf]})
    \\&\stackrel{\rm{(e)}}{=} 0, %
\end{align*}
where the equality follows because:
(a) $X$ in~\eqref{eq:LayerMDSmessages:X} is a function of $(F_{[\Nsf]}, \qv_\Ic)$;
(b) $Z_\Bc$ in~\eqref{eq:LMDC:Zjall} is function of $(F_{[\Nsf]}, \pv_\Bc)$; 
(c) since $\qv_{\Bc} = \ev_{d_\Bc}+\pv_{\Bc}$ and thus can be dropped; 
(d) the user demands are independent, and independent of the local randomness; and
(e) $(d_\Bc, \pv_\Bc,d_{\Ic \setminus \Bc}, \qv_{\Ic \setminus \Bc})$ are independent.

\section{Proof of Theorem~\ref{thm:HT3+PK}}
\label{sec:proofHT3-PK}

\paragraph*{File Partition}
We partition each file into $\KsfActive$ equal-length subfiles as
\begin{align}
    F_n = (F_{n,\ell}: \ell \in [\KsfActive]), \ \forall n \in [\Nsf].
\end{align}
With a notation similar to~\eqref{eq:bilinearproduct}, we let $F_{n,\ell} = \Tbp(\ev_n, \ev_\ell), n \in [\Nsf], \ell \in [\KsfActive]$.

\paragraph*{Cache Placement}
We consider the generator matrix $\Gm_j = [\gv_{j,1}; \cdots; \gv_{j,\KsfActive-1}]$ of size $(\KsfActive-1) \times \KsfActive$, and a vector $\xi_j$ of size $\KsfActive$ for every user~$j \in [\Ksf]$.
For each user~$j \in [\Ksf]$, we also generate i.i.d uniformly at randomly a vector $\pv_j \in \mathbb{F}_{\qsf}^{\Nsf}$ such that $\sum_{i\in[\Nsf]} p_{j,i} = \qsf-1$.
We populate the caches as
\begin{subequations}
\begin{align}
  Z_j &= \left\{\Tbp(\ev_n,\gv_{j,\ell}): n \in [\Nsf], \ell \in [\KsfActive-1]\right\} 
  \label{eq:BMDC:ZjMAN}
  \\ & \cup \left\{ \Tbp(\pv_j, \xi_j)\right\} \cup \left\{\pv_j\right\}, \ \forall j \in [\Ksf],
  \label{eq:BMDC:ZjPK}
\end{align}
\end{subequations}
and the memory size is
\begin{align}
    \Msf 
    = \Nsf \frac{\KsfActive-1}{\KsfActive} + \frac{1}{\KsfActive} 
    = 1 + \frac{\KsfActive-1}{\KsfActive}(\Nsf-1).
\end{align}
More conditions on $\{\Gm_1, \ldots, \Gm_\Ksf, \xi_1, \ldots, \xi_\Ksf\}$ will be needed to insure decodability and will be listed next.

\paragraph*{Delivery Phase} 
Given the set of active users $\Ic \in \Omega_{[\Ksf]}^{\KsfActive}$ where user~$j \in \Ic$ demands $F_{d_j}$, let $\qv_j = \pv_j + \ev_{d_j}$ as in PK scheme. 
The server sends %
\begin{subequations}
    \begin{align}
      X &=  ( X_\Ic ,\qv_{\Ic}, \Ic ), \label{eq:BMDSmessage}\\ %
      X_\Ic &= \sum_{j \in \Ic} \Tbp(\qv_j, \phi^{(\Ic)}_{j}),
      \label{eq:BMDSmessages:X}
    \end{align}
\end{subequations}
where the vectors $\phi^{(\Ic)}_{j}$ are to be chosen so as to insure users can recover their demanded linear combination of files from $X_\Ic$, as we shall specify later.

\paragraph*{Proof of Correctness}
As for the PK scheme, we rewrite $X_\Ic$ as
\begin{subequations}
\begin{align}
      X_\Ic 
      &= \underbrace{ \Tbp(\ev_{d_v}, \phi^{(\Ic)}_{v}) }_{\text{desired by user~$v$}}
       + \underbrace{ \Tbp(\pv_v, \phi^{(\Ic)}_{v}) }_{\text{from cached content in $Z_v$}}
\label{eq:BMDSmessages:Xsplit1}
    \\&+ \sum_{j \in \Ic \setminus \{v\}} \underbrace{ \Tbp(\qv_j, \phi^{(\Ic)}_{j}) }_{\text{from cached content in $Z_v$ and $\qv_{\Ic}$}}, 
    \ \forall v \in \Ic,
\label{eq:BMDSmessages:Xsplit2}
\end{align}
\label{eq:BMDSmessages:Xsplit}
\end{subequations}
where in the underlines we clarified how we intend to remove from $X_\Ic$ the `interfering' terms for any user~$v \in \Ic$. We will describe next how to remove the `interfering' terms, thus also identifying additional properties we require of the generator matrices; assuming that it is indeed possible,
the load is $\Rsf = 1/\KsfActive$.

\paragraph*{Proof of Privacy} 
Privacy can be proved similarly to the proof of Theorem~\ref{thm:HT1+PK} in Appendix~\ref{sec:proofHT1-PK}.

\paragraph*{Conditions for Decodability}
\begin{itemize}

\item
Condition~1: $[\Gm_v; \xi_v]$ is full rank for all $v\in[\Ksf]$. 

Reason: The second term in~\eqref{eq:BMDSmessages:Xsplit1} can be computed from the cache content $Z_v$ as follows:  if $[\Gm_v; \xi_v]$ is full rank, then any $\tilde{\av}\in \mathbb{F}_\qsf^{\KsfActive}$ can be obtained as 
\begin{align}
\tilde{\av} &= \sum_{\ell\in[\KsfActive-1]} a_\ell\gv_{j,\ell} + a_{\KsfActive}\xi_j, 
\end{align}
for some $\av=[a_1,\ldots,a_{\KsfActive}]$, %
thus user~$v$ can compute
\begin{align}
\Tbp(\pv_v, \tilde{\av}) = \sum_{\ell\in[\KsfActive-1]} a_\ell \left(\sum_{n\in[\Nsf]} p_{v,n} \Tbp(\ev_n,\gv_{j,\ell})\right) 
+ a_{\KsfActive} \Tbp(\pv_v, \xi_v), \label{eqQ:whatUcancompute}
\end{align}
for any $\tilde{\av}$, and thus in particular for $\tilde{\av}=\phi^{(\Ic)}_{v}$.

\item
Condition~2: $[\Gm_v; \phi^{(\Ic)}_{v}]$ is full rank for all $v\in[\Ksf]$ and all $\Ic\in\Omega_{[\Ksf]}^{\KsfActive}$.
 
Reason:  Let us assume for now that the term in~\eqref{eq:BMDSmessages:Xsplit2} can also be computed from the cache content $Z_v$, thus user~$v\in\Ic$ has recovered $\Tbp(\qv_v, \phi^{(\Ic)}_{v})$. From Condition~1, user~$v$ further computes $\Tbp(\pv_v, \phi^{(\Ic)}_{v})$ and thus attains $\Tbp(\ev_{d_j}, \phi^{(\Ic)}_{v})$
Thus we need $[\Gm_v; \phi^{(\Ic)}_{v}]$ to be full rank as well, in order for the recovered $F_{d_v}$.

\item
Condition~3: For $i, j \in [\Ksf], i \neq j$, $[\Gm_i,\Gm_j]$ is a full rank matrix.

Reason: We finally need to show under which conditions the term in~\eqref{eq:BMDSmessages:Xsplit2} can be computed from the cache content of user~$v\in\Ic$. 
The server sents $\qv_\Ic$ to all users as shows in~\eqref{eq:BMDSmessage}, equivalently, user~$v \in [\Ic]$ can compute $\{\Tbp(\qv_j, \gv_{v,\ell}): \ell \in [\KsfActive-1]\}$ for all $j \in \Ic$.
Then $\{\phi^{(\Ic)}_{j}: j \in \Ic \setminus \{v\}\}$ in~\eqref{eq:BMDSmessages:Xsplit2} should be computable from the cache of user~$v$, that is, there should exist a vector $\xv_{v,j}$ of length $\KsfActive-1$ such that 
\begin{align}
\phi^{(\Ic)}_{j} = \Gm_v \xv_{v,j}, \ \forall j,v\in\Ic, v\not=j.
\label{eqQ:whatUmust be able to compute}
\end{align}

As an example consider $\Ic = [\KsfActive]$ %
and rewrite the condition in~\eqref{eqQ:whatUmust be able to compute} for $j=\KsfActive$  as 
\begin{align}
    \underbrace{
    \begin{bmatrix}
        \Gm_1   & \Gm_2   &        &  \\
        \vdots   &         & \ddots &  \\
        \Gm_1   &         &        & \Gm_{\KsfActive-1} \\
    \end{bmatrix}
    }_{= \widetilde{\Gm}_{j} \in \mathbb{F}_\qsf^{ \KsfActive (\KsfActive-2) \times (\KsfActive-1)^2 } }
    \underbrace{
    \begin{bmatrix}
        -\xv_{1,j} \\ \vdots \\ \xv_{\KsfActive-1,j} 
    \end{bmatrix}
    }_{= \widetilde{\xv}_{j} }
    = 0,  \ j=\KsfActive,
\end{align}
where $\widetilde{\Gm}_{j}$ is a `fat matrix', i.e., $\KsfActive (\KsfActive-2) < (\KsfActive-1)^2$;
to ensure a non-zero solution for $\widetilde{\xv}_{j}$ we need $\widetilde{\Gm}_{j}$ to be full rank.
For~\eqref{eqQ:whatUmust be able to compute} to hold, it suffices that any matrix $[\Gm_j : j\in \Ic]$ is an MDS matrix for all  $\Ic\in\Omega_{[\Ksf]}^{\KsfActive}$.

\item
For sufficiently large enough finite field size $\qsf$, if $[\Gm_1,\ldots,\Gm_{\Ksf}]$ is an MDS code, then the above three conditions are satisfied. 

\end{itemize}

\section{Proof of Theorem~\ref{thm:optimality}}
\label{sec:converseproof}
In this section we are going to use the following results.

\begin{lem}[{Cut-set Bound from \cite[Theorem 2]{maddah2014fundamental}}] \label{lem: converse lemma MAN} \rm
For the classical coded caching system with with $\KsfActive$ users and $\Nsf$ files, the memory-load pair $(\Msf, \Rsf)$ is lower bounded by
    \begin{align} \label{eq: load lower bound MAN}
        \Rsf\geq s - \frac{s}{\lfloor \Nsf/s \rfloor} \Msf, \quad \forall s \in [\min\{\Nsf, \KsfActive\}].
    \end{align}
\end{lem}

\begin{lem}[{\cite[Appendix]{maddah2014fundamental}}] \label{lem: converse lemma MAN K'=N=2} \rm
For the classical coded caching system with with $\KsfActive = 2$ users and $\Nsf=2$ files, the optimal memory-load is 
\begin{align} \label{eq: load lower bound MAN K'=N=2}
    \Rsf^\star = \max\left\{2-2\Msf, \frac{3}{2}-\Msf, 1-\frac{\Msf}{2}\right\}.
\end{align}
\end{lem}

\begin{lem}[{\cite[Theorem 3]{tian2018symmetry}}] \label{lem: converse lemma ChaoTian} \rm
For the classical coded caching system with with $\KsfActive = 2$ users and $\Nsf \geq 3$ files, the optimal memory-load is %
\begin{align} \label{eq: load lower bound ChaoTian}
    \Rsf^\star = \max\left\{2-\frac{3\Msf}{\Nsf}, 1-\frac{\Msf}{\Nsf}\right\}.
\end{align}
\end{lem}

\begin{lem}[{\cite[Theorem 2]{yu2018characterizing}}] \label{lem: converse lemma YMA}\rm
    For the classical coded caching system with with $\KsfActive$ users and $\Nsf$ files, the memory-load pair $(\Msf, \Rsf)$ is lower bounded by
    \begin{align} \label{eq: load lower bound YMA}
        \Rsf \geq s - 1 + \alpha - \frac{s(s-1)-\ell(\ell-1)+2\alpha s}{2(\Nsf - \ell +1)} \Msf,
    \end{align}
    for any $s \in [\min\{\Nsf, \KsfActive\}]$, $\alpha \in [0, 1]$, and where $\ell \in [s]$ is the minimum value such that
    \begin{align} \label{ieq: l condition}
        \frac{s(s-1)-\ell(\ell-1)}{2} + \alpha s \leq (\Nsf - \ell +1) \ell.
    \end{align}
\end{lem}

\subsection{Case~\ref{item:opt:r=1}: $\min\{\Nsf, \KsfActive\}=1$}
In Theorem~\ref{thm:HT1}, there are two (trivial) corner points $(0, 1)$ and $(\Nsf, 0)$. 
The line connecting these two points meets the lower bound in~\eqref{eq: load lower bound MAN} in Lemma~\ref{lem: converse lemma MAN} for $s=1$, namely $\Rsf \geq 1-\Msf/\Nsf$.
This optimal performance does not depend on the value of $\Ksf$.

\subsection{Case~\ref{item:opt:gap is constant}: constant gap}\label{sec:gap hot no pri}
By Theorem~\ref{thm:extensionYMAdecentralized}, we have
\begin{align}
\Rsf^\star(\Msf)
      &%
       \leq \Rsf^\text{\rm (YMA+)}(\Msf)
       \leq \Rsf^\text{\rm(decen+)}(\Msf) 
    \\&=\left.\frac{1-\mu}{\mu}\Big( 1-(1-\mu)^{\min(\Nsf,\KsfActive)}\Big)\right|_{\mu:={\Msf}/{\Nsf}} 
    \\&\stackrel{\text{ \cite[Lemma 1]{yu2018characterizing} }}{\leq}  2.00884 \ \Rsf^\text{Lemma~\ref{lem: converse lemma YMA}},
\end{align}
where $\Rsf^\text{Lemma~\ref{lem: converse lemma YMA}}$ denotes the lower convex envelope of the region identified by Lemma~\ref{lem: converse lemma YMA}. 

\subsection{Case~\ref{item:opt:K'=2N=2}: $\Nsf = 2=\KsfActive = 2$}
The achievability is proved in Section~\ref{sec:K'=N=2 NO privacy} and the converse is Lemma~\ref{lem: converse lemma MAN K'=N=2}.

\subsection{Case~\ref{item:opt:K'=2N>=3}: $\Nsf \geq 3,\KsfActive=2$}
In Theorem~\ref{thm:HT1}, there are three corner points points $(0, 2)$, $(\Nsf, 0)$ and $(\Nsf/2, 1/2)$. 
Their lower convex envelop equals the lower bound in~\eqref{eq: load lower bound ChaoTian} in Lemma~\ref{lem: converse lemma ChaoTian}.
This optimal performance does not depend on the value of $\Ksf$.

\subsection{Case~\ref{item:opt:NlargeMsmall-firstsegment}: small memory and many files ($\Nsf > \KsfActive(\KsfActive+1)/2$)}
When $\Nsf \geq \KsfActive(\KsfActive+1)/2$, the inequality in~\eqref{ieq: l condition} holds for any $s \in [\KsfActive]$, $\alpha=1$ and $\ell=1$; so, if we let $s = \KsfActive$ and $\Msf = \Nsf/\KsfActive$ in~\eqref{eq: load lower bound YMA}, we obtain the lower bound
\begin{align}
    \Rsf \geq \KsfActive - \frac{\KsfActive(\KsfActive+1)}{2} \ \frac{\Msf}{\Nsf}.
\label{eq:YMAconverse}
\end{align} 
The segment connecting the corner points $( \Msf^\text{\rm HT1}_0, \Rsf^\text{\rm HT1}_0 )=(0,\KsfActive)$ and $( \Msf^\text{\rm HT1}_1, \Rsf^\text{\rm HT1}_1 ) =\biggl( \frac{\Nsf}{\KsfActive}, \frac{\KsfActive-1}{2} \biggr)$ satisfies~\eqref{eq:YMAconverse} with equality.
This optimal performance does not depend on the value of $\Ksf$.

\subsection{Case~\ref{item:optNsmallMsmall-firstsegment}: small memory and much fewer files than users}
The corner point 
$
(\Msf^\text{\rm HT2},\Rsf^\text{\rm HT2}) = (1/\KsfActive, \ \Nsf(1-1/\KsfActive)),
$
satisfies with equality the cut-set bound in Lemma~\ref{lem: converse lemma MAN} with $s=\Nsf$, namely  $\Rsf \geq \Nsf(1-\Msf)$, thereby showing that the segment connecting the trivial corner point $(\Msf,\Rsf) = (0,\Nsf)$ with this corner point is optimal.
This optimal performance does not depend on the value of $\Ksf$.

\subsection{Case~\ref{item:opt:lastsegment}: large memory}
The segment connecting the corner points $( \Msf, \Rsf)^\text{HT3}=(\Nsf\frac{\Ksf-1}{\Ksf}, \frac{1}{\Ksf})$ and $(\Msf, \Rsf) = (\Nsf, 0)$ satisfies $\Rsf \geq 1-\Msf/\Nsf$, which is Lemma~\ref{lem: converse lemma MAN} for $s=1$.

\section{Proof of the general virtual-users idea for hotplug with privacy}
\label{app:VU for private hotplug from NONprivate hotplug}

For a given $(\Nsf\KsfActive,\Nsf\Ksf,\Nsf)$-hotplug-NONprivate scheme, let 
\begin{itemize}
\item
${C}_k^{(np)}$ be the cache encoding function for user $k$, that is
	$$Z_k^{(np)} = {C}_k^{(np)}\big(F_{[\Nsf]}\big), \forall k \in [\Nsf\Ksf];$$
\item
${E}^{(np)}$ be the server encoding function, that is, for every active user set $\Ic^{(np)} \in \Omega_{[\Nsf\Ksf]}^{\Nsf\KsfActive}$, and their demands $d^{(np)}_{\Ic} \in[\Nsf]^{\Nsf\KsfActive}$ we have
    $$X^{(np)} = {E}^{(np)}\big(\Ic^{(np)}, d^{(np)}_{\Ic}, F_{[\Nsf]}\big);$$ 
\item
${G}_k^{(np)}$ be the decoding function for user $k\in [\Nsf\Ksf]$, that is, %
for every $k \in \Ic^{(np)}$ we have
    $$F_{d_k} = {G}_k^{(np)}\big(d^{(np)}_k, Z_k^{(np)}, X^{(np)}\big), \forall k \in \Ic^{(np)}, \ d^{(np)}_k \in [\Nsf];$$
\item
this non-private scheme achievs $(\Msf^{(np)},\Rsf^{(np)})$ where $H(Z_k^{(np)}) \leq \Bsf \Msf^{(np)}$ and $H(X^{(np)}) \leq \Bsf \Rsf^{(np)}$ for $\Bsf$ the file length.
\end{itemize}

We will now present a construction of an 
$(\KsfActive,\Ksf,\Nsf)$-hotplug-private scheme that achieves $(\Msf^{(np)},\Rsf^{(np)})$ from the given $(\Nsf\KsfActive,\Nsf\Ksf,\Nsf)$-hotplug-NONprivate scheme. 

\paragraph*{Cache Placement}
For each user $k$ with local i.i.d uniform randomness $\tau_k\in [\Nsf]$, the cache content is
\begin{align*}
Z_k = \left( {C}_{(k-1)\Nsf+\tau_k}^{(np)} (F_{[\Nsf]}), \, \tau_k \right), \forall k\in [\Ksf].
\end{align*}
Since storing $\tau_k$ requires $\log_\qsf(\Nsf)$ symbols, which is not a function of file length $\Bsf$, the case size in the private scheme is the same as the cache size in the NONprivate scheme.

\paragraph*{Delivery}
Let $\Psi: [\Nsf]^\Nsf \rightarrow [\Nsf]^\Nsf$ denote the {\it cyclic shift operator}, such that $\Psi(t_1,t_2,\cdots,t_\Nsf)=(t_\Nsf,t_1,\cdots,t_{\Nsf-1})$.
Let us denote a vector $\mathbb{I}_\Nsf:=(1,\cdots,\Nsf)$.
For a given active user set 
$$\Ic = \{j_1, \ldots, j_\KsfActive\} \in \Omega_{[\Nsf]}^{\KsfActive}$$ 
and their demands $d_{\Ic} \in [\Nsf]^{\KsfActive}$, we define:
an `expanded demand vector' of length $\Nsf\KsfActive$ as
\begin{align*}
\widetilde{\dv} %
:= \left( \Psi^{\tau_{j_1}\ominus d_{j_1}}(\mathbb{I}_\Nsf), 
                         \cdots, \Psi^{\tau_{j_\KsfActive}\ominus d_{j_\KsfActive}}(\mathbb{I}_\Nsf) \right),
\end{align*}
where the operator $\Psi^i$ denotes the $i$-times cyclic shift operator, which is the operator $\Psi$ applied $i$ times, and $\tau_j \ominus d_j$ denotes the difference of $\tau_j$ and $d_j$ modulo $\Nsf$;
an `expanded active user set' of size $\Nsf\KsfActive$ as
\begin{align*}
\widetilde{\Ic} 
:= \left( (j-1)\Nsf + \ell: j \in \Ic, \ell \in [3] \right); 
\end{align*}
and
\begin{align*}
\tau_\Ic \ominus d_\Ic := \left( \tau_{j_1} \ominus d_{j_1}, \cdots, \tau_{j_{\KsfActive}}\ominus d_{j_\KsfActive} \right).
\end{align*}
The server sends 
\begin{align*}
X  &= \big( X_1, X_2 \big) : \\
&X_1 = {E}^{(np)}\big( \widetilde{\Ic}, \widetilde{\dv}, F_{[\Nsf]}\big), \\
&X_2 = \Big( \tau_\Ic \ominus d_\Ic , \Ic \Big).
\end{align*}
Since sending $X_2$ does not scale with the file length $\Bsf$, thus the load in the private scheme is the same as the load size in the NONprivate scheme for sending $X_1$.

\paragraph*{Proof of Correctness}
Active user $k \in \Ic$ computes $\widetilde{\dv}$ and $\widetilde{\Ic}$ as a function of $X_2$,
and uses the $( (k-1)\Nsf+\tau_k )$-th decoding function in the NONprivate scheme to compute
\begin{align*}
F_{d_k} = {G}_{(k-1)\Nsf+\tau_k}^{(np)}(d_k,Z_k, X_1).
\end{align*}
It is clear that the decoder of the $k$-th user outputs the same file requested by the $\tau_k$-th virtual user of the $k$-th stack in the NONprivate scheme, so the desired file is retrieved.

\paragraph*{Proof of Privacy}
The proof follows from the fact that, for every $i \in [\Ksf]$, $\tau_i$ acts as one-time pad for $d_i$, which prevents any user $j \in \Ic$ but $j \neq i$ from getting any information about $d_i$.
Fix $\Bc \subseteq \Ic$, then
\begin{align*}
    &I(d_{\Ic\setminus\Bc}; X, d_{\Bc},Z_\Bc \mid {\Ic}, F_{[\Nsf]}) \\
    &\stackrel{\rm(a)}{=} I(d_{\Ic\setminus\Bc}; X, d_{\Bc} \mid {\Ic}, F_{[\Nsf]})\\
    &\stackrel{\rm(b)}{=} I(d_{\Ic\setminus\Bc}; \tau_\Ic \ominus d_\Ic, d_{\Bc} \mid {\Ic}, F_{[\Nsf]}) \\
    &= I(d_{\Ic\setminus\Bc}; d_{\Bc} \mid {\Ic}, F_{[\Nsf]}) + I(d_{\Ic\setminus\Bc}; \tau_\Ic \ominus d_\Ic \mid d_{\Bc}, {\Ic}, F_{[\Nsf]}) \\
    &\stackrel{\rm(c)}{=} 0
\end{align*}
where the equality follows because (a) $Z_\Bc$ is a function of $F_{[\Nsf]}$; (b) $X$ is a function of $(F_{[\Nsf]}, \tau_\Ic \ominus d_\Ic, \Ic)$; (c) the user demands and user randomness are independent.
This shows that the derived scheme satisfies the privacy condition.

\section{Proof of Constant Gap for Hotplug with Privacy}
\label{app:constantgaponhotplugprivacy}

The main idea is to leverage the following chain of inequalities
\begin{align*}
\frac{1}{\mathsf{ClassicalGap}} \cdot  \underbrace{\Rsf(\Msf;\KsfActive,\KsfActive,\Nsf)}_{\text{achievable classical $\KsfActive$ users}}
\leq \underbrace{\Rsf^\star(\Msf;\KsfActive,\KsfActive,\Nsf)}_{\text{optimal classical $\KsfActive$ users}}
\leq \underbrace{\Rsf^\star(\Msf;\KsfActive,\Ksf,\Nsf)}_{\text{optimal hotplug}}
\leq \underbrace{\Rsf(\Msf;\KsfActive,\Ksf,\Nsf)}_{\text{achievable hotplug}},
\end{align*}
and bound 
\begin{align*}
\mathsf{HotplugGap} = 
\sup \frac{\Rsf(\Msf;\KsfActive,\Ksf,\Nsf)}{\Rsf^\star(\Msf;\KsfActive,\Ksf,\Nsf)}
\leq \mathsf{ClassicalGap} \cdot
\sup \frac{\Rsf(\Msf;\KsfActive,\Ksf,\Nsf)}{\Rsf(\Msf;\KsfActive,\KsfActive,\Nsf)}.
\end{align*}
If we can find a hotpulog scheme for which the achevable load $\Rsf(\Msf;\KsfActive,\Ksf,\Nsf)$ is not a function of $\Ksf$, then we conclude $\mathsf{HotplugGap}=\mathsf{ClassicalGap}$.

For the case without privacy, we used "decen+" $\Rsf(\Msf;\KsfActive,\Ksf,\Nsf) = \frac{1-\mu}{\mu}\Big( 1-(1-\mu)^{\min(\KsfActive,\Nsf)} \Big)$, to prove that the gap with hotplug is the same as in the classical (not hotplug) case, as in Appendix~\ref{sec:gap hot no pri}.

For the case with privacy we can use PK+" in Theorem~\ref{thm:extensionPK} to show  to prove that the gap with hotplug is the same as in the classical (not hotplug) case in Theorem~\ref{thm:PK}. To notice is that the load with hotplug is increasing in $\KsfActive$, so the classical case is the ``worst case''. Details are not reported for brevity.

\bibliographystyle{ieeetr}
\bibliography{2023J-ym-v5}

\end{document}